\documentclass[12pt,draftclsnofoot,onecolumn]{IEEEtran}
\pdfoutput=1
\usepackage{cite}
\usepackage{amsmath,amssymb,amsfonts}
\usepackage{mathtools}
\usepackage{algorithmic}
\usepackage{algorithm2e}
\usepackage{graphicx}
\usepackage{textcomp}
\usepackage{gensymb}
\usepackage{xcolor}
\usepackage{subfigure}
\usepackage{xcolor}
\usepackage{float}
\usepackage{multicol}
\usepackage{cuted}
\usepackage{mathtools,amssymb,lipsum}
\usepackage[belowskip=-18pt,aboveskip=0pt]{caption}

\newtheorem{theorem}{Theorem}
\newtheorem{lemma}{Lemma}
\newtheorem{corollary}{Corollary}

\begin{document}

\title{Uncoordinated Spectrum Sharing in Millimeter Wave Networks Using Carrier Sensing}     
\author{Shamik Sarkar,~\IEEEmembership{Student Member,~IEEE},
        Xiang~Zhang,~\IEEEmembership{Student Member,~IEEE},
        Arupjyoti Bhuyan,~\IEEEmembership{Senior Member,~IEEE},
        Mingyue Ji,~\IEEEmembership{Member,~IEEE},
        and Sneha~Kumar Kasera,~\IEEEmembership{Senior Member,~IEEE}
\thanks{Part of this work was presented in the IEEE $54^{\text{th}}$ Asilomar Conference on Signals, Systems, and Computers, 2020}
\thanks{S. Sarkar and S. Kasera are with the School of Computing, University of Utah, Salt Lake City, UT 84112, USA (E-mail: Shamik.Sarkar@utah.edu, kasera@cs.utah.edu).}
\thanks{X. Zhang and M. Ji are with the Department of Electrical and Computer Engineering, University of Utah, Salt Lake City, UT 84112, USA (E-mail:~\{xiang.zhang, mingyue.ji\}@utah.edu)}
\thanks{A. Bhuyan is with the INL Wireless Security Institute, Idaho National Laboratory, Idaho Falls, ID 83415, USA (E-mail:arupjyoti.bhuyan@inl.gov).}
\thanks{This work is supported through the INL Laboratory Directed Research \& Development (LDRD) Program under DOE Idaho Operations Office Contract DE-AC07-05ID14517. The work of M. Ji and S. Kasera is also supported 
by NSF Award 1824558.}
}

\maketitle

\begin{abstract}
We propose using Carrier Sensing (CS) for distributed interference management in millimeter-wave (mmWave) cellular networks where spectrum is shared by multiple operators that do not coordinate among themselves. In addition, even the base station sites can be shared by the operators. 
We describe important challenges in using traditional CS in this setting and propose enhanced CS protocols to address these challenges. Using stochastic geometry, we develop a general framework for downlink coverage probability analysis of our shared mmWave network in the presence of CS and derive the downlink coverage probability expressions for 
several
CS protocols. To the best of our knowledge, our work is the first to investigate and analyze (using stochastic geometry) CS for mmWave networks with spectrum and BS sites shared among non-coordinating operators. We evaluate the downlink coverage probability of our shared mmWave network using simulations as well as 
numerical examples based on our analysis.
Our evaluations show that our proposed enhancements lead to 
an improvement in downlink coverage probability, 
compared to the downlink coverage probability with no CS,
for higher values of signal-to-interference and noise ratio (SINR). Interestingly, our evaluations also reveal that for lower values of SINR, not using any CS is the best strategy in terms of the downlink coverage probability.
\end{abstract}


\section{Introduction} \label{intro}
The abundance of available spectrum beyond 24 GHz, commonly known as the millimeter-wave (mmWave) band 
\cite{pi2011introduction},
is a major contributor to the significant bandwidth improvements that 5G brings. 
Importantly, the necessity of beam-based transmissions for mmWave, instead of the traditional sector-based transmissions, engenders the opportunity of spatial spectrum sharing \cite{boccardi2016spectrum}. Spectrum sharing in mmWave networks is essential for enabling \emph{private cellular networks} \cite{
private5g}, which is one of the crucial use cases of 5G. Operators of private 5G networks, e.g., airports, factories, would be sharing spectrum with the mobile operators (public 5G cellular operators). 
Multiple co-existent private 5G networks can also share spectrum among themselves. 

In some scenarios, licensed bands can be used for spectrum sharing, e.g., a private 5G network using a dedicated slice of the co-existent public mobile operator's licensed spectrum \cite{private5g}.
However, from a general viewpoint, the problem of spectrum sharing using unlicensed bands is of much broader interest.  
Sharing of unlicensed spectrum would be widespread among operators of the same radio access technology (RAT), e.g., in the cases of spectrum sharing between operators of private-public, private-private, and public-public cellular networks. Thus, unlike various existing works \cite{lagen2018listen,
mosleh2020dynamic,song2019cooperative,
nekovee2016distributed
} that have investigated unlicensed spectrum sharing between WiGig and 5G, we investigate the problem of unlicensed spectrum sharing among different operators of the same RAT, e.g., 5G. Henceforth, we use the term `operator' in a general sense, without caring about whether it is private or public, unless explicitly stated. The very nature of unlicensed usage of the spectrum creates new opportunities for uncoordinated sharing, that has two important advantages.
First, uncoordinated spectrum sharing has the flexibility of ad hoc spectrum usage without going through an extensive and time-consuming process via any central coordination. Second, there is no vulnerable central point of attack. The use of 
a central coordinator
creates a vulnerable central point of attack that could be targeted by adversaries for unfair provisioning or denial of service. Thus, we focus on uncoordinated sharing of unlicensed mmWave spectrum.

However, along with the advantages, uncoordinated spectrum sharing brings in the significant challenge of \emph{distributed interference management}. In our shared mmWave network, interference management is essential because, in the absence of coordination, different operators may simultaneously use the same portions of the unlicensed spectrum. Additionally, different operators may share the same strategically important base station (BS) sites/towers. 
BS site sharing by operators increases the possibility of strong interferers; for a user equipment (UE), there can be an interferer (above/below its associated BS) with interference as strong as the signal power.

In this paper, we investigate Carrier Sensing (CS) for distributed interference management in a mmWave network, with the spectrum and BS sites shared among non-coordinating operators. 
We determine that the choice regarding whether 
CS is incorporated at the transmitter or at the receiver is of prime importance. Specifically, we observe that when CS is used at the transmitter (CST) 
in our shared mmWave network, 
the directionality of mmWave signals amplifies the hidden terminal problem. 
Furthermore, multiple BSs at different heights on the same tower cannot sense each other’s  
transmission. 
Hence, we propose using CS at the receiver (CSR) to overcome the limitations of CST. 
Intuitively, with directional transmissions, the receivers are in the best positions to assess whether or not they experience interference during reception. 
However, even CSR cannot prevent interference that may start during the data transmission phase, right after the CS phase is over. 
To tackle this problem, we propose an enhanced version of CSR, which we call directional CSR with announcements (\textit{d}CSRA), that enables interference protection beyond CS. The main idea in \textit{d}CSRA is that, if a receiver senses the channel to be free, it also sends out a few broadcast announcements to prevent BSs (that may cause interference) from starting their downlink transmissions, while the announcing node is receiving downlink signals.


Using stochastic geometry \cite{elsawy2013stochastic}, we develop a general framework for downlink coverage probability analysis of our shared mmWave network in the presence of CS. Our framework is not specific to any particular CS protocol, and it can be used for any 
CS schemes that we discuss in this work. Due to the suitability of CSR in our mmWave network, we use our framework to derive the analytical expressions for downlink coverage probability with different CSR schemes.
Through extensive evaluations, in Section~\ref{results}, we show that the superiority of a particular CS protocol, over other CS protocols, is dependent on the signal-to-interference and noise ratio (SINR). Hence, our coverage probability analysis serves as a useful tool for deciding which protocol to use under a particular situation, without running extensive simulations.

Using simulations, we validate our coverage probability analysis for the CSR schemes. We also evaluate the impact of various factors (sensing threshold, BS site overlap) on the coverage probability of our shared mmWave network. Our evaluation results show that CSR schemes are always better than the CST schemes. In the high SINR regime, CSR schemes are advantageous over not using any CS. However, interestingly, our evaluations also reveal that in the low SINR regime not using any CS is the best strategy for achieving higher coverage probability. In summary, although CS should improve the performance of a shared spectrum network by avoiding interference, we find that CS may not result in optimal behavior under all circumstances because it avoids interference at the cost of reduced transmissions. 
CS is advantageous only when the benefit of avoiding interference outweighs the disadvantage of reduced transmissions.

\textbf{Related Work}: Research around mmWave spectrum sharing has progressed primarily in the following two categories.

\textit{Spectrum sharing between same RATs:} The idea of mmWave spectrum pooling among mobile operators has been explored in various works \cite{boccardi2016spectrum,shokri2016spectrum,
gupta2016feasibility,rebato2017hybrid,jurdi2018modeling}. These works have demonstrated that spectrum pooling among mobile operators can significantly boost their downlink throughput. 
In fact, this advantage of spectrum pooling can be achieved even without any coordination among the mobile operators, as long as the individual networks have a comparable density of BSs \cite{gupta2016feasibility}.
A recent work along this line of research has proposed a game-theoretic approach for distributed beam scheduling for uncoordinated sharing of 
mmWave spectrum among different operators \cite{zhang2020non}.
However, if spectrum pooling is performed using licensed bands, then coordination among the mobile operators is essential because, a licensed mobile operator would be willing to share part of its spectrum only when its licensed spectrum is unused \cite{boccardi2016spectrum}; otherwise, the quality of service of its subscribed users may be affected.
Various centralized and distributed strategies have been proposed for coordination among the mobile operators \cite{rebato2018spectrum,boccardi2016spectrum,shokri2016spectrum}.

\textit{Spectrum sharing across different RATs:} Unlicensed mmWave bands can be shared by WiFi and mobile operators \cite{nekovee2016distributed,
lagen2018listen,
song2019cooperative
}.
However, due to the use of directional beams in mmWave, the spectrum sharing solutions for LTE-LAA cannot be applied directly in the mmWave bands. In general, unlicensed spectrum sharing among different RATs requires distributed interference management because coordination between operators of different RATs is unlikely. The interference can be managed by running distributed algorithms on the BSs, for scheduling the downlink time slots of the associated UEs \cite{nekovee2016distributed
}. Alternately, distributed interference management can be performed by adding intelligence/adaptability in the CS protocol \cite{lagen2018listen,
song2019cooperative}.

In our work, we draw insights from these existing works and build upon their contributions. However, 
our work is the first to investigate and analyze 
CS in a mmWave network with shared BS sites and allows spectrum sharing among the operators without any coordination. 

In summary, we make the following important contributions in this paper:
\begin{itemize}
    \item We investigate CS protocols for distributed interference management in a mmWave network, with the spectrum and BS sites shared among operators having no coordination. We describe several drawbacks of traditional CST in our setup and propose the use of CSR.
    \item We describe that in the absence of coordination among BSs,
    two types of interferers may exist and introduce the notion of hidden interferers and deaf interferers. We explain that neither CST nor CSR can prevent interference from the deaf interferers. 
    We propose an enhanced version of CSR to 
    reduce the interference from the deaf interferers. 
    \item Based on stochastic geometry, we develop a general framework for downlink coverage probability analysis of our shared mmWave network in the presence of CS. Using our framework, we derive the coverage probability expressions for the non CS scheme, where no CS is used, and
    for different CSR schemes, including the proposed dCSRA.
    \item We validate our coverage probability analysis using simulations. We show that our analytical results and the simulations results are very close to each other for different values of SINR. 
    Using our evaluation results, we demonstrate that our 
    protocols lead to an
    improvement in downlink coverage probability,
    over no CS, 
    for higher values of SINR. 

\end{itemize}


\textit{Organization of the remaining paper}: In Section~\ref{system_model}, we describe the system model of our considered mmWave network. In Section~\ref{methodology}, we introduce the ideas of hidden and deaf interferers, and explain the capabilities of CST and CSR in terms of dealing with hidden and deaf interferers. In this section, we also present the details of our proposed \textit{d}CSRA protocol. In Section~\ref{coverage_prob}, we analyze the downlink coverage probability of our shared mmWave network. We present the evaluation results in Section~\ref{results}, and Section~\ref{conclusions} provides the conclusions.

\section{System model} \label{system_model}
We consider an unlicensed band of $W$ MHz that is shared by $M$ operators. There is no coordination among the operators as well as between different BSs of the same operator. 


\textit{BS site sharing model}: We consider that some BS sites are shared by multiple operators. For modeling shared BS sites, we adopt the approach used in \cite{jurdi2018modeling}. We use $\mathcal{O}= \{1,2,...,M\}$ to denote the set of operators, and $\mathcal{P(O)}$ is the power set of $\mathcal{O}$. We use $\Phi_{\mathcal{S}}$, a homogeneous Poisson Point Process (PPP) with density $\lambda_{\mathcal{S}}$, to represent the locations of BS sites that are shared by the elements of $\mathcal{S} \in \mathcal{P(O)}$. $\{\Phi_{\mathcal{S}}\}$ represent the collection of all the different $\Phi_{\mathcal{S}}$, where the elements of $\{\Phi_\mathcal{S}\}$ are independent homogeneous PPPs. 
BSs of operator $m$ form the point process $\Phi_m$ with density $\lambda_m$, where $\Phi_m = \bigcup_{\mathcal{S}: m \in \mathcal{S}} \Phi_\mathcal{S}$ and $\lambda_m = \sum_{\mathcal{S}: m \in \mathcal{S}} \lambda_\mathcal{S}$. Due to the superposition property of PPPs, $\Phi_m$ is also PPP  \cite{jurdi2018modeling}. For any $\mathcal{S}, \mathcal{S'} \in \mathcal{P(O)}, \Phi_{\mathcal{S}} \cap \Phi_{\mathcal{S'}} = \phi$, almost surely, because a collection of independent PPPs have no point in common \cite{jurdi2018modeling}. Thus, in $\Phi_m = \bigcup_{\mathcal{S}: m \in \mathcal{S}} \Phi_\mathcal{S}$, we do not double count the BSs of network $m$.
We use $X_{j,m}$ for the location of the $j^{th}$ BS of network $m$. With a slight abuse of notation, we use $X_{j,m}$ for the BS itself, if there is no confusion. 
We use $U_{x,y}$ to indicate the location of a UE having Cartesian coordinate $(x, y)$.

\textit{Blocking model}: 
The blocking of each link is independent and identically distributed (i.i.d) as: a link is in line-of-sight (LoS) with probability $p_{L}(r) = e^{-\beta r}$, and in non line-of-sight (NLoS) with probability $p_{N}(r) = 1 - p_{L}(r)$, where $r$ is the link distance in meters \cite{andrews2016modeling}. Here, $\beta$ is the blocking parameter of mmWave signals.

\textit{Path loss model}: Free space path loss at a distance of $r$ meters from a transmitter is modeled as $C_{\tau}r^{-\alpha_{\tau}}$, where $C_{\tau}$ is the path loss at a reference distance of 1 meter, $\alpha_{\tau}$ is the path loss exponent (PLE) of mmWave signals \cite{rappaport2013millimeter}, and $\tau \in \{L \text{ (LoS)}, N \text{ (NLoS)}\}$. I.e., if $\tau = L$, then $C_{\tau} = C_L$, $\alpha_{\tau} = \alpha_L$, and if $\tau = N$, then $C_{\tau} = C_N$, $\alpha_{\tau} = \alpha_N$. In the remaining paper, we follow the convention that, for a variable associated with a link between a BS at $X_{j,m}$ and another BS at $X_{b,n}$, we use the variable with subscript $j,m$ and superscript $b,n$. In contrast, for a link between a BS at $X_{j,m}$ and a UE at $U_{x,y}$, we use the UE's coordinates, $(x,y)$, as the superscript in parentheses. However, for compactness, we omit the superscript when the UE is at $(0,0)$, i.e., $(x,y) = (0,0)$. 
Using this convention, $C_{\tau}$ is either $C_{\tau(j,m)}^{b,n}$ or $C_{\tau(j,m)}^{(x,y)}$, and $\alpha_{\tau}$ is either $\alpha_{\tau(j,m)}^{b,n}$ or $\alpha_{\tau(j,m)}^{(x,y)}$, depending on whether the link is between a pair of BSs or between a BS and a UE. 

\textit{Fading model}: 
We consider that each link undergoes independent Rayleigh fading. 
Hence, the loss in received power due to small scale fading is modeled as an exponential random variable, 
$F_{j,m}^{b,n}$ or $F_{j,m}^{(x,y)}$, depending on whether the link is between a pair of BSs or between a BS and a UE. Without loss of generality, we assume that $\mathbb{E}\big[F_{j,m}^{b,n}\big] = \mathbb{E}\big[F_{j,m}^{(x,y)}\big] = 1$.
Finally, fading from the co-located BSs are considered independent as they would be mounted at different heights. 

\textit{Association model}: We consider 
a system where a UE is served only by its subscribed operator. A UE associates with a BS, among all the BSs of its subscribed operator, that provides maximum received signal power, averaged over the fading randomness \cite{gupta2016feasibility}. The associated BS of a UE may not be the nearest BS to the UE. The nearest BS to the UE may have a NLoS link with the UE, providing lesser power than a BS having LoS link with the UE, but at a farther distance.
We assume that all the BSs transmit with power $P_X$, and the UEs transmit with power $P_U$.

\textit{Antenna gain model}: We consider uniform linear antenna arrays (ULA)\cite{andrews2016modeling} with $n_{BS}$ and $n_{UE}$ antenna elements at each of the BSs and UEs, respectively. Beam steering is done only in the horizontal direction while the vertical steering angle is always fixed (assuming all UEs are at ground level). We consider codebook based analog beamforming \cite{andrews2016modeling}, i.e., the beam direction is chosen as the one that provides maximum signal strength during beam training, among a set of predefined beams specified in the codebook. For analytical tractability, we assume that an antenna array's radiation pattern follows a step function with a constant gain, $M_{BS}$ (for BS), $M_{UE}$ (for UE), in the main lobe, and a constant gain, $m_{BS}$ (for BS), $m_{UE}$ (for UE), in the side lobe \cite{andrews2016modeling}. For the gains,
we use the following expressions, based on \cite{rebato2019stochastic}: $M_{BS} = 10^{0.8}n_{BS}$, $M_{UE} = 10^{0.8}n_{UE}$, $m_{BS} = 1/\sin^{2}(\frac{3\pi}{2\sqrt{n_{BS}}})$, and $m_{UE} = 1/\sin^{2}(\frac{3\pi}{2\sqrt{n_{UE}}})$.
We consider single stream downlink transmissions, i.e., a BS serves only one UE in a time slot.
From a UE's viewpoint, the misalignment 
between its main lobe and an interfering BS's main lobe is uniformly random in $[0, 2\pi]$, in the azimuth. Thus, the combined antenna gain from an interfering BS at $X_{j,m}$ to a UE at $U_{x,y}$ is modeled as a random variable, $G_{j,m}^{(x,y)}$, distributed as:
\begin{equation} \label{eq:antenna_gain}
    G_{j,m}^{(x,y)} = \begin{cases}
    M_{BS}M_{UE} \text{   w.p.   }  \big(\frac{\theta_{BS}}{2\pi}\big) \big(\frac{\theta_{UE}}{2\pi}\big) \\[-7pt]
    M_{BS}m_{UE} \text{   w.p.   }  \big(\frac{\theta_{BS}}{2\pi}\big) \big(1 - \frac{\theta_{UE}}{2\pi}\big) \\[-7pt]
    m_{BS}M_{UE} \text{   w.p.   }  \big(1 - \frac{\theta_{BS}}{2\pi}\big) \big(\frac{\theta_{UE}}{2\pi}\big) \\[-7pt]
    m_{BS}m_{UE} \text{   w.p.   }  \big(1 - \frac{\theta_{BS}}{2\pi}\big) \big(1 - \frac{\theta_{UE}}{2\pi}\big) 
    \end{cases}
\end{equation}
Here $\theta_{BS}$ and $\theta_{UE}$ are the main lobe beamwidth (in radians) for the BSs and the UEs, respectively. While computing the downlink SINR, we assume that a UE and its associated BS have gone through the beam training, and their antennas are aligned for maximum gain, which is $M_{BS}M_{UE}$.

\textit{Performance metric}: Our focus 
is to investigate the performance of various CS protocols in a shared mmWave network. For that, we consider a UE's downlink coverage probability, $P_c(Z); Z > 0$, which is the probability that a UE's SINR is above $Z$, as the performance metric. 
$P_c(Z)$ is the complimentary cumulative distribution function (CCDF) of the UE's SINR.

\begin{table*}[t] 
\caption{Usage of symbols}
\label{tab:symbols}
    \centering
    {
    \begin{tabular}
    {|l|l|}
        \hline
        \textbf{Symbol} & \textbf{Meaning} \\
        \hline
        $M$, $\mathcal{O}$, $\mathcal{P(O)}$ & Number of operators, set of operators, and power set of $\mathcal{O}$ \\
        \hline
        $\Phi_{\mathcal{S}}$, $\lambda_{\mathcal{S}}$  & PPP with density $\lambda_{\mathcal{S}}$, representing BS sites shared by elements of $\mathcal{S} \in \mathcal{P(O)}$ \\
        \hline
        $\Phi_m$, $\lambda_m$ & PPP with density $\lambda_m$, representing BS sites of operator $m$ \\
        \hline
        $X_{j,m}$ and $U_{x,y}$ & Location of the $j^{th}$ BS of network $m$, and location of a UE with Cartesian coordinate $(x,y)$ \\
        \hline
        $p_{L}(r)$, $p_{N}(r)$; $\beta$ & Probability of an $r$ meters link being LoS, NLoS, respectively; mmWave blocking parameter
        \\
        \hline
        $C_{L}$, $C_{N}$ & mmWave path loss at 1 meter from the transmitter, for LoS and NLoS links, respectively; \\
        $C_{\tau(j,m)}^{b,n}$ or $C_{\tau(j,m)}^{(x,y)}$ & Notation for $C_{\tau}$; $\tau \in \{L, N\}$ when link is between a pair of BSs or between a BS and a UE  \\
        \hline
        $\alpha_{L}$, $\alpha_{N}$ & Propagation path loss exponent of mmWave signals for LoS and NLoS links, respectively; \\
        $\alpha_{\tau(j,m)}^{b,n}$ or $\alpha_{\tau(j,m)}^{(x,y)}$ & 
        Notation for $\alpha_{\tau}$; $\tau \in \{L, N\}$ when link is between a pair of BSs or between a BS and a UE \\
        \hline
        $F_{j,m}^{b,n}$ or $F_{j,m}^{(x,y)}$  & Fading random variable; link is between a pair of BSs or between a BS and a UE \\
        \hline
        $P_X$, $P_U$ & Transmit power of the BSs and UEs, respectively \\
        \hline
        $n_{BS}$, $n_{UE}$ & Number of antenna elements at the BSs and UEs, respectively \\
        \hline
        $M_{BS}, M_{UE}; m_{BS}, m_{UE}$ & Main lobe gain for BSs, UEs, respectively; side lobe gain for BSs, UEs, respectively  \\
        \hline
        $\theta_{BS}$, $\theta_{UE}$ & Main lobe beam width (radians) for the BSs and UEs, respectively \\
        \hline
        $G_{j,m}^{(x,y)}$ & Combined antenna gain from a BS at $X_{j,m}$ to a UE at $U_{x,y}$ \\
        \hline
        $t_s$, $p_{\mathbb{T}}$ & Duration of a downlink time slot, transmission probability of a BS in a time slot  \\
        \hline
        $P_{th}$, $P_{th}^{A}$ & Sensing thresholds for any active downlink transmissions and announcements, respectively\\
        \hline
        $A_{j,m}^{b,n}$, $A_{j,m}^{(x,y)}$  & Antenna gain during sensing between a BS at $X_{j,m}$ and a sensing node. \\
        & The sensing node is a BS at $X_{j,m}$, and a UE at $U_{x,y}$, respectively \\
        \hline
        $N_0$, $W$, $N_F$, $N_f$  & Noise PSD per Hz, receiver bandwidth, noise figure, and receiver noise floor\\
        \hline
        $B_{x,y}(r)$, $B_0(r)$  & Ball of radius $r$ centered at $U_{x,y}$ and $(0,0)$, respectively \\
        \hline
        $B_0(D_{L}(r))$, $B_0(D_{N}(r))$ & Interference exclusion zones due to association rule; association distance is $r$ meters \\
        \hline
        $\mathcal{R}_{L,h}$, $\mathcal{R}_{L,d}$,$\mathcal{R}_{N,h}$, $\mathcal{R}_{N,d}$  & Interference exclusion zones due to CS from LoS hidden, LoS deaf, NLoS hidden, and \\
        & NLoS deaf interferers, respectively. \\
        \hline
        $\bar{N}_c$ & Average number of contenders to a CS node \\
        \hline
    \end{tabular}
    }
\end{table*}
\section{Carrier Sensing for distributed interference management} \label{methodology}


Since there is no coordination among the BSs, the usage of random medium access protocols is a promising solution in managing the interference in a distributed manner. Among the well known random access protocols, we choose CS over ALOHA and slotted ALOHA because ALOHA suffers from low throughput and slotted ALOHA requires synchronization of time slots among all the BSs. Thus, we consider CS as the random access protocol in our considered mmWave network, i.e., all the BSs belonging to the different operators use CS. 

In this section, first, we present a brief overview of the various CS schemes. Although CS is the appropriate choice, it cannot eliminate all the undesired interference. This unavoidable interference is characterized next, in Section~\ref{interference_types}, by introducing the notion of hidden and deaf interferers. Then, in Section~\ref{protocols}, we describe the capabilities of different CS protocols, in terms of avoiding interference from hidden and deaf interferers. Finally, we propose a new CS protocol, \textit{d}CSRA, that is more capable than the other schemes in terms of avoiding interference.


\subsection{Carrier Sensing overview} \label{cs_overview}
Traditionally (WiFi, 
LTE-LAA), 
CS is done at the transmitter, where the transmitter listens for any ongoing transmission before transmitting its own signals. If the transmitter identifies the channel to be occupied, it postpones its transmission; otherwise, it transmits its signal. 
We explain in Section~\ref{protocols} that CST has several drawbacks
and propose using 
CSR. 
CSR has been investigated for mmWave networks in \cite{lagen2018listen}; however, unlike our problem of spectrum sharing between the same RATs, this work considers spectrum sharing across different RATs. Additionally, the scenario of co-located BSs is not considered in \cite{lagen2018listen}.
Both CST and CSR can be performed omnidirectionally or directionally. 
With omnidirectional CS, the sensing node listens for any ongoing transmission in all directions. 
In contrast, with directional CS, the sensing node measures the channel power only in the direction of its main lobe. Considering both the choices of sensing location and direction, we have four possibilities: omnidirectional CST (\textit{o}CST), directional CST (\textit{d}CST), omnidirectional CSR (\textit{o}CSR), and directional CSR (\textit{d}CSR). We assume that all the BSs in the shared mmWave network use the same variant of CS, with the same sensing threshold.

\subsection{Interference in spite of Carrier Sensing} \label{interference_types}
While the purpose of CS is to avoid interference, it cannot eliminate interference completely. To characterize this unavoidable interference, we introduce the notion of hidden interferers and deaf interferers. Our characterization of the interferers is defined with respect to a UE. For the purpose of explaining hidden and deaf interferers, we consider a \emph{typical UE}, located at the center of the considered region, $(0, 0)$, as the reference UE.
The difference between hidden and deaf interferers arises due to the timing characteristics of our mmWave network. 
We first explain these timing characteristics and then present the ideas of hidden and deaf interferers.

\begin{figure}[t] 
    \centering
        {\includegraphics[scale=0.32]{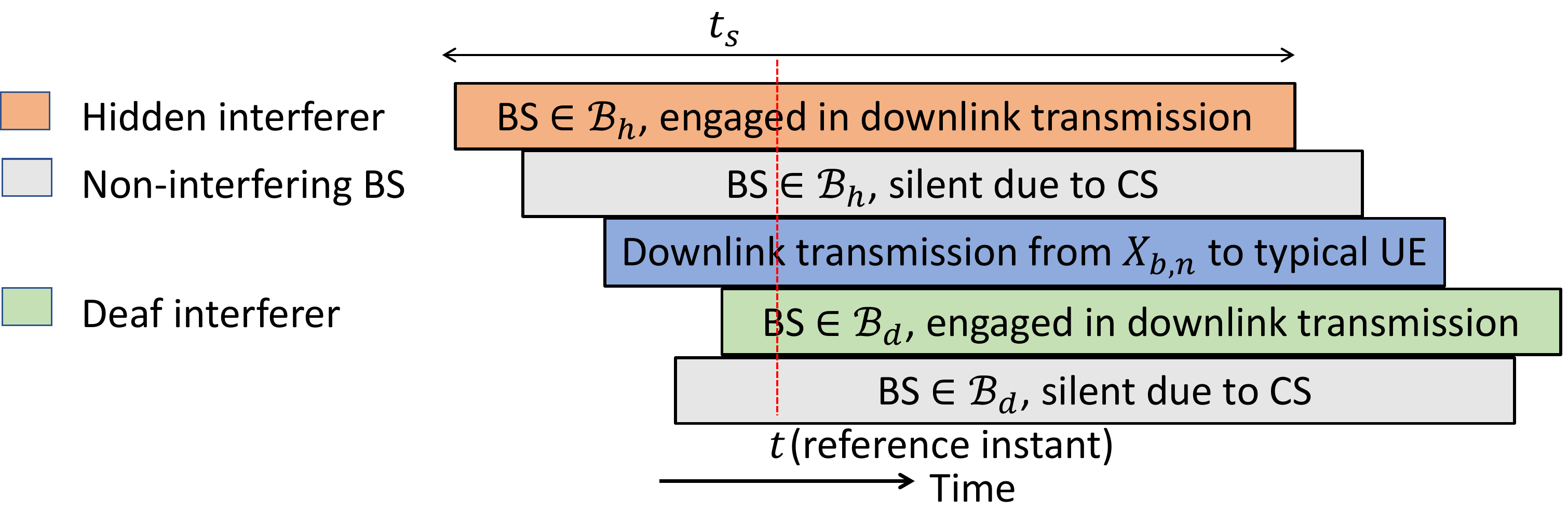}}
    \caption{Timing characteristics of our mmWave network.
    }
    \label{fig:hidden_and_deaf}
\end{figure}

\subsubsection{Timing characteristics}
Since the BSs use single-stream downlink channels, they serve their associated UEs using a time division multiple access (TDMA) scheme with a round-robin scheduling. We assume that all the BSs use the same duration for their downlink time slots, denoted by $t_s$. However, the time slots of different BSs may not be aligned, as shown with the help of an example in Fig.~\ref{fig:hidden_and_deaf}. This figure shows that the time slots of all the BSs are the same, but the beginning of their time slots are not aligned. At the beginning of the time slots, the sensing nodes perform CS, with sensing time negligible compared to $t_s$. If a sensing node (a scheduled UE or its associated BS) measures the channel power (time-averaged over fading randomness) to be below the sensing threshold, $P_{th}$, then the BS initiates downlink transmission to the scheduled UE; otherwise, the transmission is deferred. The sensing node senses the channel again after a duration of $t_s$, at the beginning of the next time slot. Note that, if the sensing is performed by the UEs, then at each time slot only the scheduled UE for that time slot performs the CS. To summarize, a BS uses TDMA for resource sharing among all the UEs associated with it, and CS is used for contention-based resource sharing among different BSs.
We assume that a BS can be silent only because of CS, 
not due to the lack of downlink load/requests. 

Now, consider a downlink time slot of the typical UE, where the sensing node has assessed the channel to be free, and the typical UE is receiving downlink signals from its associated BS, located at $X_{b,n}$. In the context of downlink transmissions, the BSs are the transmitters and the UEs are the receivers\footnote{While our analysis can be extended for uplink transmissions, in this paper we consider downlink transmissions only.}. Thus, the sensing node is the typical UE itself or its associated BS, depending on whether CSR or CST is used. At any point of time, $t$, during this downlink time slot of the typical UE, all the BSs (except the typical UE's associated BS) can be divided into two classes: $\mathcal{B}_h$ and $\mathcal{B}_d$. $\mathcal{B}_h$ is the set of BSs whose last CS phase, prior to $t$, precedes the CS of BS at $X_{b,n}$. In contrast, $\mathcal{B}_d$ is the set of BSs whose last CS phase, prior to $t$, succeeds the CS of $X_{b,n}$. For example, in Fig.~\ref{fig:hidden_and_deaf}, the top two rows are time slots of BSs belonging to $\mathcal{B}_h$, and the bottom two rows are time slots of BSs belonging to $\mathcal{B}_d$.
Although we define $\mathcal{B}_h$ and $\mathcal{B}_d$ in terms of CS by BSs, in reality, the CS is performed by the BSs themselves or their scheduled UEs depending on whether CST or CSR is used. We use the phrase `CS by BS' for concisely defining $\mathcal{B}_h$ and $\mathcal{B}_d$. Additionally, the distinction between CST and CSR is not relevant for distinguishing $\mathcal{B}_h$ from $\mathcal{B}_d$. 
Next, we use $\mathcal{B}_h$ and $\mathcal{B}_d$ to define hidden and deaf interferers, respectively.


\begin{figure}[t]
    \centering
    \begin{subfigure}[Hidden and deaf interferers with CST]
    {\label{fig:hidden_cst}
    \includegraphics[scale=0.36]{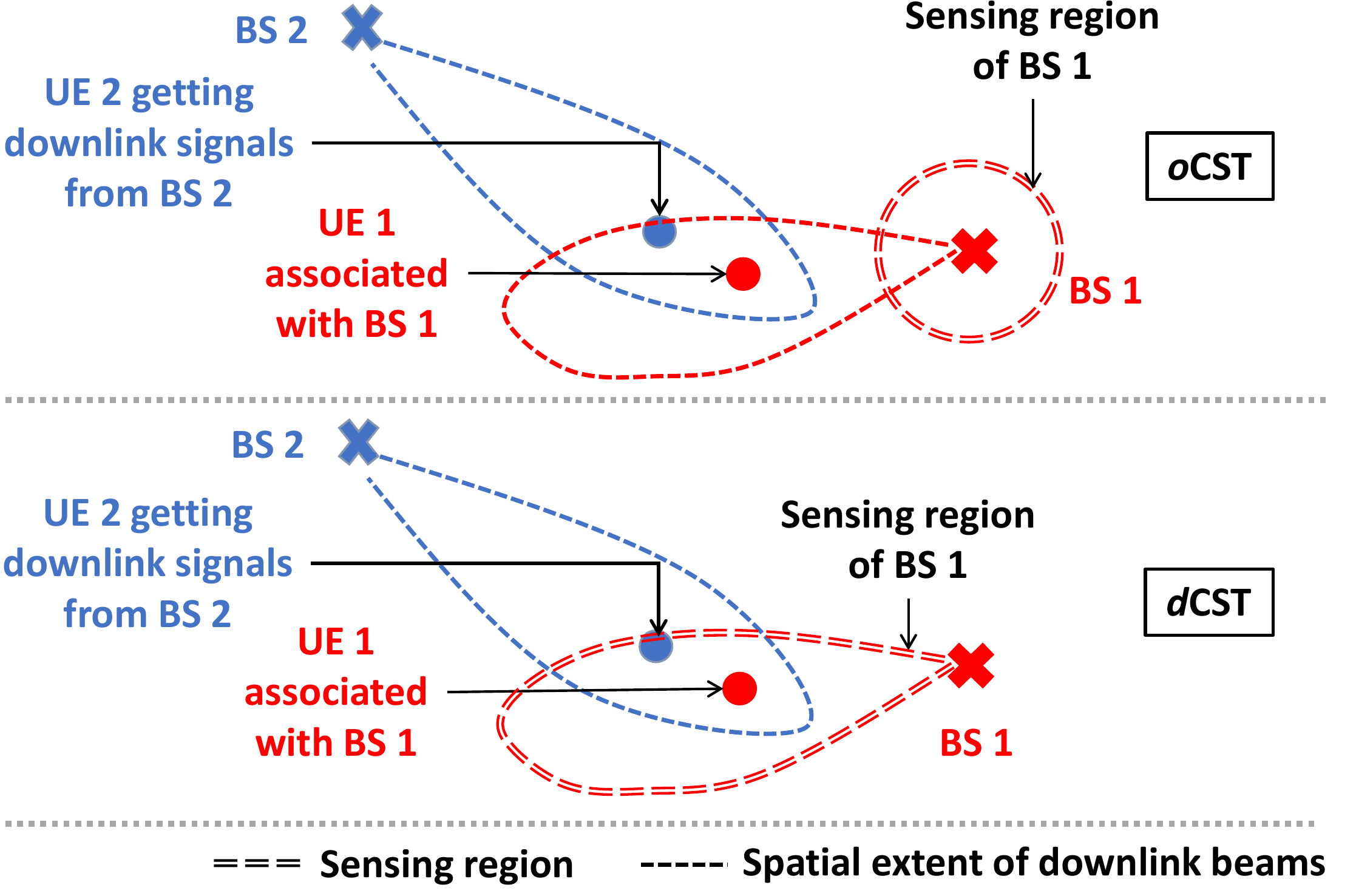}}
    \end{subfigure}
    \hspace{0.25in}
    \begin{subfigure}[Exposed terminal problem in CST]
    {\label{fig:exposed_cst}
    \includegraphics[scale=0.34]{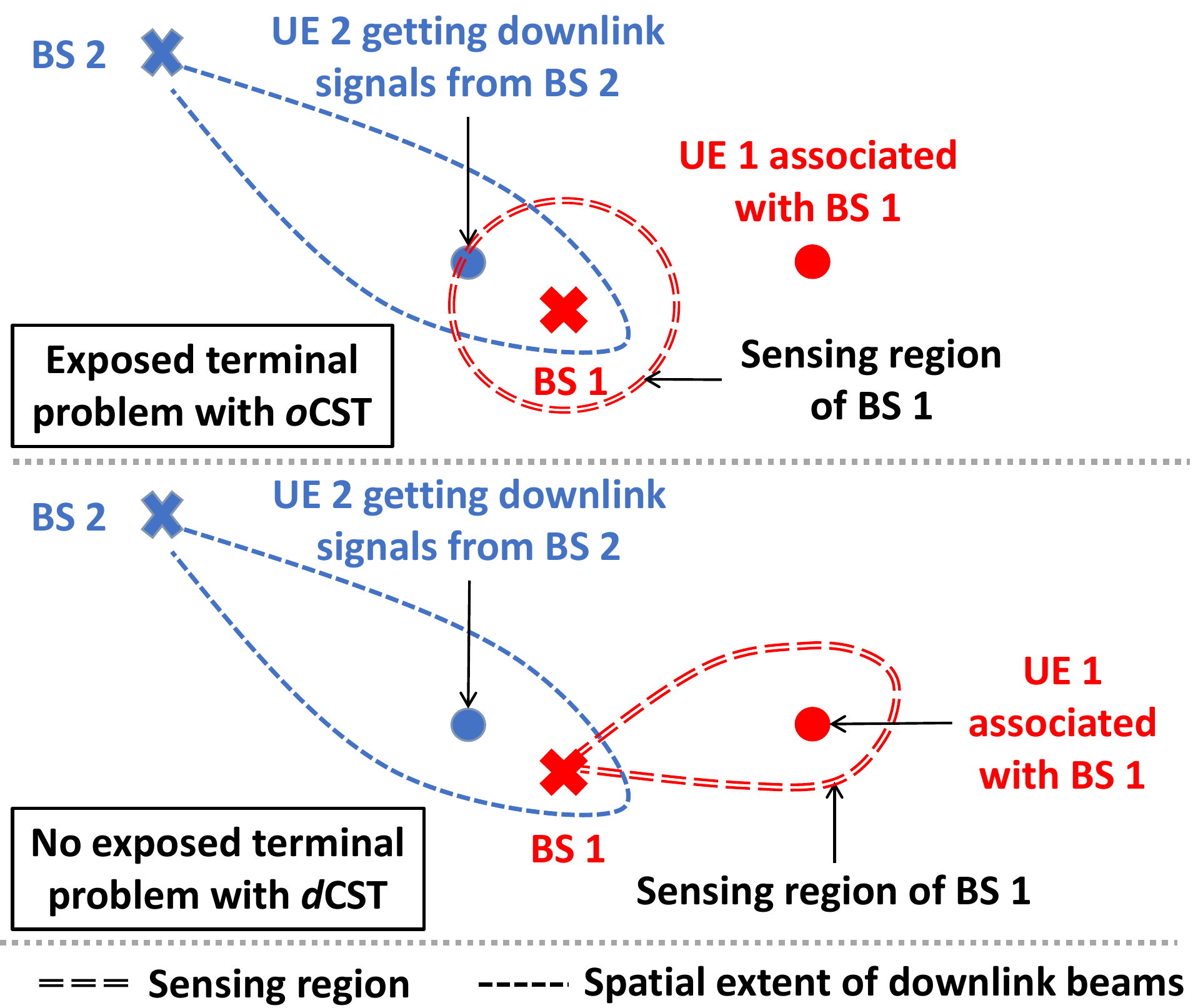}}
    \end{subfigure}
    \caption{
    (a) shows that neither \textit{o}CST nor \textit{d}CST can eliminate hidden and deaf interferers. (b) shows that \textit{d}CST can alleviate the exposed terminal problem significantly, compared to \textit{d}CST. 
    } 
\end{figure}

\subsubsection{Hidden interferers} \label{hidden_def}
Among the set of BSs in $\mathcal{B}_h$, a subset of it would not be engaged in downlink transmission due to their own respective CS. Let us denote this subset as $\mathcal{B}_{h,0}$, where the $0$ indicates no transmission. The remaining BSs in $\mathcal{B}_h$, denoted by $\mathcal{B}_{h,1}$,
would be engaged in downlink transmissions. For example, among the two BSs belonging to $\mathcal{B}_h$ in Fig.~\ref{fig:hidden_and_deaf}, one is silent and the other is active. We define the hidden interferers based on $\mathcal{B}_{h,1}$. Among the BSs belonging to $\mathcal{B}_{h,1}$, we call a BS to be a hidden interferer if it causes interference to the typical UE. We present the 
mathematical 
conditions for a BS to be a hidden interferer in the following.
\begin{lemma} \label{lemma:hidden_def}
A BS at $X_{j,m}$, belonging to $\mathcal{B}_{h,1}$, is a hidden interferer to the typical UE, if:
\begin{equation}\label{eq:hidden_bs_condition}
    \begin{aligned}
      & \text{\textbf{CST: }} C_{\tau(j,m)}^{b,n}A_{j,m}^{b,n} ||X_{j,m} - X_{b,n}||^{-\alpha_{\tau(j,m)}^{b,n}} < \frac{P_{th}}{P_{X}} \text{ \textbf{and} } C_{\tau(j,m)}F_{j,m}G_{j,m} ||X_{j,m}||^{-\alpha_{\tau(j,m)}} > \frac{N_{f}}{P_{X}} \\[-5pt]
      & \text{\textbf{CSR: }} C_{\tau(j,m)}A_{j,m} ||X_{j,m}||^{-\alpha_{\tau(j,m)}} < \frac{P_{th}}{P_{X}} \text{ \textbf{and} } C_{\tau(j,m)}F_{j,m}G_{j,m} ||X_{j,m}||^{-\alpha_{\tau(j,m)}} > \frac{N_{f}}{P_{X}}
    \end{aligned}
\end{equation}
where $A_{j,m}^{b,n}$ is the antenna gain during sensing between the BS at $X_{j,m}$ and the sensing BS at $X_{b,n}$. $A_{j,m}$ is the antenna gain during sensing between the BS at $X_{j,m}$ and the typical UE acting as the sensing node.  Recall from Section~\ref{system_model} that for any variable associated a link, we omit the superscript when the UE is at $(0,0)$. Hence, $C_{\tau(j,m)}^{(0,0)} = C_{\tau(j,m)}$, $A_{j,m}^{(0,0)} = A_{j,m}$, $F_{j,m}^{(0,0)} = F_{j,m}$, and $G_{j,m}^{(0,0)} = G_{j,m}$. $N_{f} = N_0 W + N_F$ is the noise floor. $N_0$ is the power spectral density (PSD) of the thermal noise, and $N_F$ is the noise figure of the UE's receiver.
\end{lemma}

\begin{IEEEproof}
Since the channel has been assessed to be free for downlink transmission to the typical UE, the sum of the received power (averaged over fading randomness) from all the BSs belonging to $\mathcal{B}_{h,1}$ is below the sensing threshold, $P_{th}$, at the typical UE or its associated BS at $X_{b,n}$, depending on whether CSR or CST is used. Mathematically,
\begin{equation} \label{eq:hidden_conditions1}
\begin{aligned}
      & \textbf{CST: }\sum_{m \in \mathcal{O}} \sum_{\substack{j: X_{j,m} \in \mathcal{B}_{h,1} \\ (j,m) \neq (b,n) }} P_X  C_{\tau(j,m)}^{b,n}A_{j,m}^{b,n} ||X_{j,m} - X_{b,n}||^{-\alpha_{\tau(j,m)}^{b,n}} < P_{th} \\[-5pt]
     & \textbf{CSR: }\sum_{m \in \mathcal{O}}  \sum_{\substack{j: X_{j,m} \in \mathcal{B}_{h,1} \\ (j,m) \neq (b,n) }} P_{X}C_{\tau(j,m)}A_{j,m} ||X_{j,m}||^{-\alpha_{\tau(j,m)}} < P_{th}
\end{aligned}
\end{equation}
Since the received power from one BS, belonging to $\mathcal{B}_{h,1}$, is less than the sum of the received power from all the BSs in $\mathcal{B}_{h,1}$, we can write,
\begin{equation*} \label{eq:hidden_conditions2}
\begin{aligned}
      & \textbf{CST:}P_{X}C_{\tau(j,m)}^{b,n}A_{j,m}^{b,n} ||X_{j,m} - X_{b,n}||^{-\alpha_{\tau(j,m)}^{b,n}}  < \sum_{m \in \mathcal{O}} \sum_{\cramped{\substack{j: X_{j,m} \in \mathcal{B}_{h,1} \\ (j,m) \neq (b,n) }}} P_X  C_{\tau(j,m)}^{b,n}A_{j,m}^{b,n} ||X_{j,m} - X_{b,n}||^{-\alpha_{\tau(j,m)}^{b,n}} \\[-6pt]
     & \textbf{CSR: }  P_{X}C_{\tau(j,m)}A_{j,m} ||X_{j,m}||^{-\alpha_{\tau(j,m)}} < \sum_{m \in \mathcal{O}} \sum_{\substack{j: X_{j,m} \in \mathcal{B}_{h,1} \\ (j,m) \neq (b,n) }}  P_{X}C_{\tau(j,m)}A_{j,m} ||X_{j,m}||^{-\alpha_{\tau(j,m)}}
    \end{aligned}
\end{equation*}
Now, combining the above inequality for CST with the CST inequality in~\eqref{eq:hidden_conditions1}, we get the first inequality for CST in~\eqref{eq:hidden_bs_condition}. Similarly, combining the above inequality for CSR with the CSR inequality in~\eqref{eq:hidden_conditions1}, we get the first inequality for CSR in~\eqref{eq:hidden_bs_condition}. 
The second inequality in~\eqref{eq:hidden_bs_condition}, which is same for both CST and CSR, indicates that the BS at $X_{j,m}$ causes interference to the typical UE. 
By interference, we imply undesired power that is above the noise floor. Hence, a BS located at $X_{j,m}$ causes interference to the typical UE if, $P_{X}C_{\tau(j,m)}F_{j,m}G_{j,m} ||X_{j,m}||^{-\alpha_{\tau(j,m)}} > N_{f}$.
\end{IEEEproof}

\subsubsection{Deaf interferers} \label{deaf_def}
In contrast to the hidden interferers, deaf interferers are a subset of $\mathcal{B}_d$. Similar to $\mathcal{B}_h$, a subset of $\mathcal{B}_d$ would be silent and the rest active, i.e., $\mathcal{B}_d = \mathcal{B}_{d,0} \cup \mathcal{B}_{d,1}$, with $ \mathcal{B}_{d,0} \cap \mathcal{B}_{d,1} = \phi$.
In Fig.~\ref{fig:hidden_and_deaf}, among the two BSs belonging to $\mathcal{B}_d$ one is silent and the other is active. We define deaf interferers based on $\mathcal{B}_{d,1}$.
We call a BS, belonging to $\mathcal{B}_{d,1}$, a deaf interferer if it causes interference to the typical UE.
The mathematical conditions for a BS to be a deaf interferer are given in the following lemma.
\begin{lemma} \label{lemma:deaf_defn}
A BSs at $X_{j,m}$, belonging to $\mathcal{B}_{d,1}$, is a deaf interferer to the typical UE, if:
\begin{equation}\label{eq:deaf_bs_condition}
\begin{aligned}
    & \textbf{CST: } C_{\tau(b,n)}^{j,m}A_{b,n}^{j,m} ||X_{b,n} - X_{j,m}||^{-\alpha_{\tau(b,n)}^{j,m}} < \frac{P_{th}}{P_{X}} \textbf{ and } C_{\tau(j,m)}H_{j,m}G_{j,m} ||X_{j,m}||^{-\alpha_{\tau(j,m)}} > \frac{N_{f}}{P_{X}} \\[-3pt]
    & \textbf{CSR: } C_{\tau(b,n)}^{(x,y)}A_{b,n}^{(x,y)} ||X_{b,n} - U_{x,y}||^{-\alpha_{\tau(b,n)}^{(x,y)}} < \frac{P_{th}}{P_{X}} \textbf{ and } C_{\tau(j,m)}H_{j,m}G_{j,m} ||X_{j,m}||^{-\alpha_{\tau(j,m)}} > \frac{N_{f}}{P_{X}}
    \end{aligned}
\end{equation}
where $U_{x,y}$ is the location of the scheduled UE of the BS at $X_{j,m}$. 
\end{lemma}

\begin{IEEEproof}
The first inequality in~\eqref{eq:deaf_bs_condition}, for both CST and CSR, can be derived using a similar procedure as in the proof of Lemma~\ref{lemma:hidden_def}. However, in this case, the sensing node is not the typical UE or its associated BS; rather the sensing node is a UE located at $U_{x,y}$ (or the UE's associated BS) whose CS phase starts while the typical UE is already receiving downlink signals. Hence, in~\eqref{eq:deaf_bs_condition}, both for CST and CSR, the first inequality is based on the fact that the sensing node for $X_{j,m}$ could not detect the ongoing downlink transmission from the BS at $X_{b,n}$ to the typical UE. In case of CST, the sensing node is the BS itself at $X_{j,m}$, and in case of CSR, the sensing node is the UE at $U_{x,y}$. Similar to~\eqref{eq:hidden_bs_condition}, the second inequality in~\eqref{eq:deaf_bs_condition} is same for both CST and CSR and indicates that the BS at $X_{j,m}$ causes interference to the typical UE.
\end{IEEEproof} 

\textit{Remark:} Both hidden interferers and deaf interferers fall under the general class of hidden terminals 
In our context, the distinction between hidden and deaf interferers is crucial as neither CST nor CSR can 
tackle the deaf interferers, as explained in the next section.


\subsection{Tackling hidden and deaf interferers} \label{protocols}
In this section, we describe the CS protocols' capability in tackling hidden and deaf interferers. 

\subsubsection{CS at Transmitter (CST)} \label{cst} 
CST can neither eliminate hidden interferers, nor deaf interferers, irrespective of whether the sensing is done omnidirectionally (\textit{o}CST) or directionally (\textit{d}CST), as shown with the help of an example in Fig.~\ref{fig:hidden_cst}. For both the cases in Fig.~\ref{fig:hidden_cst}, UE 1 is within the main lobe of BS 2, but still BS 1 transmits downlink signals to UE 1, because BS 1 cannot sense the ongoing transmissions of BS 2. Thus, the downlink signals of both the UEs experience interference. In this example, BS 2 is a hidden interferer to UE 1, and BS 1 is a deaf interferer to UE 2. 
For CST, the sensing antenna gains (combining transmitter and sensing node) are: 
\begin{equation} \label{eq:ocst_gain}
    \textbf{ \textit{o}CST: } A_{j,m}^{b,n} = A_{b,n}^{j,m} = \begin{cases}
    M_{BS} \big(M_{BS} \times 10^{-0.7}\big) \text{ w.p. } \frac{\theta_{BS}}{2\pi}  \\[-7pt]
    m_{BS} \big(M_{BS} \times 10^{-0.7}\big) \text{ w.p. } \big(1 - \frac{\theta_{BS}}{2\pi}\big)
    \end{cases}
\end{equation}
\begin{equation} \label{eq:dcst_gain}
    \textbf{ \textit{d}CST: } A_{j,m}^{b,n} = A_{b,n}^{j,m} = \begin{cases}
    M_{BS}M_{BS} \text{   w.p.   }  \big(\frac{\theta_{BS}}{2\pi}\big) \big(\frac{\theta_{BS}}{2\pi}\big) \\[-7pt]
    M_{BS}m_{BS} \text{   w.p.   }  \big(\frac{\theta_{BS}}{2\pi}\big) \big(1 - \frac{\theta_{BS}}{2\pi}\big) \\[-7pt]
    m_{BS}M_{BS} \text{   w.p.   }  \big(1 - \frac{\theta_{BS}}{2\pi}\big) \big(\frac{\theta_{BS}}{2\pi}\big) \\[-7pt]
    m_{BS}m_{BS} \text{   w.p.   }  \big(1 - \frac{\theta_{BS}}{2\pi}\big) \big(1 - \frac{\theta_{BS}}{2\pi}\big) 
    \end{cases}
\end{equation}
In case of \textit{o}CST, the CS is performed by the BS using an omnidirectional antenna pattern. Since we consider ULAs, each BS uses three sectors/panels, 
each of $\frac{2 \pi}{3}$ radians,
for uniform coverage across the azimuth. For realising an omnidirectional sensing pattern, each antenna array panel must be able to sense across $\frac{2 \pi}{3}$ radians.
However, realising a $\frac{2 \pi}{3}$ radians
beam width with high fidelity is not possible for mmWave ULAs \cite{ramachandran2010adaptive}. Consequently, we assume that in case of \textit{o}CST, the CS is 
done via a quasi-omnidirectional antenna pattern \cite{nitsche2014ieee}, i.e., the gain across a $\frac{2 \pi}{3}$ radians wide beam is not constant, but it may vary randomly. To model this quasi-omnidirectional antenna pattern, we use the main lobe gain but penalise it by 7 dB to accommodate the randomness in the gain\footnote{We choose the value of penalty due to quasi-omnidirectional antenna pattern to be 7 dB because it has been reported that the fluctuations in main lobe gain could be around 7-10 dB due to the lack of high fidelity \cite{ramachandran2010adaptive}.}. For this reason, the sensing antenna gain for \textit{o}CST is $M_{BS} \times 10^{-0.7}$.
The probabilities in~\eqref{eq:ocst_gain} are based on whether the sensing BS is within the main lobe of the transmitting BS or not.
In case of \textit{d}CST, the sensing antenna gain in~\eqref{eq:dcst_gain} is self explanatory, based on our step function antenna model described in Section~\ref{system_model}.

\textit{d}CST has an advantage over \textit{o}CST. We explain this with the help of an example in Fig.~\ref{fig:exposed_cst}. In this figure, BS 2 would cause no interference to UE 1, but still, due to omnidirectional sensing (\textit{o}CST), BS 1 is going to defer its downlink transmission to UE 1. This problem, known as the exposed terminal problem, causes under utilization of the shared spectrum.
In contrast, due to the use of directional sensing (\textit{d}CST), the problem of exposed terminal is significantly reduced, as shown in the bottom part of Fig.~\ref{fig:exposed_cst}. In this figure, the main lobe beams of BS 1 (sensing) and BS 2 (transmitting) are not aligned. Hence, BS 1 senses low channel power, and starts downlink transmission to UE 1. 
BS 2 does not cause interference to the 
signals from BS 1 to UE 1.

When multiple BSs are at different heights on the same tower, they cannot sense each other's ongoing transmission via \textit{o}CST or \textit{d}CST. When these undetectable co-located BSs act as hidden interferers, they may produce interference as strong as the desired signal at a UE. We show an example scenario in Fig.~\ref{fig:colocated_bs_cst} with two co-located BSs. In this figure, BS 1 is unable to sense the ongoing transmissions from BS 2. Thus, BS 1 starts transmitting downlink signals to UE 1, and BS 2 becomes a hidden interferer for UE 1.

\subsubsection{Carrier Sensing at Receiver (CSR)} \label{csr} 

\begin{figure}[t]
    \centering
    \begin{subfigure}[Operation of CST in presence of co-located BSs]
    {\label{fig:colocated_bs_cst}
    \includegraphics[scale=0.36]{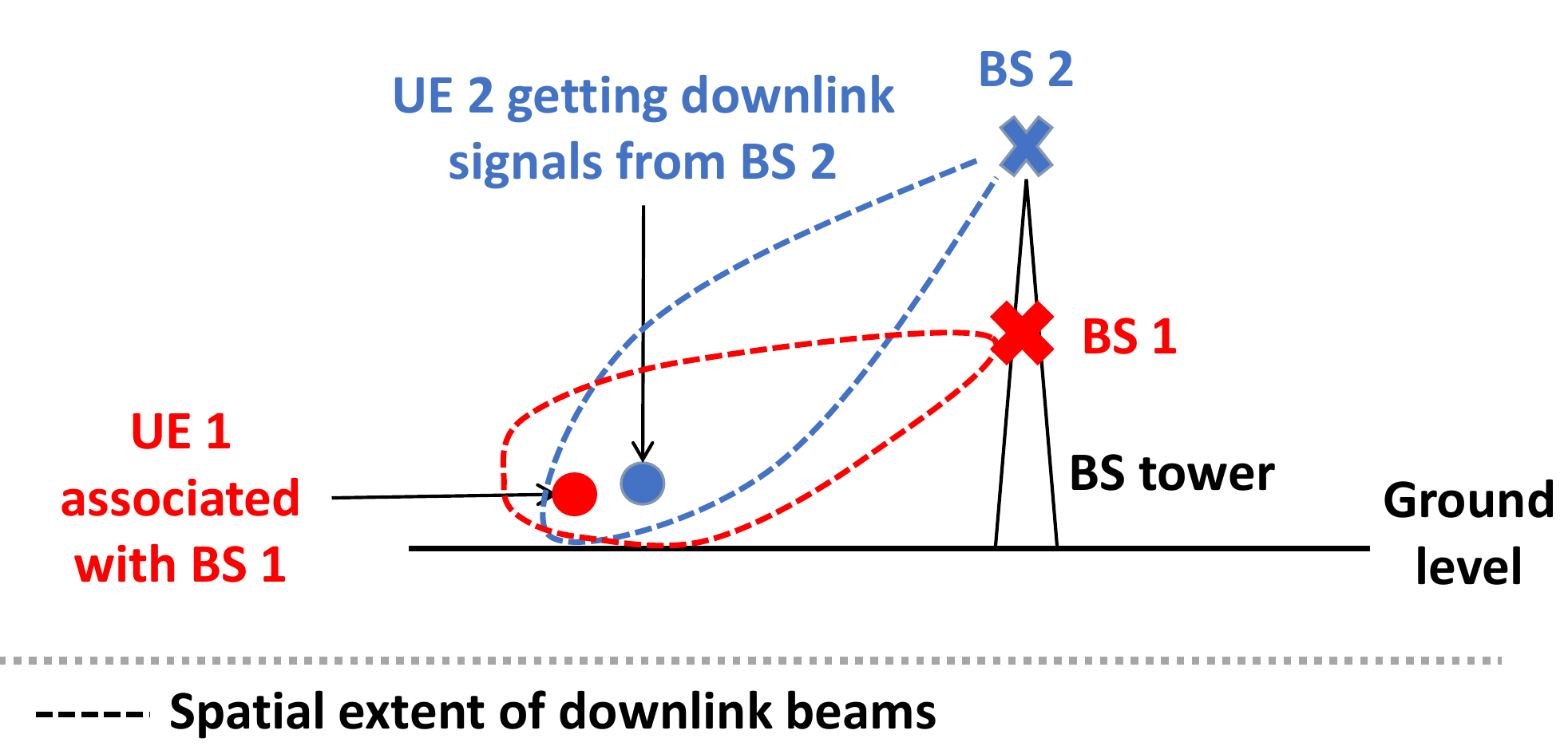}}
    \end{subfigure}
    \hspace{0.2in}
    \begin{subfigure}[Operation of CSR in presence of co-located BSs]
    {\label{fig:colocated_bs_csr}
    \includegraphics[scale=0.36]{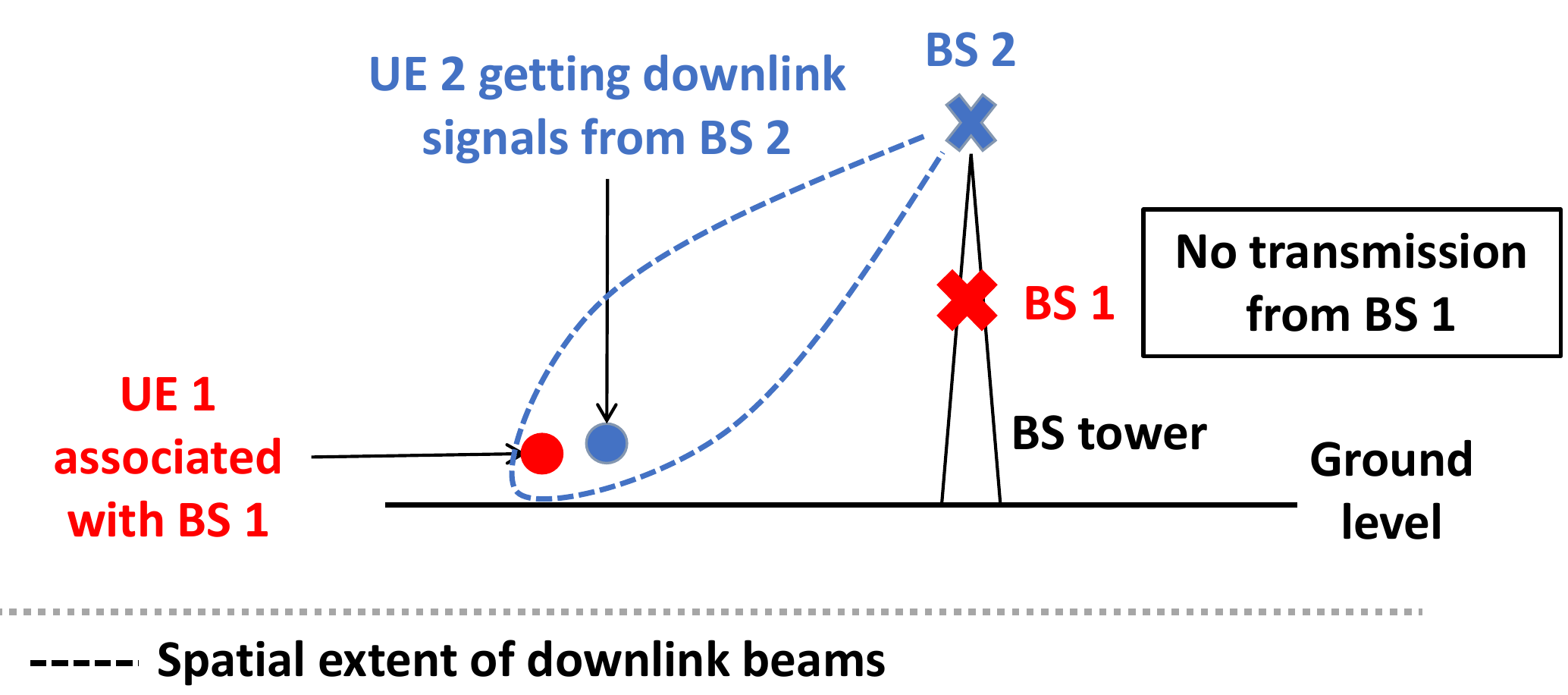}}
    \end{subfigure}
    \caption{Figure showing that CST cannot avoid interference from co-located BSs, but CSR can. 
    }
\end{figure}

Due to the above-described disadvantages with CST, we explore the option of CS at the receivers, i.e., at the UEs, in the context of downlink transmissions. If a scheduled UE senses the channel to be free, it informs its associated BS. Only then, the associated BS transmits downlink signals to the UE. Similar to CST, CSR can be directional (\textit{d}CSR) or omnidirectional (\textit{o}CSR). Now, let us revisit the problem of interference from co-located BSs using Fig.~\ref{fig:colocated_bs_csr}. If UE 1 performs CS while BS 2 is transmitting downlink signals to UE 2, UE 1 will measure a significant amount of power in the channel. Hence, it would not inform BS 1 that the channel is free, and BS 1 will not transmit any signal. Thus, CSR can resolve the problem of co-located BSs acting as hidden interferers. More generally, CSR can eliminate interference from 
all the hidden interferers, as long as the sensing threshold, $P_{th}$, is not 
much higher than the noise floor, $N_f$. 
With \textit{o}CSR, the sensing antenna gain is:
\begin{equation} \label{eq:ocsr_gain}
    A_{j,m} = A_{b,n}^{(x, y)} = \begin{cases}
    M_{BS} \big(M_{UE} \times 10^{-0.7} \big) \text{ w.p. } \frac{\theta_{BS}}{2\pi}  \\[-7pt]
    m_{BS} \big(M_{UE} \times 10^{-0.7} \big) \text{ w.p. } \big(1 - \frac{\theta_{BS}}{2\pi}\big)
    \end{cases}
\end{equation}
Similar to~\eqref{eq:ocst_gain}, here we add a penalty term due to the omnidirectional sensing of the UE. In case of \textit{d}CSR, the sensing antenna gains are $A_{j,m} = G_{j,m}$ and $A_{b,n}^{(x, y)} = G_{b,n}^{(x, y)}$, where $G_{j,m}$ and $G_{b,n}^{(x, y)}$ have the same distribution as $G_{j,m}^{(x,y)}$ (defined in~\eqref{eq:antenna_gain}).

 \begin{figure*}[t]
    \centering
    \begin{subfigure}[\textit{d}CSR]
    {\includegraphics[scale=0.34]{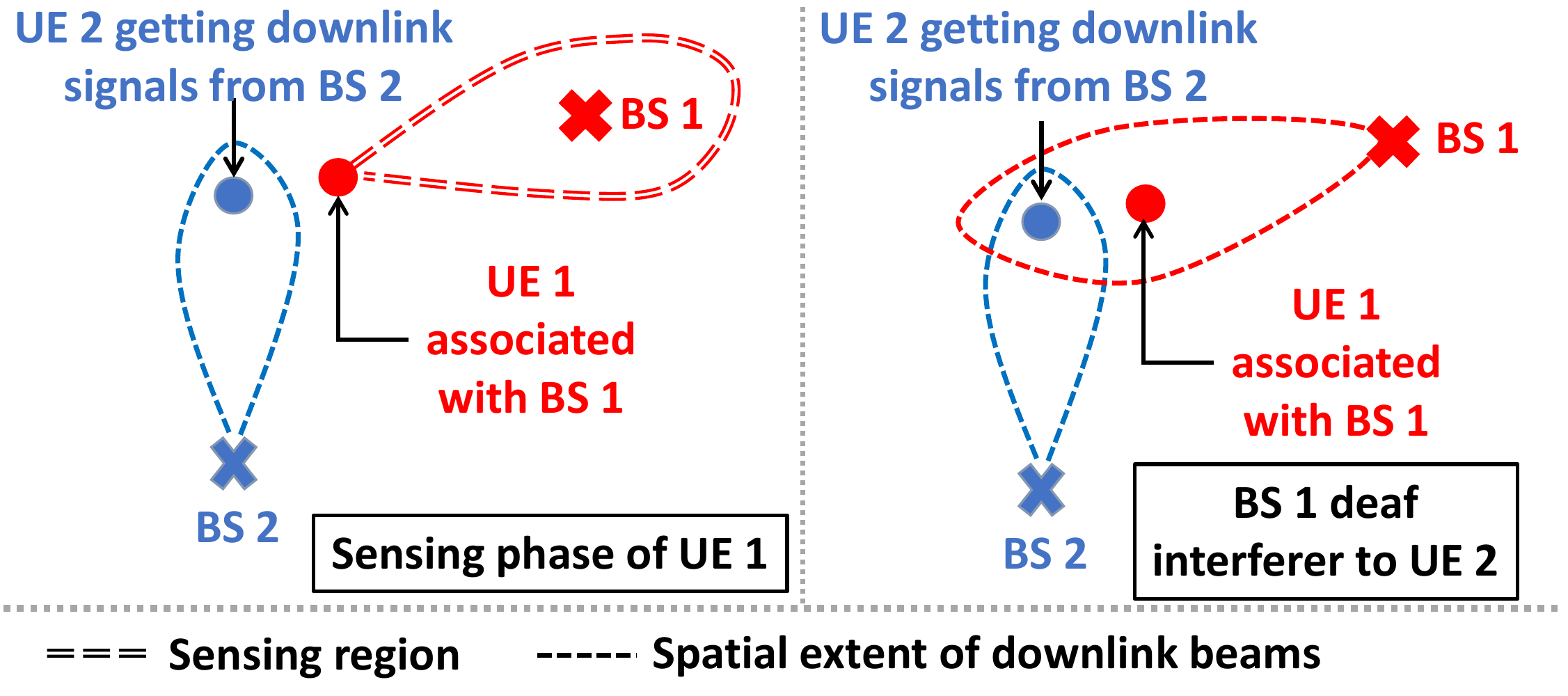}}
    \end{subfigure}
    \hspace{0.0in}
    \begin{subfigure}[\textit{d}CSRA]
    {\includegraphics[scale=0.34]{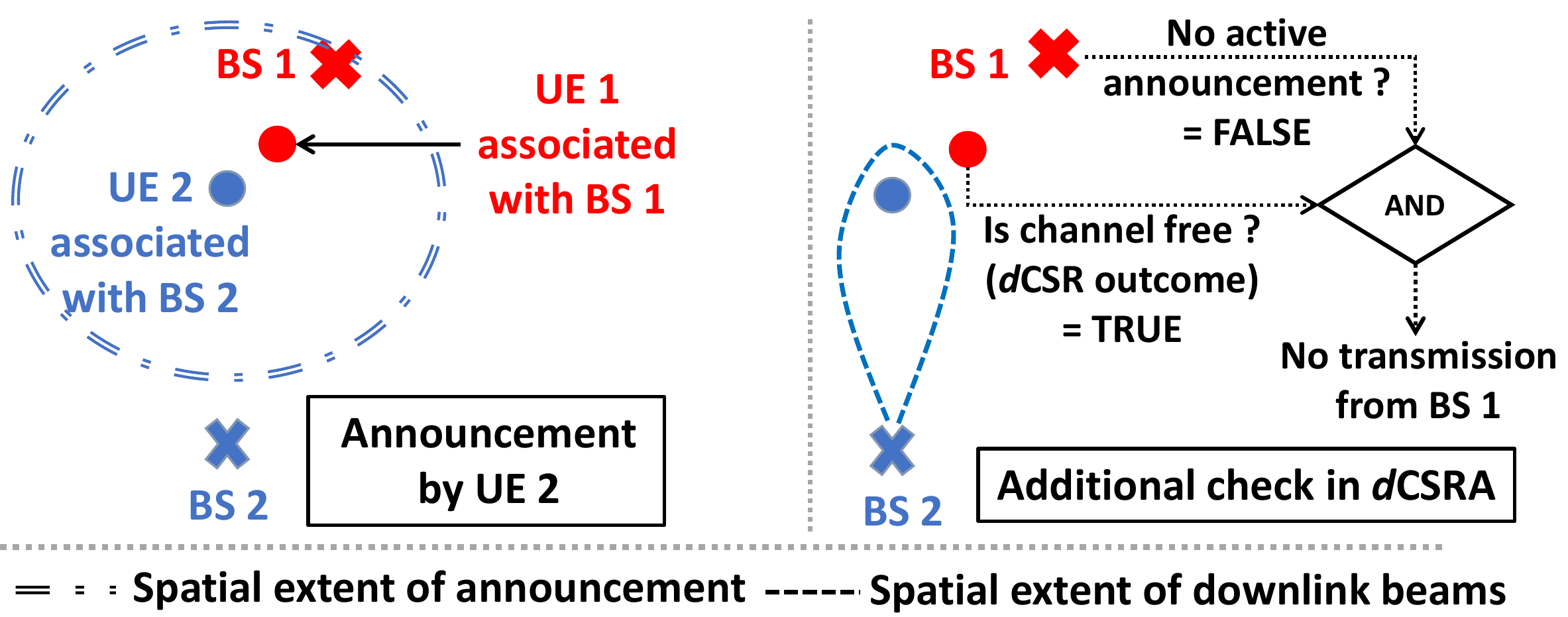}}
    \end{subfigure}
    \caption{Figures showing deaf interferers. (a) shows the deaf interference problem associated with \textit{d}CSR. (b) shows protection from deaf interferers by announcement. 
    } 
    \label{fig:csra}
\end{figure*}

CSR can tackle the hidden interferers, but it cannot prevent interference from the deaf interferers as shown in Fig.~\ref{fig:csra}(a). In this figure, BS 2 is transmitting downlink signals to UE 2, which has assessed the channel to be free via CSR. During the timeslot of downlink transmission to UE 2, UE 1 performs CSR (Fig.~\ref{fig:csra}(a) is for \textit{d}CSR, but a similar figure can be drawn for \textit{o}CSR) and fails to sense the ongoing transmissions from BS 2. Hence, UE 1 informs its associated BS, BS 1, to transmit downlink signals. This results in BS 1 becoming a deaf interferer to UE 2. Note that, the imperfect CS by UE 1 does not affect itself, rather it affects UE 2.


\subsubsection{\textit{d}CSR with Announcements (\textit{d}CSRA)} \label{csra} 
To address the problem of deaf interferers, we propose \textit{d}CSR with announcements and explain its working using Fig.~\ref{fig:csra}(b). Tackling deaf interferers requires interference protection beyond CS. Thus, in \textit{d}CSRA, if a UE (UE 2 in Fig.~\ref{fig:csra}(b)) assesses the channel to be unoccupied via \textit{d}CSR, it sends out a few broadcast announcements. If a BS (BS 1 in Fig.~\ref{fig:csra}(b)) hears this announcement, it would detect the presence of a UE in its vicinity. Now, even if this BS’s scheduled UE (UE 1 in Fig.~\ref{fig:csra}(b)) incorrectly assesses the channel to be free, this BS would refrain from transmission and prevent being a deaf interferer to the UE that had made the announcement previously.

\textit{Remark:} There are a few important points to note about \textit{d}CSRA. First, we use \textit{d}CSRA instead of \textit{o}CSRA to reduce the problem of exposed terminals. 
Second, for a BS to hear the announcements, it must be silent (not transmitting). The BSs that are close to an announcing UE are assured to be silent; otherwise, the UE would not have assessed the channel to be free in the first place. In contrast, the BSs that are farther away from an announcing UE may not hear the announcements and may cause deaf interference. However, such interference would not be severe as the interferers are farther away from the UE. 
Thus, \textit{d}CSRA cannot get rid of all the deaf interferers, but it prevents the strong ones. 
Third, while listening for the announcements from a UE, the BSs use a sensing threshold, $P_{th}^{A}$, that can be different from $P_{th}$, 
as the UEs transmit at a different power than the BSs.
Finally, the announcements by a UE are sent out omnidirectionally, as shown in Fig.~\ref{fig:csra}(b), because a deaf interferer can be anywhere around a UE. 
However, the BSs listen for the announcements in a directional way because their interference would be strongest along their main lobe while they act as deaf interferers. Hence, it is logical for the BSs to use higher sensing gain along their main lobe (directed towards the next scheduled UE) when listening for the UEs' announcements. 
Thus, the antenna gain while listening for announcements is $A_{(x,y)}^{j,m}$, whose distribution is same as that of $A_{j,m}^{(x,y)}$ in~\eqref{eq:ocsr_gain}.

\section{Coverage Probability Analysis} \label{coverage_prob}
In this section, we analyze the downlink coverage probability of a UE in our considered mmWave network using stochastic geometry. We first build a general framework for a UE's downlink coverage probability analysis in the presence of CS. Then, we use this framework for deriving a UE's coverage probability for the different CS protocols discussed in Section~\ref{methodology}.

For our analysis, we consider a typical UE,
and assume that the typical UE is a subscriber of network $n$, and 
its associated BS is located at $X_{b,n}$. 
We define $\mathbb{T}$ as the event that the typical UE receives downlink signals from its associated BS, and $p_{\mathbb{T}}$ as the probability of $\mathbb{T}$.
Essentially, $p_{\mathbb{T}}$ is the probability that a CS node
assesses the channel to be free. 
We define $L_{\mathcal{S}}$ and $N_{\mathcal{S}}$ as two sub-processes of $\Phi_{\mathcal{S}}; \mathcal{S} \in \mathcal{P(O)}; n \in \mathcal{S}$, such that $L_{\mathcal{S}}$ and $N_{\mathcal{S}}$ consist of the elements of $\Phi_{\mathcal{S}}$ that have LoS and NLoS links, respectively, with the typical UE. We define $R$ as a random variable that denotes the link distance between the typical UE and its associated BS, belonging to $L_{\mathcal{S}}$ or $N_{\mathcal{S}}$. $f_R \big(r, L_{\mathcal{S}} \big)$ and $f_R \big(r, N_{\mathcal{S}} \big)$ are the probability density functions (PDF) of $R$ when $X_{b,n} \in L_{\mathcal{S}}$ and $X_{b,n} \in N_{\mathcal{S}}$, respectively. 

\begin{theorem} \label{theorem:general_cov_prob}
In the presence of CS, the downlink coverage probability of a typical UE, during its scheduled time slot, is: 
\begin{equation} \label{eq:cov_prob_final}
    P_c(Z) = p_{\mathbb{T}} \int\limits_{r = 0}^{\infty} \sum_{\substack{\mathcal{S} \in \mathcal{P(O)} \\  n \in \mathcal{S}}} \Big[ e^{-\sigma^2 s_L} \mathcal{L}_{I(L_{\mathcal{S}}) \big{|} R, \mathbb{T}}(s_L) f_R (r, L_{\mathcal{S}}) + e^{-\sigma^2 s_N} \mathcal{L}_{I(N_{\mathcal{S}}) \big{|} R, \mathbb{T}}(s_N) f_R (r, N_{\mathcal{S}}) \Big] dr
\end{equation}
where $s_L = \frac{r^{\alpha_{L(b,n)}}Z}{C_{L(b,n)} G_{b,n}}$ and $s_N = \frac{r^{\alpha_{N(b,n)}}Z}{C_{N(b,n)} G_{b,n}}$, with $G_{b,n} = M_{BS}M_{UE}$. $I(L_{\mathcal{S}}) \big{|} R, \mathbb{T}$ and $I(N_{\mathcal{S}}) \big{|} R, \mathbb{T} $ represent the normalized interference (normalized by $P_X$) to the typical UE when $X_{b,n} \in L_{\mathcal{S}}$ and $X_{b,n} \in N_{\mathcal{S}}$, respectively, conditioned on $R = r$ and $\mathbb{T}$. Finally, $\mathcal{L}_{I(L_{\mathcal{S}}) \big{|} R, \mathbb{T}}$ and $\mathcal{L}_{I(N_{\mathcal{S}}) \big{|} R, \mathbb{T}}$ are the Laplace transform of  $I(L_{\mathcal{S}})\big{|} R, \mathbb{T}$, and $I(L_{\mathcal{S}})\big{|} R, \mathbb{T}$, respectively. 
\end{theorem}

\begin{IEEEproof}
Due to our PPP based modelling and Slivnyak's theorem \cite{elsawy2013stochastic}, the CCDF of any UE's SINR in the considered region would be same as the CCDF of the typical UE's SINR. The typical UE's coverage probability can be written as:
\begin{equation} \label{eq:disjoint_sum_of_pc}
\begin{aligned}
    P_c(Z) = \sum_{\substack{\mathcal{S} \in \mathcal{P(O)} \\  n \in \mathcal{S}}} P_c(Z, \Phi_{\mathcal{S}}) = \sum_{\substack{\mathcal{S} \in \mathcal{P(O)} \\  n \in \mathcal{S}}} P_c(Z, L_{\mathcal{S}}) +  P_c(Z, N_{\mathcal{S}})
\end{aligned}
\end{equation}
where $P_c(Z, \Phi_{\mathcal{S}})$ is the probability that the typical UE is in SINR coverage of $Z$, while $X_{b,n} \in \Phi_{\mathcal{S}}$. $P_c(Z, L_{\mathcal{S}})$ and $P_c(Z, N_{\mathcal{S}})$ are the probabilities that the typical UE is in SINR coverage of $Z$, while $X_{b,n} \in L_{\mathcal{S}}$ and $X_{b,n} \in N_{\mathcal{S}}$, respectively. In \eqref{eq:disjoint_sum_of_pc}, we use the independence of the elements of $\{\Phi_{\mathcal{S}}\}$. Thus, the events of the typical UE's association with different elements of $\{\Phi_{\mathcal{S}}\}$ are disjoint. Further, for any $\Phi_{\mathcal{S}}$, the events of the typical UE's association with $L_{\mathcal{S}}$ and $N_{\mathcal{S}}$ are also disjoint. Next, we analyze the terms $P_c(Z, L_{\mathcal{S}})$ and $P_c(Z, N_{\mathcal{S}})$. Towards that goal, let us consider the term $P_c(Z, \tau_{\mathcal{S}})$ where $\tau_{\mathcal{S}} \in \{L_{\mathcal{S}},N_{\mathcal{S}}\}$. Thus, $P_c(Z, \tau_{\mathcal{S}}) = P_c(Z, L_{\mathcal{S}})$, if $\tau_{\mathcal{S}} = L_{\mathcal{S}}$, and $P_c(Z, \tau_{\mathcal{S}}) = P_c(Z, N_{\mathcal{S}})$, if $\tau_{\mathcal{S}} = N_{\mathcal{S}}$. We can write $P_c(Z, \tau_{\mathcal{S}})$ as,
\begin{equation}\label{eq:pc_in_three_parts}
\begin{aligned}
    P_c(Z, \tau_{\mathcal{S}}) 
    & = \int_{r = 0}^{\infty} P \Big(\text{SINR} > Z \cap \mathbb{A}_{\tau_{\mathcal{S}}} \cap \mathbb{T} \Big{|} R = r \Big) f_R \big(r, \tau_{\mathcal{S}} \big) dr \\[-1pt]
    & = \int_{r = 0}^{\infty} P \Big(\text{SINR} > Z \cap \mathbb{A}_{\tau_{\mathcal{S}}} \Big{|} \mathbb{T}, R = r \Big) P \Big(\mathbb{T}\Big{|}R=r \Big) f_R \big(r, \tau_{\mathcal{S}} \big) dr\\[-1pt]
    & = p_{\mathbb{T}} \int_{r = 0}^{\infty} P \Big(\text{SINR} > Z \cap \mathbb{A}_{\tau_{\mathcal{S}}} \Big{|} \mathbb{T}, R = r \Big) f_R \big(r, \tau_{\mathcal{S}} \big) dr
\end{aligned}
\end{equation}
where $\mathbb{A}_{\tau_{\mathcal{S}}}$ is the event of the typical UE's association with a BS belonging to $\tau_{\mathcal{S}}$. 
We call $P(\mathbb{T} \big{|} R = r)$ the transmission probability as downlink transmission to the UE happens only when the sensing node finds the channel to be free. $P(\mathbb{T} \big{|} R = r)$ depends on the number of contending transmitters (BSs) in the sensing region of the CS node. 
Due to our association criteria (see Section~\ref{system_model}), there is an interference exclusion zone (dependent on the association distance, $r$) around the typical UE.
If this interference exclusion zone overlaps with the sensing region of the CS node for the typical UE, then the number of contenders become dependent on the association distance, $r$. However, in~\eqref{eq:pc_in_three_parts}, we assume that the CS outcome does not depend on the association distance, i.e., $P(\mathbb{T} \big{|} R = r) = P(\mathbb{T}) = p_{\mathbb{T}}$. We make this assumption because we also need the transmission probability of the interferers to the typical UE for quantifying the interference, and it would be very difficult (if at all possible) to separately analyze the transmission probabilities of all the BS based on the distances of their associated UEs. Hence, instead of the association distance, $r$, we use the average association distance, $\bar{R}$, for analyzing the transmission probability as described later in Section~\ref{tx_prob}. 
This way, the transmission probability becomes independent of $r$
and the same $p_{\mathbb{T}}$ can be used for all the BSs.

Let us now focus on $P \Big(\text{SINR} > Z \cap \mathbb{A}_{\tau_{\mathcal{S}}} \Big{|} \mathbb{T}, R = r \Big)$, the conditional coverage probability.

\begin{lemma} \label{lemma:conditional_cp}
The conditional coverage probability of the typical UE is,
\begin{equation} \label{eq:convert_to_laplace}
 P \Big(\text{SINR} > Z \cap \mathbb{A}_{\tau_{\mathcal{S}}} \Big{|} \mathbb{T}, R = r \Big) = e^{-\sigma^2 s_{\tau}} \mathcal{L}_{I(\tau_{\mathcal{S}}) \big{|}R,\mathbb{T}} (s_{\tau})
\end{equation}
where 
$\sigma^2 = \frac{N_{f}}{P_X}$. In~\eqref{eq:convert_to_laplace}, and subsequently is this paper, we use the following convention. If $\tau_{\mathcal{S}} = L_{\mathcal{S}}$, then $\tau = L$ and $\tau^c = N$; otherwise, if $\tau_{\mathcal{S}} = N_{\mathcal{S}}$, then $\tau = N$ and $\tau^c = L$.  Accordingly, if $\tau_{\mathcal{S}} = L_{\mathcal{S}}$, then $s_{\tau} = s_{L}$ and $\mathcal{L}_{I(\tau_{\mathcal{S}}) \big{|}R,\mathbb{T}} = \mathcal{L}_{I(L_{\mathcal{S}}) \big{|}R,\mathbb{T}}$; otherwise, if $\tau_{\mathcal{S}} = N_{\mathcal{S}}$, then $s_{\tau} = s_{N}$ and $\mathcal{L}_{I(\tau_{\mathcal{S}}) \big{|}R,\mathbb{T}} = \mathcal{L}_{I(N_{\mathcal{S}}) \big{|}R,\mathbb{T}}$.
\end{lemma}

\begin{IEEEproof}
See Appendix~\ref{appendix:conditional_cp} for the proof of Lemma~\ref{lemma:conditional_cp}.
\end{IEEEproof}

Finally, using~\eqref{eq:convert_to_laplace} in~\eqref{eq:pc_in_three_parts}, and the resulting expression in~\eqref{eq:disjoint_sum_of_pc}, we get~\eqref{eq:cov_prob_final}.
\end{IEEEproof}

To evaluate the coverage probability in~\eqref{eq:cov_prob_final}, we need the expressions for $f_R \big(r, \tau_{\mathcal{S}} \big)$, $\mathcal{L}_{I(\tau_{\mathcal{S}}) \big{|}R,\mathbb{T}}$, and $p_{\mathbb{T}}$.
First, we present the expression for $f_R \big(r, \tau_{\mathcal{S}} \big)$. Then, we analyze $\mathcal{L}_{I(\tau_{\mathcal{S}}) \big{|}R,\mathbb{T}}$ and $p_{\mathbb{T}}$ in Section~\ref{laplace} and Section~\ref{tx_prob}, respectively.
$f_R \big(r, \tau_{\mathcal{S}} \big)$ can be obtained as in \cite{jurdi2018modeling}, where authors investigate sharing of mmWave spectrum and BS sites among operators without any CS protocol. Since the association happens before CS, the CS protocols have no impact on $f_R \big(r, \tau_{\mathcal{S}} \big)$. Thus, the expression for $f_R \big(r, \tau_{\mathcal{S}} \big)$ is \cite{jurdi2018modeling}:
\begin{equation} \label{eq:link_prob}
    f_R \big(r, \tau_{\mathcal{S}} \big) = 2 \pi \lambda_{\mathcal{S}} r p_{\tau}(r) e^ {- v_{\mathcal{S},\tau}(r) - v_{\mathcal{S},\tau^c}(D_{\tau^c}(r))} \times \prod_{\substack{\mathcal{S}' \in \mathcal{P(O)} \backslash \mathcal{S} \\ n \in \mathcal{S}'}} e^ {- v_{\mathcal{S}',\tau}(r) - v_{\mathcal{S}',\tau^c}(D_{\tau^c}(r))}
\end{equation}
Following our convention regarding $\tau$,
if $\tau_{\mathcal{S}} = L_{\mathcal{S}}$, then $p_{\tau}(r) = p_L(r)$, $v_{\mathcal{S},\tau}(r) = v_{\mathcal{S}, L}(r)$, $v_{\mathcal{S},\tau^c}(r) = v_{\mathcal{S},N}(r)$, and $D_{\tau^c}(r) = D_N(r)$. If $\tau_{\mathcal{S}} = N_{\mathcal{S}}$, then $p_{\tau}(r) = p_N(r)$, $v_{\mathcal{S},\tau}(r) = v_{\mathcal{S}, N}(r)$, $v_{\mathcal{S},\tau^c}(r) = v_{\mathcal{S},L}(r)$, and $D_{\tau^c}(r) = D_L(r)$. The expressions for $v_{\mathcal{S}, L}(r)$ and $v_{\mathcal{S}, N}(r)$ are $2 \pi \lambda_{\mathcal{S}} \int\limits_0^r p_L(t)tdt$ and $2 \pi \lambda_{\mathcal{S}} \int\limits_0^r p_N(t)tdt$, respectively.
Lastly,
\begin{equation} \label{eq:exclusion_zone}
    D_{L}(r) = \bigg(\frac{C_L}{C_N}\bigg) ^ \frac{1}{\alpha_L} \times r ^ \frac{\alpha_N}{\alpha_L} \text{ and } 
    D_{N}(r) = \bigg(\frac{C_N}{C_L}\bigg) ^ \frac{1}{\alpha_N} \times r ^ \frac{\alpha_L}{\alpha_N}
\end{equation}
$D_{L}(r)$ is the radius of the interference exclusion zone (circular) of LoS interferers belonging to network $n$ when the typical UE and its associated BS has a NLoS link of distance $r$ meters. $D_{N}(r)$ is the radius of the interference exclusion zone (circular) of NLoS interferers belonging to network $n$ when the typical UE and its associated BS has a LoS link of distance $r$ meters. 

\subsection{Laplace transform of interference} \label{laplace}
In this section we analyze the Laplace transform of interference, $\mathcal{L}_{I(\tau_{\mathcal{S}})\big{|}R,\mathbb{T}} (s_{\tau})$. For compactness, we remove $\cdot\big{|}R,\mathbb{T}$ from all the terms with the understanding that all the following analysis is for interference to the typical UE, conditioned on $R = r$ and $\mathbb{T}$. Thus we use, $I(\tau_{\mathcal{S}})$ for $I(\tau_{\mathcal{S}}\big{|}R,\mathbb{T})$ and $\mathcal{L}_{I(\tau_{\mathcal{S}})} (s_{\tau})$ for $\mathcal{L}_{I(\tau_{\mathcal{S}})\big{|}R,\mathbb{T}} (s_{\tau})$. 
In the following two lemmas, we first present the expression for $\mathcal{L}_{I(L_{\mathcal{S}})} (s_L)$, and then for $\mathcal{L}_{I(N_{\mathcal{S}})} (s_N)$. 
Before that, we define $\mathcal{R}_{L,h}$, $\mathcal{R}_{L,d}$, $\mathcal{R}_{N,h}$, and $\mathcal{R}_{N,d}$. $\mathcal{R}_{L,h}$ and $\mathcal{R}_{L,d}$ are the interference exclusion zone around the typical UE where LoS hidden interferers and LoS deaf interferers, respectively, cannot be present due to CS. Similarly, $\mathcal{R}_{N,h}$ and $\mathcal{R}_{N,d}$ are the interference exclusion zone around the typical UE where NLoS hidden interferers and NLoS deaf interferers, respectively, cannot be present due to CS.

\begin{lemma} \label{lemma:laplace_los_asscn}
The Laplace transform of $I(L_{\mathcal{S}})$ is given by:
\begin{equation} \label{eq:product_of_laplace_los_asso}
    \begin{aligned}
    & \mathcal{L}_{I(L_{\mathcal{S}})} (s_L) = u_{L,h}(s_L,r)^{|\mathcal{S}| - 1} . u_{L,d}(s_L,r)^{|\mathcal{S}| - 1} \\[-6pt]
    & . \prod_{\mathclap{\substack{\mathcal{S}^{''} \in \mathcal{P(O)} \\ n \notin \mathcal{S}}}}  e^{\Big(-2\pi \lambda_{\mathcal{S}^{''}} \displaystyle \int\limits_{t = 0}^{\infty} \big(1 - u_{L,h}(s_L,t) ^ {|\mathcal{S}^{''}|}\big) t p_L(t) dt \Big)} . \prod_{\mathclap{\substack{\mathcal{S}^{'} \in \mathcal{P(O)} \\ n \in \mathcal{S}^{'}}}}  e^{\Big(- 2\pi \lambda_{\mathcal{S}^{'}} \displaystyle \int\limits_{t = r}^{\infty} \big(1 - u_{L,h}(s_L,t) ^ {|\mathcal{S}^{'}|}\big) t p_L(t) dt \Big)} \\[-14pt]
    & . \prod_{\mathclap{\substack{\mathcal{S}^{''} \in \mathcal{P(O)} \\ n \notin \mathcal{S}}}}  e^{\Big(- 2\pi \lambda_{\mathcal{S}^{''}} \displaystyle \int\limits_{t = 0}^{\infty} \big(1 - u_{L,d}(s_L,t) ^ {|\mathcal{S}^{''}|}\big) t p_L(t) dt \Big)} . \prod_{\mathclap{\substack{\mathcal{S}^{'} \in \mathcal{P(O)} \\ n \in \mathcal{S}^{'}}}}  e^{\Big(- 2\pi \lambda_{\mathcal{S}^{'}} \displaystyle \int\limits_{t = r}^{\infty} \big(1 - u_{L,d}(s_L,t) ^ {|\mathcal{S}^{'}|}\big) t p_L(t) dt \Big)} \\[-14pt]
    & . \prod_{\mathclap{\substack{\mathcal{S}^{''} \in \mathcal{P(O)} \\ n \notin \mathcal{S}^{''}}}}  e^{\Big(- 2\pi \lambda_{\mathcal{S}^{''}} \displaystyle \int\limits_{t = 0}^{\infty} \big(1 - u_{N,h}(s_L,t) ^ {|\mathcal{S}^{''}|}\big) t p_N(t) dt \Big)} . \prod_{\mathclap{\substack{\mathcal{S}^{'} \in \mathcal{P(O)} \\ n \in \mathcal{S}^{'}}}}  e^{\Big(- 2\pi \lambda_{\mathcal{S}^{'}} \displaystyle \int\limits_{\mathclap{t = D_N(r)}}^{\infty} \big(1 - u_{N,h}(s_L,t) ^ {|\mathcal{S}^{'}|}\big) t p_N(t) dt \Big)} \\[-14pt]
    & . \prod_{\mathclap{\substack{\mathcal{S}^{''} \in \mathcal{P(O)} \\ n \notin \mathcal{S}^{''}}}}  e^{\Big(- 2\pi \lambda_{\mathcal{S}^{''}} \displaystyle \int\limits_{t = 0}^{\infty} \big(1 - u_{N,d}(s_L,t) ^ {|\mathcal{S}^{''}|}\big) t p_N(t) dt \Big)} . \prod_{\mathclap{\substack{\mathcal{S}^{'} \in \mathcal{P(O)} \\ n \in \mathcal{S}^{'}}}}  e^{\Big(- 2\pi \lambda_{\mathcal{S}^{'}} \displaystyle \int\limits_{\mathclap{t = D_N(r)}}^{\infty} \big(1 - u_{N,d}(s_L,t) ^ {|\mathcal{S}^{'}|}\big) t p_N(t) dt \Big)}
    \end{aligned}
\end{equation}
where $u_{L,h}(s_L,t) = 1$, if $t \in \mathcal{R}_{L,h}$, otherwise $u_{L,h}(s_L,t) = u_L(s_L,t)$; $u_{L,d}(s_L,t) = 1$ if $t \in \mathcal{R}_{L,d}$, otherwise $u_{L,d}(s_L,t) = u_L(s_L,t)$;
$u_{N,h}(s_L,t) = 1$ if $t \in \mathcal{R}_{N,h}$, otherwise $u_{N,h}(s_L,t) = u_N(s_L,t)$; $u_{N,d}(s_L,t) = 1$ if $t \in \mathcal{R}_{N,d}$, otherwise $u_{N,d}(s_L,t) = u_N(s_L,t)$. Finally, $u_{L}(s_L,t)$ is given by, 
\begin{equation*}
\begin{aligned}
    & u_{L}(s_L,t) = \big(1 - \frac{p_{\mathbb{T}}}{2}\big)  \Big[\frac{\theta_{BS}}{2 \pi}  \frac{\theta_{UE}}{2 \pi} + \frac{\theta_{BS}}{2 \pi} \Big(1 - \frac{\theta_{UE}}{2 \pi}\Big) + \Big(1 - \frac{\theta_{BS}}{2 \pi}\Big) \frac{\theta_{UE}}{2 \pi} + \Big(1 - \frac{\theta_{BS}}{2 \pi}\Big) \Big(1 - \frac{\theta_{UE}}{2 \pi} \Big) \Big] \\[0pt]
    & \hspace{0.8in} + \frac{p_{\mathbb{T}}}{2} \bigg[ \frac{(\theta_{BS}/2 \pi)(\theta_{UE}/2 \pi)}{1 + s_L C_L M_{BS}M_{UE} t^{-\alpha_L}} + \frac{(\theta_{BS}/2 \pi)(1 - \theta_{UE}/2 \pi)}{1 + s_L C_L M_{BS}m_{UE} t^{-\alpha_L}}
\end{aligned}
\end{equation*}
\begin{equation*}
    \hspace{1.3in} + \frac{(1 - \theta_{BS}/2 \pi)(\theta_{UE}/2 \pi)}{1 + s_L C_L m_{BS}M_{UE} t^{-\alpha_L}} + \frac{(1 - \theta_{BS}/2 \pi)(1 - \theta_{UE}/2 \pi) }{1 + s_L C_L  m_{BS}m_{UE} t^{-\alpha_L}} \bigg]
\end{equation*}
$u_{N}(s_L,t)$ is same as $u_{L}(s_L,t)$, but $C_L$ replaced $C_N$ and $\alpha_L$ replaced by $\alpha_N$.
\end{lemma}

\begin{IEEEproof}
See Appendix~\ref{appendix:laplace_los} for the proof.
\end{IEEEproof}

\begin{lemma} \label{lemma:laplace_nlos_asscn}
The Laplace transform of $I(N_{\mathcal{S}})$ is given by:
\begin{equation} \label{eq:product_of_laplace_nlos_asso}
    \begin{aligned}
    & \mathcal{L}_{I(N_{\mathcal{S}})} (s_N) = u_{N,h}(s_N,r)^{|\mathcal{S}| - 1} . u_{N,d}(s_N,r)^{|\mathcal{S}| - 1} \\[-5pt]
    & . \prod_{\mathclap{\substack{\mathcal{S}^{''} \in \mathcal{P(O)} \\ n \notin \mathcal{S}^{''}}}}  e^{\Big(- 2\pi \lambda_{\mathcal{S}^{''}} \displaystyle \int\limits_{t = 0}^{\infty} \big(1 - u_{N,h}(s_N,t) ^ {|\mathcal{S}^{''}|}\Big) t p_N(t) dt \big)} . \prod_{\mathclap{\substack{\mathcal{S}^{'} \in \mathcal{P(O)} \\ n \in \mathcal{S}^{'}}}}  e^{\Big(- 2\pi \lambda_{S^{'}} \displaystyle \int\limits_{t = r}^{\infty} \big(1 - u_{N,h}(s_N,t) ^ {|\mathcal{S}^{'}|}\big) t p_N(t) dt \Big)} \\[-14pt]
    & . \prod_{\mathclap{\substack{\mathcal{S}^{''} \in \mathcal{P(O)} \\ n \notin \mathcal{S}^{''}}}}  e^{\Big(- 2\pi \lambda_{\mathcal{S}^{''}} \displaystyle \int\limits_{t = 0}^{\infty} \big(1 - u_{N,d}(s_N,t) ^ {|\mathcal{S}^{''}|}\big) t p_N(t) dt \Big)} . \prod_{\mathclap{\substack{\mathcal{S}^{'} \in \mathcal{P(O)} \\ n \in \mathcal{S}^{'}}}}  e^{\Big(- 2\pi \lambda_{\mathcal{S}^{'}} \displaystyle \int\limits_{t = r}^{\infty} \big(1 - u_{N,d}(s_N,t) ^ {|\mathcal{S}^{'}|}\big) t p_N(t) dt \Big)} \\[-14pt]
    & . \prod_{\mathclap{\substack{\mathcal{S}^{''} \in \mathcal{P(O)}  \\ n \notin \mathcal{S}^{''}}}}  e^{\Big(- 2\pi \lambda_{\mathcal{S}^{''}} \displaystyle  \int\limits_{t = 0}^{\infty} \big(1 - u_{L,h}(s_N,t) ^ {|\mathcal{S}^{''}|}\Big) t p_L(t) dt \big)} . \prod_{\mathclap{\substack{\mathcal{S}^{'} \in \mathcal{P(O)} \\ n \in \mathcal{S}^{'}}}}  e^{\Big(- 2\pi \lambda_{\mathcal{S}^{'}} \displaystyle  \int\limits_{\mathclap{t = D_L(r)}}^{\infty} \big(1 - u_{L,h}(s_N,t) ^ {|\mathcal{S}^{'}|}\big) t p_L(t) dt \Big)} \\[-14pt]
    & . \prod_{\mathclap{\substack{\mathcal{S}^{''} \in \mathcal{P(O)} \\ n \notin \mathcal{S}^{''}}}}  e^{\Big(- 2\pi \lambda_{\mathcal{S}^{''}} \displaystyle \int\limits_{t = 0}^{\infty} \big(1 - u_{L,d}(s_N,t) ^ {|\mathcal{S}^{''}|}\big) t p_L(t) dt \Big)} . \prod_{\mathclap{\substack{\mathcal{S}^{'} \in \mathcal{P(O)} \\ n \in \mathcal{S}^{'}}}}  e^{\Big(- 2\pi \lambda_{\mathcal{S}^{'}} \displaystyle \int\limits_{\mathclap{t = D_L(r)}}^{\infty} \big(1 - u_{L,d}(s_N,t) ^ {|\mathcal{S}^{'}|}\big) t p_L(t) dt \Big)}
    \end{aligned}
\end{equation}
where $u_{L,h}(s_N,t)$, $u_{L,d}(s_N,t)$, $u_{N,h}(s_N,t)$, and $u_{N,d}(s_N,t)$ are same as $u_{L,h}(s_L,t)$, $u_{L,d}(s_L,t)$, $u_{N,h}(s_L,t)$, and $u_{N,d}(s_L,t)$, respectively, with $s_L$ replaced by $s_N$ in the respective expressions.
\end{lemma}

\begin{IEEEproof}
See Appendix~\ref{appendix:laplace_nlos} for the proof.
\end{IEEEproof}

\subsection{Transmission probability} \label{tx_prob}
In this section, we analyze the transmission probability, $p_{\mathbb{T}}$. As mentioned before, we assume that all the BSs use the same CS protocol and the same sensing threshold, $P_{th}$. Additionally, we assume that all the operators have the same density of BSs (a reasonable assumption as all the operators in our problem are for the same RAT). Thus, we can use the same average association distance, $\bar{R}$ (introduced in the proof of Theorem~\ref{theorem:general_cov_prob}), for any UE belonging to any operator. So, based on the above factors, $p_{\mathbb{T}}$ is same for all the BSs.

If there are $N_c$ contending BSs within the sensing region of a CS node, then the node will find the channel to be free with probability $(1 - p_{\mathbb{T}}) ^ {N_c}$, i.e., none of the contenders are active. Hence, we can obtain $p_{\mathbb{T}}$ by solving 
$p_{\mathbb{T}} = (1 - p_{\mathbb{T}}) ^ {N_c}$. To do so, first, we have to find $N_c$. However, $N_c$ is random because, whether a BS is a contender or not depends on the contender's 
link type with the sensing node and
its directionality towards the sensing node, that are not deterministic. To circumvent the randomness of $N_c$
and use a deterministic value of $N_c$, 
we use the average number of contenders, $\bar{N}_c$, in place of $N_c$ and find $p_{\mathbb{T}}$ as $p_{\mathbb{T}} = (1 - p_{\mathbb{T}}) ^ {\bar{N}_c}$.  We show in Section~\ref{results}, that this approximation does not cause the analytically obtained $p_{\mathbb{T}}$ to be very different from the $p_{\mathbb{T}}$ obtained by simulations. The following lemma presents the expression for $\bar{N}_c$.

\begin{lemma} \label{lemma:tx_prob}
The average number of contenders to a CS node is given by $\bar{N}_c = \bar{N}_{c,L} + \bar{N}_{c,N} + \mathbf{1}_{A} . \bar{N}_c^A$, where $\bar{N}_{c,L}$ and $\bar{N}_{c,N}$ are the average number of LoS and NLoS contenders, respectively. $\bar{N}_c^A$ is the average number of contenders due to any active announcement. $\mathbf{1}_{A}$ is 1 if \textit{d}CSRA is used; otherwise, it is 0. 
Assuming that a sensing node belongs to operator $n$, i.e., a BS of operator $n$, or a UE subscribed to operator $n$, $\bar{N}_{c,L}$, $\bar{N}_{c,N}$, and $\bar{N}_c^A$ are given by:
\begin{equation} \label{eq:los_contenders}
    \begin{aligned}
    \bar{N}_{c,L} = &  \sum_{\mathclap{\substack{\mathcal{S}^{''} \in \mathcal{P(O)} \\ n \notin \mathcal{S}^{''}}}} \lambda_{\mathcal{S}''} \bigg( \mathbb{E}_{R_{cs,L}} \Big[ \int_{\theta = 0} ^ {\theta_{cs}} \int_{t = 0} ^ {R_{cs,L}}  p_L(t) t dt d\theta \Big] +  \mathbb{E}_{r_{cs,L}} \Big[ \int_{\theta = \theta_{cs}} ^ {2\pi} \int_{t = 0} ^ {r_{cs,L}}  p_L(t) t dt d\theta \Big] \bigg)   \\[-5pt]
    + & \sum_{\mathclap{\substack{\mathcal{S}^{'} \in \mathcal{P(O)} \\ n \in \mathcal{S}^{'}}}} \lambda_{\mathcal{S}'} \bigg( \mathbb{E}_{R_{cs,L}} \Big[ \int_{\theta = 0} ^ {\theta_{cs}} \int_{t = D(\bar{R})} ^ {R_{cs,L}}  p_L(t) t dt d\theta \Big] +   \mathbb{E}_{r_{cs,L}} \Big[ \int_{\theta = \theta_{cs}} ^ {2\pi} \int_{t = D(\bar{R})} ^ {r_{cs,L}}  p_L(t) t dt d\theta \Big]  \bigg)
    \end{aligned}
\end{equation}
\begin{equation} \label{eq:nlos_contenders}
    \begin{aligned}
    \bar{N}_{c,N} = &  \sum_{\mathclap{\substack{\mathcal{S}^{''} \in \mathcal{P(O)} \\ n \notin \mathcal{S}^{''}}}} \lambda_{\mathcal{S}''} \bigg( \mathbb{E}_{R_{cs,N}} \Big[ \int_{\theta = 0} ^ {\theta_{cs}} \int_{t = 0} ^ {R_{cs,N}}  p_N(t) t dt d\theta \Big]  +  \mathbb{E}_{r_{cs,N}} \Big[  \int_{\theta = \theta_{cs}} ^ {2\pi} \int_{t = 0} ^ {r_{cs,N}}  p_N(t) t dt d\theta \Big]  \bigg)\\[-5pt]
    + &  \sum_{\mathclap{\substack{\mathcal{S}^{'} \in \mathcal{P(O)} \\ n \in \mathcal{S}^{'}}}} \lambda_{\mathcal{S}'} \bigg( \mathbb{E}_{R_{cs,N}} \Big[ \int_{\theta = 0} ^ {\theta_{cs}} \int_{t = D(\bar{R})} ^ {R_{cs,N}}  p_N(t) t dt d\theta \Big] +  \mathbb{E}_{r_{cs,N}} \Big[  \int_{\theta = \theta_{cs}} ^ {2\pi} \int_{t = D(\bar{R})} ^ {r_{cs,N}}  p_N(t) t dt d\theta \Big] \bigg)
    \end{aligned}
\end{equation}
\begin{equation} \label{eq:announcement_contenders}
\begin{aligned}
    \bar{N}_{c}^{A} = &
     \sum_{\mathcal{S}^{'''} \in \mathcal{P(O)}} \lambda_{\mathcal{S}'''} \bigg( \int_{\theta = 0} ^ {\theta_{BS}} \int_{t = 0} ^ {R_{A,L}}  p_L(t) t dt d\theta +  \int_{\theta = \theta_{BS}} ^ {2\pi} \int_{t = 0} ^ {r_{A,L}}  p_L(t) t dt d\theta   \\[-5pt]
     & \hspace{1.0in} +  \int_{\theta = 0} ^ {\theta_{BS}} \int_{t = 0} ^ {R_{A,N}}  p_N(t) t dt d\theta + \int_{\theta = \theta_{BS}} ^ {2\pi} \int_{t = 0} ^ {r_{A,N}}  p_N(t) t dt d\theta \bigg)
\end{aligned}
\end{equation}
where $\theta_{cs}$ is the main lobe beam width of the CS node's sensing antenna. $D(\bar{R})$ is the radius of the interference exclusion zone (circular) around the CS node where LoS and NLoS contenders belonging to $\mathcal{S}^{'} \in \mathcal{P(O)}; n \in \mathcal{S}^{'}$ cannot be present.
$D(\bar{R}) = \bar{R}$ for CSR and $D(\bar{R}) = 0$ for CST. 
$R_{cs,L}$ and $R_{cs,N}$ are the distances of the farthest LoS and NLoS contenders, respectively, for a CS node along its main lobe. $r_{cs,L}$ and $r_{cs,N}$ are the distances of the farthest LoS and NLoS contenders, respectively, for a CS node along its side lobe. 
$R_{A,L}$ and $R_{A,N}$ are the distances of the farthest LoS and NLoS contenders (announcing UE), respectively, along the main lobe of a BS that is listening for announcements in \textit{d}CSRA. $r_{A,L}$ and $r_{A,N}$ are the distances of the farthest LoS and NLoS contenders (announcing UE), respectively, along the side lobe of a BS that is listening for announcements in \textit{d}CSRA. 
$R_{cs,L}$, $R_{cs,N}$, $r_{cs,L}$ and $r_{cs,N}$ are random variables and their distribution depends on the CS protocol. In contrast, $R_{A,L}$, $R_{A,N}$, $r_{A,L}$ and $r_{A,N}$ are deterministic.
The distributions of $R_{cs,L}$, $R_{cs,N}$, $r_{cs,L}$ and $r_{cs,N}$ for different protocols, and the expressions for $R_{A,L}$, $R_{A,N}$, $r_{A,L}$ and $r_{A,N}$ are given in Appendix~\ref{appendix:tx_prob_dist}. 
\end{lemma}

\begin{IEEEproof}
See Appendix~\ref{appendix:tx_prob} for the proof.
\end{IEEEproof}

\subsection{Coverage probability with different protocols} \label{cov_prob_for_protos}
In this section, we describe how the coverage probability expression of~\eqref{eq:cov_prob_final} varies for different protocols. As in \cite{jurdi2018modeling}, we consider a spectrum sharing system with two operators, i.e., $M=2$. Thus, $\mathcal{P(O)} = \{\{1\}, \{2\}, \{1,2\} \}$, and we assume $n$ is network operator 1. Thus, in~\eqref{eq:cov_prob_final}, $S^{''} =\{2\}$, $S^{'} \in \{\{1\}, \{1,2\}\}$, and $\mathcal{S}  \in  \{\{1\}, \{1,2\}\}$.

\subsubsection{Non CS (nonCS) scheme}
In this case, no CS is performed before the transmission of downlink signals; thus $p_{\mathbb{T}} = 1$. Among the remaining terms in~\eqref{eq:cov_prob_final}, the Laplace transforms are protocol dependent.
For the nonCS scheme, the Laplace transforms in~\eqref{eq:product_of_laplace_los_asso} and~\eqref{eq:product_of_laplace_nlos_asso} are given by the following corollary.
Here, we use the the two operator model of \cite{jurdi2018modeling}. Specifically, we use $a=\frac{\lambda_1}{\lambda}$, $b=1-\frac{\lambda_2}{\lambda}$, where $\lambda_1$, $\lambda_2$ are the BS density of the two operators, and $\lambda = \lambda_1 + \lambda_2 -\rho\lambda$. The parameter $\rho$ $(0 \leq \rho \leq 1)$ is the overlap coefficient, which is a measure of spatial correlation between the BS sites of the two operators.

\begin{corollary} \label{corollary:laplace_noncs}
When $\mathcal{S} = \{1\}$, the Laplace transform in~\eqref{eq:product_of_laplace_los_asso} is given by $\mathcal{L}_{I (L_{\mathcal{S}})}^{\{1\}} (s_L) = $
\begin{equation*} \label{eq:cov_prob_noncs}
    \begin{aligned}
    & e^{- 4\pi \lambda \Big[ \displaystyle \int\limits_{t = 0}^{\infty} (1 - a) \big(1 - u_{L}(s_L,t) \big) t p_L(t) dt + \displaystyle \int\limits_{t = r}^{\infty} \big(1 - u_{L}(s_L,t)\big) \big(a + \rho u_{L}(s_L,t)\big) t p_L(t) dt \Big] } \\[-12pt]
    & . e^{- 4\pi \lambda \Big[\displaystyle \int\limits_{t = 0}^{\infty} (1-a) \big(1 - u_{N}(s_L,t) \big) t p_N(t) dt  + \displaystyle \int\limits_{\mathclap{t = D_N(r)}}^{\infty} \big(1 - u_{N}(s_L,t)\big) \big(a + \rho u_{N}(s_L,t)\big) t p_N(t) dt \Big] }
    \end{aligned}
\end{equation*}
When $\mathcal{S} = \{1,2\}$, we have $\mathcal{L}_{I (L_{\mathcal{S}})}^{\{1,2\}} (s_L) = u_{L}(s_L,r)^{2} . \mathcal{L}_{I (L_{\mathcal{S}})}^{\{1\}} (s_L)$. The expressions for $\mathcal{L}_{I (N_{\mathcal{S}})}^{\{1\}} (s_N)$ and $\mathcal{L}_{I (N_{\mathcal{S}})}^{\{1,2\}} (s_N)$ are same as $\mathcal{L}_{I (L_{\mathcal{S}})}^{\{1\}} (s_L)$ and $\mathcal{L}_{I (L_{\mathcal{S}})}^{\{1,2\}} (s_L)$, respectively, but with $u_{L}(s_L,r)$, $u_{L}(s_L,t)$, $u_{N}(s_L,t)$, $p_L(t)$, $p_N(t)$, and $D_N(r)$ replaced by $u_{N}(s_N,r)$, $u_{N}(s_N,t)$, $u_{L}(s_N,t)$,  $p_N(t)$, $p_L(t)$, and $D_L(r)$, respectively.
\end{corollary}

\begin{IEEEproof}
In the absence of CS, the interference exclusion zones due to CS are non-existent, i.e., $\mathcal{R}_{L,h} = \mathcal{R}_{L,d} = \mathcal{R}_{N,h} = \mathcal{R}_{N,d} = B_0(0)$. Using these in our expressions of Laplace transforms in~\eqref{eq:product_of_laplace_los_asso} and~\eqref{eq:product_of_laplace_nlos_asso}, and the two operator model of \cite{jurdi2018modeling}, we get the expressions in Corollary~\ref{corollary:laplace_noncs}.
\end{IEEEproof}

\subsubsection{CSR schemes} \label{cov_prob_csr}
In case of \textit{o}CSR, \textit{d}CSR, and \textit{d}CSRA, $\theta_{cs}$ is $2\pi$, $\theta_{UE}$, and $\theta_{UE}$, respectively. Using these values of $\theta_{cs}$ and the distributions of the sensing distances in Appendix~\ref{appendix:tx_prob_dist}, we can obtain $\bar{N}_{c,L}$, $\bar{N}_{c,N}$, and $\bar{N}_c^A$ for the different CSR protocols based on~\ref{eq:los_contenders},~\ref{eq:nlos_contenders}, and~\ref{eq:announcement_contenders}.
Then, we compute $\bar{N}_c = \bar{N}_{c,L} + \bar{N}_{c,N} + \mathbf{1}_{A} . \bar{N}_c^A$, and, in turn, the transmission probability as $p_{\mathbb{T}} = (1 - p_{\mathbb{T}}) ^ {\bar{N}_c}$, for each of the CSR protocols.
Now, let us look at the Laplace transforms in~\eqref{eq:product_of_laplace_los_asso} and~\eqref{eq:product_of_laplace_nlos_asso}.
In case of CSR, the interference exclusion regions due to CS, $\mathcal{R}_{L,h}$ and $\mathcal{R}_{N,h}$ are around the typical UE. These regions are not deterministic due to the randomness associated with antenna gain during sensing. To obtain a deterministic value of coverage probability, we approximate these regions as $\mathcal{R}_{L,h} = B_0(h_L)$ and $\mathcal{R}_{N,h} = B_0(h_N)$, 
where $h_L = \mathbb{E}_{A_{j,m}} \big[ \big(\frac{P_X C_L A_{j,m}}{P_{th}}\big) ^ {\frac{1}{\alpha_L}} \big]$ and $h_N = \mathbb{E}_{A_{j,m}} \big[ \big(\frac{P_X C_N A_{j,m}}{P_{th}}\big) ^ {\frac{1}{\alpha_N}} \big]$. 
Thus, $h_L$ and $h_N$ are the average sensing distances of LoS and NLoS hidden interferers, averaged over antenna gain randomness, $A_{j,m}^{(x,y)}$. The distributions of $A_{j,m}$ for different CSR protocols were presented in Section~\ref{csr}.
We show in Section~\ref{results} that the above approximation has minimal impact on the overall coverage probability of the typical UE. 
As the distribution of $A_{j,m}$ is different for different CSR schemes, 
$h_L$ and $h_N$ are different for \textit{o}CSR and \textit{d}CSR, but same for \textit{d}CSR and \textit{d}CSRA. 
Similar to the hidden interferers, for the deaf interferers, we consider $\mathcal{R}_{L,d} = B_0(d_L)$ and $\mathcal{R}_{N,d} = B_0(d_N)$, where $d_L$ and $d_N$ are the average sensing radius for LoS and NLoS deaf interferers, respectively. 
We use $d_L = d_N = 0$ for both \textit{o}CSR and \textit{d}CSR, as they do not provide any interference protection from the deaf interferers. However, $d_L$ and $d_N$ are non-zero for \textit{d}CSRA due to the use of announcements. 
For \textit{d}CSRA, $d_L = \mathbb{E}_{A_{(x,y)}^{j,m}} \big[ \big(\frac{P_U C_L A_{(x,y)}^{j,m}}{P_{th}^A}\big) ^ {\frac{1}{\alpha_L}} \big]$ and $d_N = \mathbb{E}_{A_{(x,y)}^{j,m}} \big[ \big(\frac{P_U C_N A_{(x,y)}^{j,m}}{P_{th}^A}\big) ^ {\frac{1}{\alpha_N}} \big]$, where $A_{(x,y)}^{j,m}$ was defined in Section~\ref{csra}.
Now, for the CSR schemes, the Laplace transforms in~\eqref{eq:product_of_laplace_los_asso} and~\eqref{eq:product_of_laplace_nlos_asso} are given by the following corollary. The following expressions are given in terms of $h_L$, $h_N$, $d_L$, and $d_N$. For a specific CSR protocol, among \textit{o}CSR, \textit{d}CSR, and \textit{d}CSRA, the values of $h_L$, $h_N$, $d_L$, and $d_N$ must be modified as discussed above. 

\begin{corollary} \label{corollary:laplace_csr}
When $\mathcal{S} = \{1\}$, the Laplace transform in~\eqref{eq:product_of_laplace_los_asso} is given by $\mathcal{L}_{I (L_{\mathcal{S}})}^{\{1\}} (s_L) = $
\begin{equation*}
    \begin{aligned}
    & \exp\bigg(- 2\pi \lambda \bigg[\int\limits_{t = h_L}^{\infty} \big(1-a\big) \big(1 - u_{L}(s,t)\big) t p_L(t) dt + \int\limits_{t = d_L}^{\infty} \big(1-a\big) \big(1 - u_{L}(s,t)\big) t p_L(t) dt \\[-12pt]
    & \hspace{0.1in} + \int\limits_{\mathclap{t = \max(r,h_L)}}^{\infty} \big(1 - u_{L}(s,t) \big) \big(a + \rho u_{L}(s,t) \big) t p_L(t) dt + \int\limits_{\mathclap{t = \max(r,d_L)}}^{\infty} \big(1 - u_{L}(s,t) \big) \big(a + \rho u_{L}(s,t) \big) t p_L(t) dt \\
\end{aligned}
\end{equation*}
\begin{equation} \label{eq:cov_prob_csr}
    \begin{aligned}
    & \hspace{0.1in} +  \int\limits_{t = h_N}^{\infty} \big(1-a\big) \big(1 - u_{N}(s,t)\big) t p_N(t) dt +  \int\limits_{t = d_N}^{\infty} \big(1-a\big) \big(1 - u_{N}(s,t)\big) t p_N(t) dt \\[-5pt]
    &  + \hspace{0.25in} \int\limits_{\mathclap{t = \max(D_N(r),h_N)}}^{\infty} \big(1 - u_{N}(s,t) \big) \big(a + \rho u_{N}(s,t) \big) t p_N(t) dt  + \int\limits_{\mathclap{t = \max(D_N(r),d_N)}}^{\infty} \big(1 - u_{N}(s,t) \big) \big(a + \rho u_{N}(s,t) \big) t p_N(t) dt \bigg] \bigg)
    \end{aligned}
\end{equation}
When $\mathcal{S} = \{1,2\}$, we have $\mathcal{L}_{I(L_{\mathcal{S}})}^{\{1,2\}} (s_L) = u_{L,h}(s_L,r) . u_{L,d}(s_L,r) . \mathcal{L}_{I(L_{\mathcal{S}})}^{\{1\}} (s_L)$. The expressions for $\mathcal{L}_{I (N_{\mathcal{S}})}^{\{1\}} (s_N)$, $\mathcal{L}_{I (N_{\mathcal{S}})}^{\{1,2\}} (s_N)$ are same as $\mathcal{L}_{I (L_{\mathcal{S}})}^{\{1\}} (s_L)$, $\mathcal{L}_{I (L_{\mathcal{S}})}^{\{1,2\}} (s_L)$, respectively, but with $u_{L,h}(s_L,r)$, $u_{L,d}(s_L,r)$, $u_{L}(s_L,t)$, $u_{N}(s_L,t)$, $p_L(t)$, $p_N(t)$, $h_L$, $h_N$, $d_L$, $d_N$, and $D_N(r)$ replaced by $u_{N,h}(s_N,r)$, $u_{N,d}(s_N,r)$, $u_{N}(s_N,t)$, $u_{L}(s_N,t)$, $p_N(t)$, $p_L(t)$, $h_N$, $h_L$, $d_N$, $d_L$, and $D_L(r)$, respectively.
\end{corollary}

\begin{IEEEproof}
Using $\mathcal{R}_{L,h} = B_0(h_L)$, $\mathcal{R}_{N,h} = B_0(h_N)$, $\mathcal{R}_{L,d} = B_0(d_L)$, and $\mathcal{R}_{N,d} = B_0(d_N)$ 
in~\eqref{eq:product_of_laplace_los_asso} and~\eqref{eq:product_of_laplace_nlos_asso}, and the two operator model of \cite{jurdi2018modeling}, we get the expressions in Corollary~\ref{corollary:laplace_csr}.
\end{IEEEproof}


\subsubsection{CST schemes} We do not derive the expressions for the CST schemes because, as explained in Section~\ref{protocols}, CST has several drawbacks for our considered mmWave network, and it is always inferior than CSR. We validate this claim using simulations in Section~\ref{results}. Thus, the detailed analysis of CST schemes will not provide any additional insights.

\section{Evaluations} \label{results}
In this section, we compare the coverage probability of the typical UE for the different CS schemes, discussed in Section~\ref{methodology}. We consider the nonCS scheme as the baseline for our evaluations. 
For our evaluations, we consider a shared band of $W=600$ MHz; specifically, the 37.0-37.6 GHz band, which is currently under consideration to be designated as a shared band \cite{qualcomm37ghz}. 
Based on the propagation characterises of mmWave signals at 37 GHz \cite{rappaport2013millimeter}, we use $C_L = -60$ dB, $C_N = -70$ dB, $\alpha_L = 2$, and $\alpha_N = 4$. For transmit powers, we use $P_X = 36$ dBm and $P_{U} = 15$ dBm. For the number of antennas, we use $n_{BS} = 64$ and $n_{UE}=16$. For the main lobe beamwidth, we use $\theta_{BS} = \pi/18$ and $\theta_{UE} = \pi/6$. We use $N_F = 10$ dB for the UEs' noise figure. For the mmWave blocking parameter, we use $\beta = 0.007$. We consider $M = 2$ operators in a 10 km x 10 km area with BS density of $30/\text{km}^2$ for each of the operators.

\begin{figure*}[t]
    \centering
    \begin{subfigure}[Coverage probability]
    {\label{fig:pc}
    \includegraphics[scale=0.164]{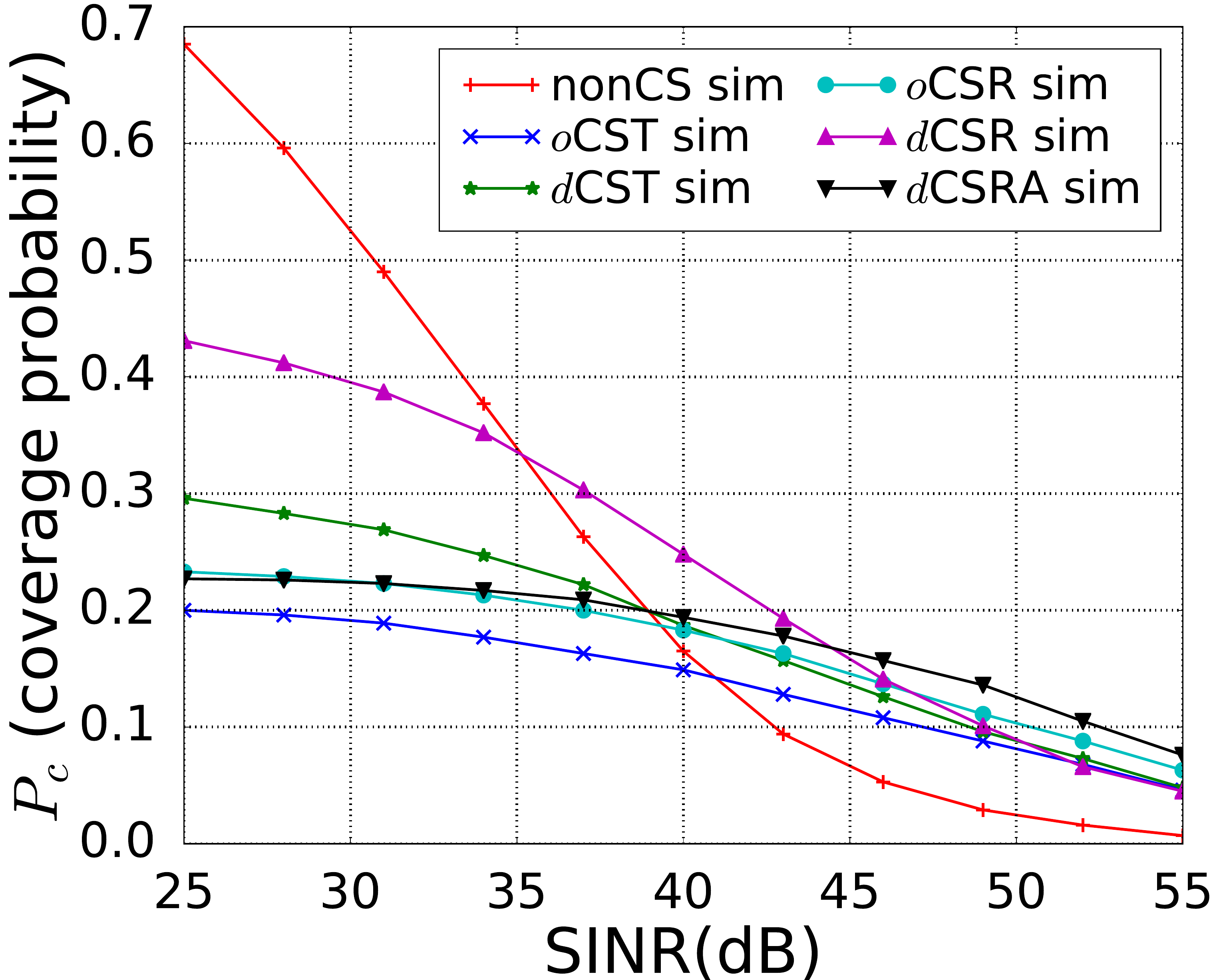}}
    \end{subfigure}
    \hspace{0.2in}
    \begin{subfigure}[Transmission probability]
    {\label{fig:tx_prob}
    \includegraphics[scale=0.174]{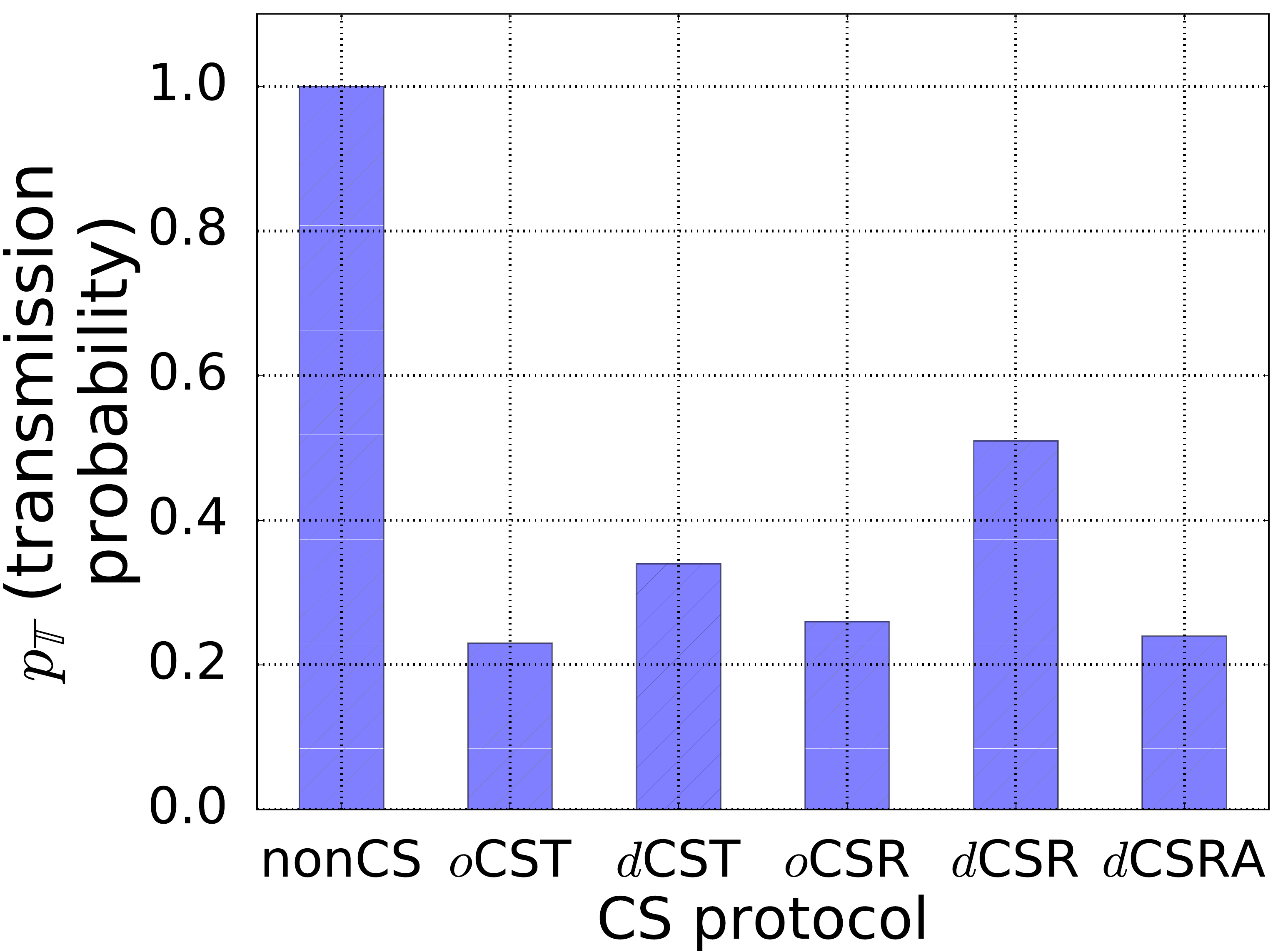}}
    \end{subfigure}
    \caption{Coverage probability and transmission probability of different protocols. For this figure, $P_{th} = N_f + 15$ dB and $P_{th}^{A} = N_f + 0$ dB. 
    The overlap coefficient is $\rho = 0.5$.}
    \label{fig:compare_pc}
\end{figure*}

We use both simulations and numerical evaluations, as required. For numerically evaluating $P_c(Z)$, we use the expressions derived in Section~\ref{cov_prob_for_protos}. For the average association distance, $\bar{R}$, required for numerical evaluation of $p_{\mathbb{T}}$, we use 100 meters. For finding $\bar{R}$, we make use of simulations and determine $\bar{R}$ by averaging the association distances of the typical UE across the different iterations in our simulations. Our simulation procedure is described below. Using $\bar{R} = 100$ meters is a reasonable choice as the average cell radius for the operators is 103 meters when their BS density is $30/\text{km}^2$. For evaluating $P_c(Z)$ based on simulations, we use a two step procedure. First, we find the transmission probability, $p_{\mathbb{T}}$ via simulations. We use 10000 different iterations, where each iteration corresponds to a different realization of operator PPPs. We generate the realizations of the operator PPPs (BS locations) with BS site sharing, as described in \cite{jurdi2018modeling}. For each iteration, $i$, we count the number of contenders $N_{c(i)}$, 
and then find a transmission probability for that iteration by solving $p_{\mathbb{T},i} = (1 - p_{\mathbb{T},i})^{N_{c(i)}}$. Then, after all the iterations, we find $p_{\mathbb{T}}$ as $p_{\mathbb{T}} = \frac{1}{10000}\sum_{i=1}^{10000} p_{\mathbb{T},i} $. At the second step, we repeat the above procedure of 10000 simulation runs, but this time, for each iteration, we calculate the SINR of the typical UE, $\text{SINR}_i$. For computing $\text{SINR}_i$, we use the following procedure. For each iteration, we sequentially check each of the BSs, other than the typical UE's serving BS at $X_{b,n}$. If a BS is a contender, we set it as active with probability $p_{\mathbb{T}}$. If any of the contenders in iteration $i$ is active, we set $\text{SINR}_i = 0$ (in linear scale) for that iteration. If none of the contenders are active, then we compute the interference for each of the remaining non-contending BSs and add them up to compute the aggregate interference $I$. For a non-contending BS at $X_{j,m}$, we assume it belongs to $\mathcal{B}_h$ with probability 0.5, and to $\mathcal{B}_d$ otherwise. Then, we compute its interference to the typical UE, $I_{X_{j,m}}$, as:
$I_{X_{j,m}} = C_{\tau(j,m)} F_{j,m} G_{j,m} ||X_{j,m}||^{-\alpha_{\tau(j,m)}}$, with probability $p_{\mathbb{T}}$; otherwise, $I_{X_{j,m}} = 0$.
If $X_{j,m} \in \mathcal{B}_d$, and \textit{d}CSRA is used, then $I_{X_{j,m}}$ is 0 if $X_{j,m}$ was able to hear the typical UE's announcements; otherwise, $I_{X_{j,m}}$ is computed as in the above equation. After computing $I$, we compute $\text{SINR}_i$ as, $\text{SINR}_i = \frac{P_X C_{\tau(b,n)} F_{b,n} M_{BS} M_{UE} r^{-\alpha_{\tau(b,n)}}}{\sigma^2 + I}$, where $r$ is the distance between the typical UE and its associated BS. $C_{\tau(b,n)}, F_{b,n}, r, \alpha_{\tau(b,n)}$, and $I$ varies with each iteration. Finally, after all the 10000 iterations, we find the fraction of iterations where $\text{SINR}_i > Z$, and use that fraction for $P_c(Z)$. 
Note that, in the above described procedure, we need $p_{\mathbb{T}}$ for deciding whether a contender or an interferer is active or not. 
For this reason, we compute the $p_{\mathbb{T}}$  in the first step and then $P_c(Z)$ in the second step.

\begin{figure*}[t]
    \centering
    \begin{subfigure}[Coverage probability]
    {\label{fig:pc_sim_vs_anl}
    \includegraphics[scale=0.163]{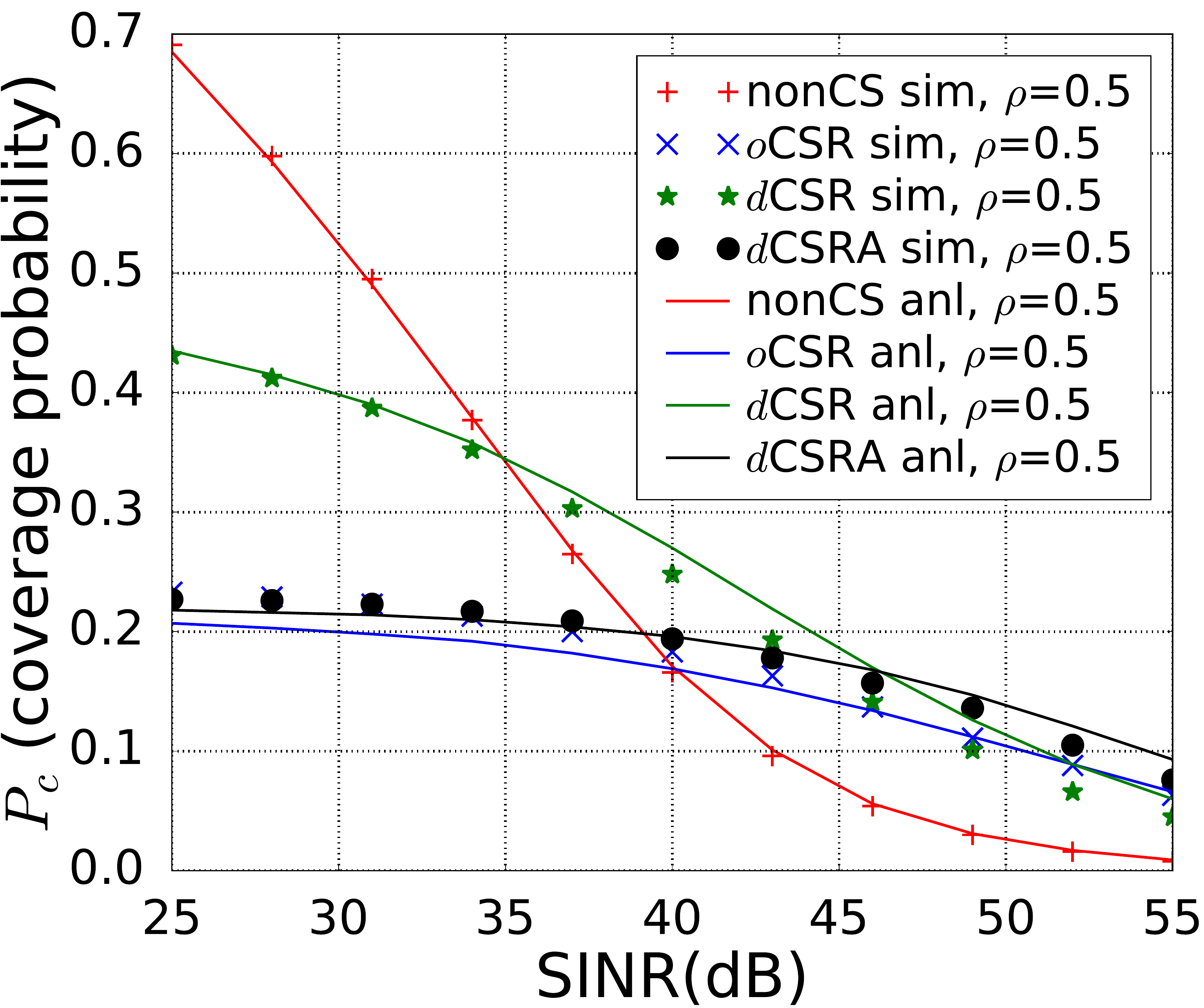}}
    \end{subfigure}
    \hspace{0.2in}
    \begin{subfigure}[Transmission probability]
    {\label{fig:tx_prob_sim_vs_anl}
    \includegraphics[scale=0.173]{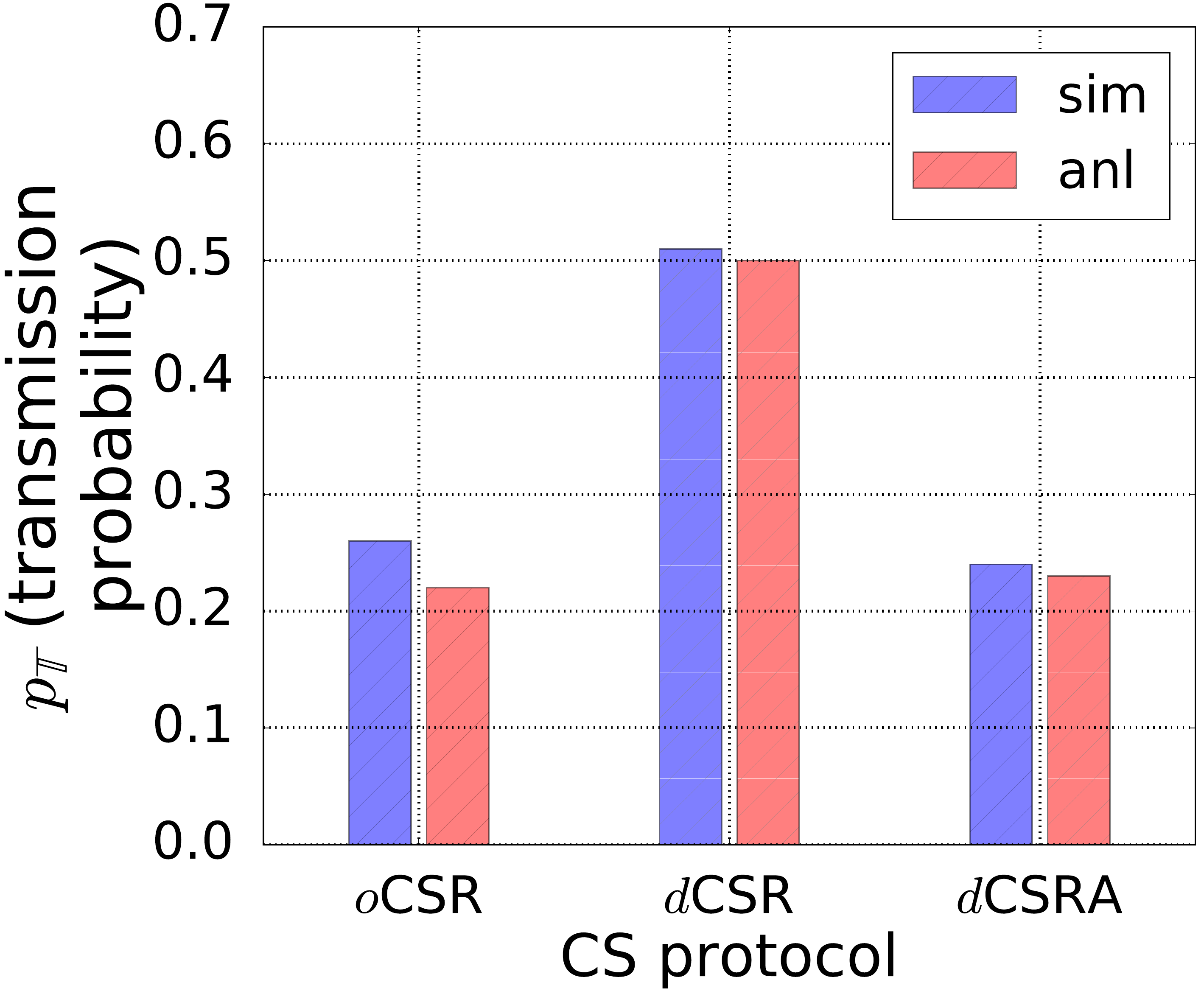}}
    \end{subfigure}
    \caption{Coverage probability and transmission probability of nonCS and CSR schemes, using both simulations and analysis. In this figure, $P_{th} = N_f + 15$ dB, $P_{th}^{A} = N_f + 0$ dB, 
    and $\rho=0.5$.}
    \label{fig:sim_vs_anl}
\end{figure*}

\subsection{Results}
\textbf{CS vs. nonCS}: In Fig.~\ref{fig:pc}, we show the downlink coverage probability of a UE in our shared mmWave network with different protocols. Since we do not have analytical expressions for CST, and we want to compare all the protocols in the same figure, we use simulations for the curves in this figure. 
We observe from Fig.~\ref{fig:pc} that, in terms of $P_c(Z)$, CS is beneficial only for higher values of SINR (above 35 dB). For lower values of SINR the $P_c(Z)$ with all the CS protocols
is inferior to the $P_c(Z)$ with the nonCS scheme, where no CS is used. Thus, for lower values of SINR (below 35 dB), not using any CS is the best strategy in terms of $P_c(Z)$.
To explain the trends of different protocols, first, we note that the coverage probability can be written as $P_c(Z) = \text{Pr}[(\text{SINR} > Z) \big{|} \mathbb{T}] \times p_{\mathbb{T}}$.
Thus, $P_c(Z)$ is upper bounded by $\min\big(\text{Pr}[(\text{SINR} > Z) \big{|} \mathbb{T}], p_{\mathbb{T}}\big)$. Now, Fig.~\ref{fig:tx_prob} shows $p_{\mathbb{T}}$ for all the protocols, obtained via simulations as described in the beginning of this section. We see from Fig.~\ref{fig:tx_prob} that the $p_{\mathbb{T}}$ for the nonCS scheme is 1.0, but the values of $p_{\mathbb{T}}$ for the CS protocols are much lower than 1.0. 
In the lower SINR region, $\text{Pr}[(\text{SINR} > Z) \big{|} \mathbb{T}]$ is high for all the protocols, but the low $p_{\mathbb{T}}$ for the CS protocols result in lower values of $P_c(Z)$ for the CS protocols, compared to the nonCS scheme.
As we move to the higher SINR region, the probability of having both high signal power and low interference power reduces for all the protocols. Consequently, in the higher SINR region, interference (in turn, $\text{Pr}[(\text{SINR} > Z) \big{|} \mathbb{T}]$) plays a more dominant role over $p_{\mathbb{T}}$. Hence the $P_c(Z)$ with the nonCS scheme is inferior to the $P_c(Z)$ with the CS schemes because, unlike the CS schemes, the nonCS scheme has no way of avoiding interference.


\begin{figure*}[t]
    \centering
    \begin{subfigure}[\textit{d}CSR]
    {\label{fig:pc_dcsr}
    \includegraphics[scale=0.161]{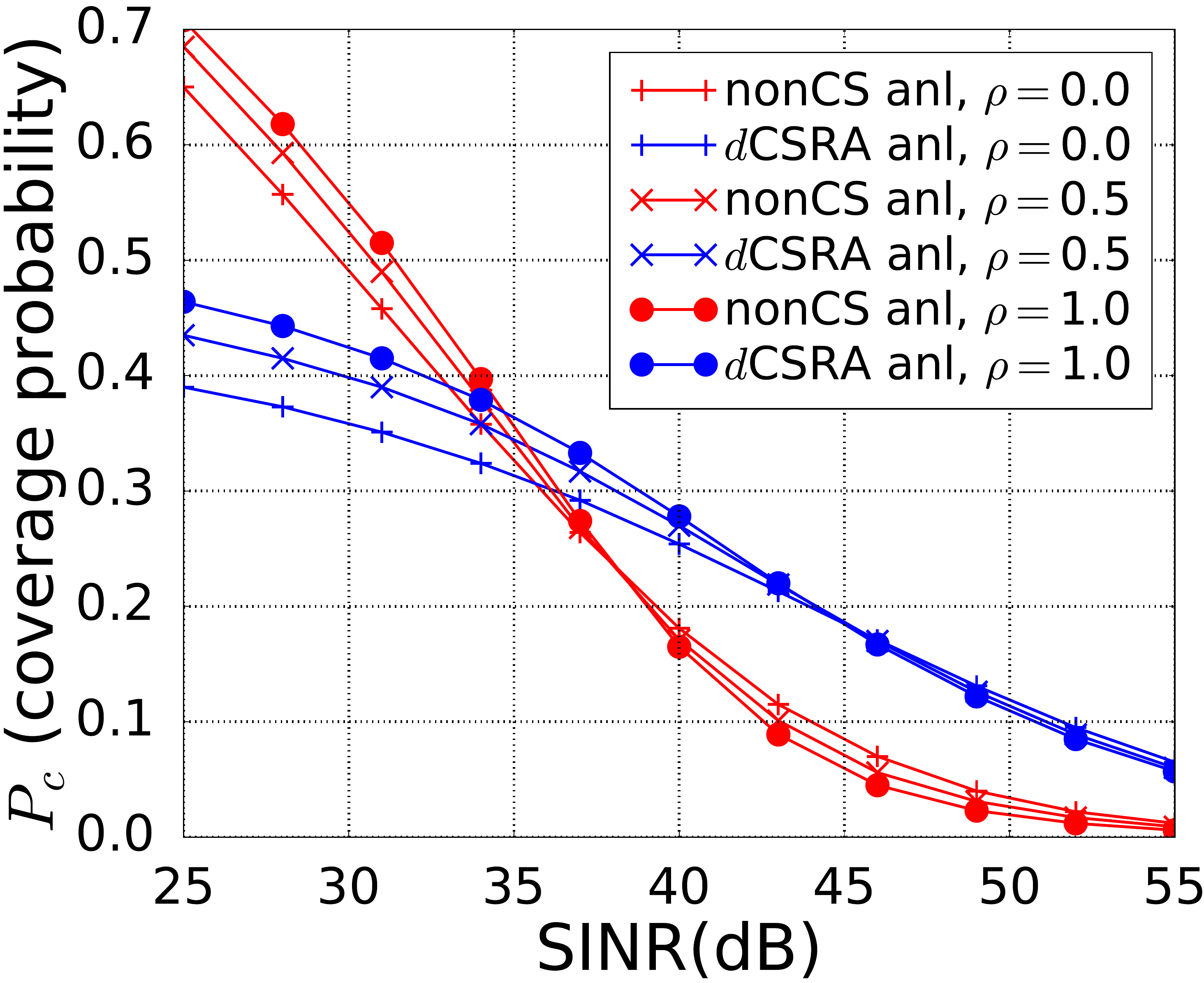}}
    \end{subfigure}
    \begin{subfigure}[\textit{d}CSRA]
    {\label{fig:pc_dcsra}
    \includegraphics[scale=0.152]{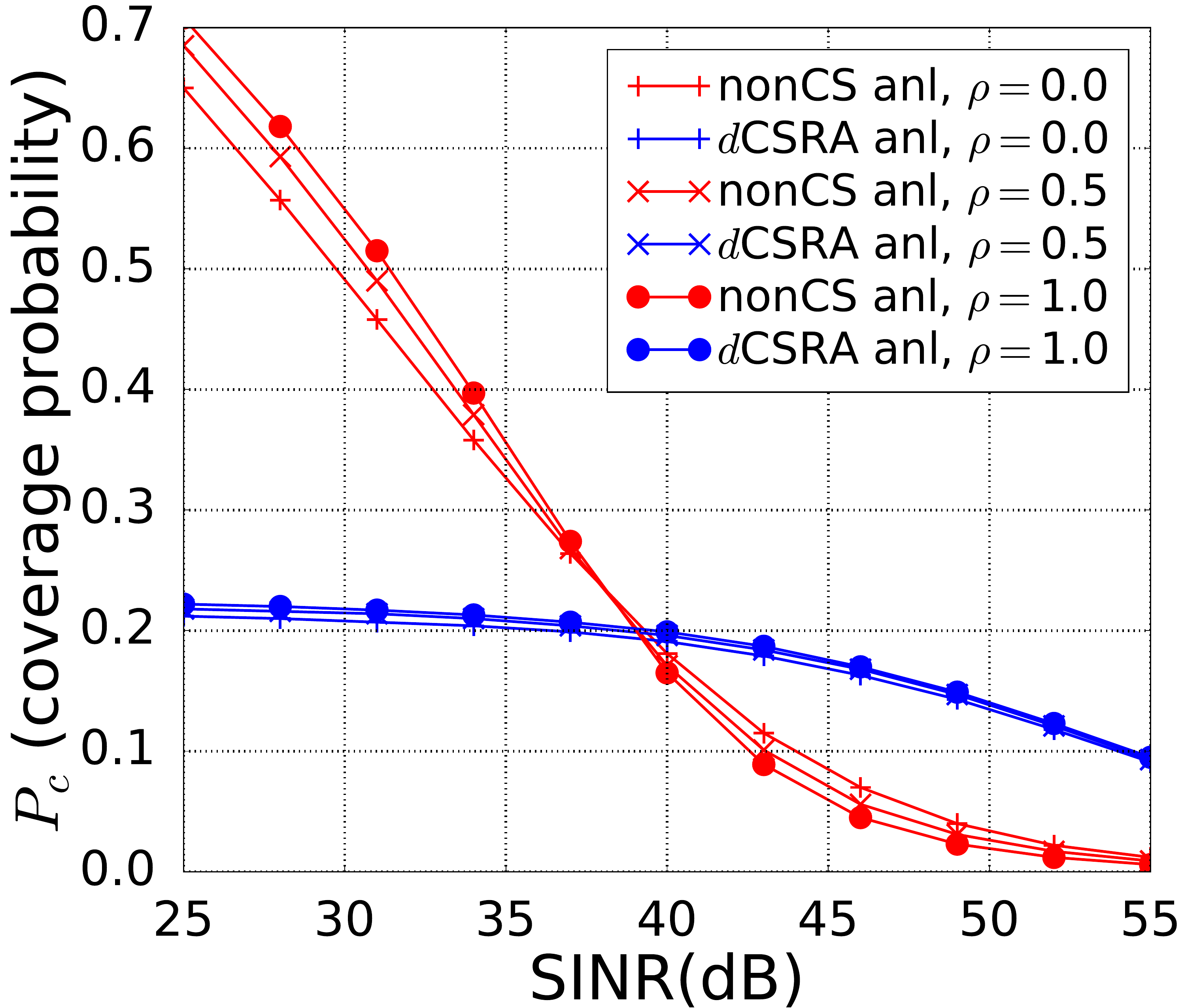}}
    \end{subfigure}
    \begin{subfigure}[Transmission probability]
    {\label{fig:tx_prob_dcsr_dcsra}
    \includegraphics[scale=0.155]{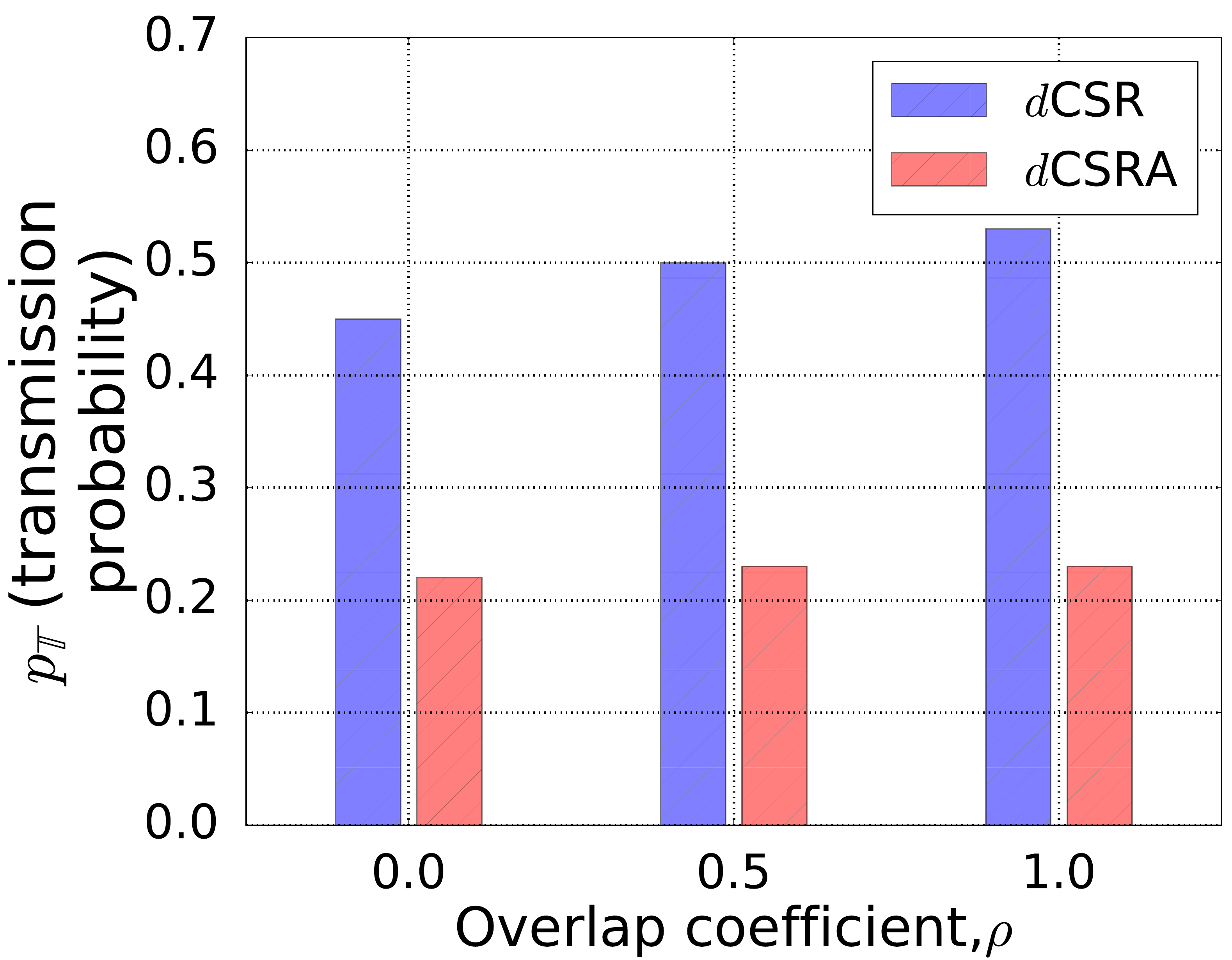}}
    \end{subfigure}
    \caption{(a), (b): Coverage probability with nonCS, \textit{d}CSR, and \textit{d}CSRA for different overlap coefficient, $\rho$. (c) Transmission probability of \textit{d}CSR and \textit{d}CSRA for different $\rho$. In this figure, $P_{th} = N_f + 15$ dB, $P_{th}^{A} = N_f + 0$ dB, and the results are obtained via analysis.}
    \label{fig:pc_vs_overlap}
\end{figure*}
\begin{figure*}[t]
    \centering
    \begin{subfigure}[Effect of $P_{th}$ on \textit{d}CSR]
    {\label{fig:pth_dcsr}
    \includegraphics[scale=0.149]{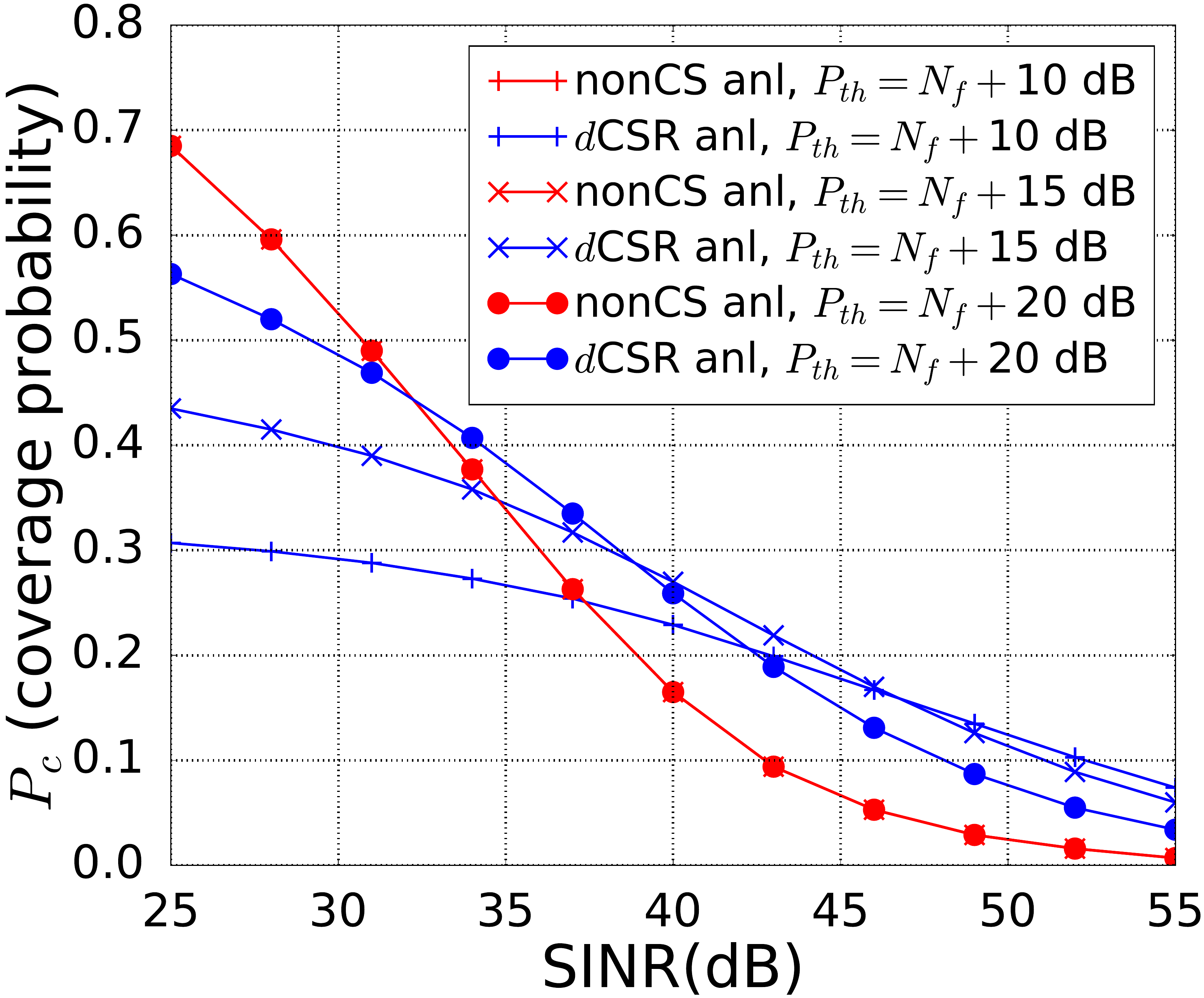}}
    \end{subfigure}
    \begin{subfigure}[Effect of $P_{th}$ on \textit{d}CSRA]
    {\label{fig:pth_dcsra}
    \includegraphics[scale=0.150]{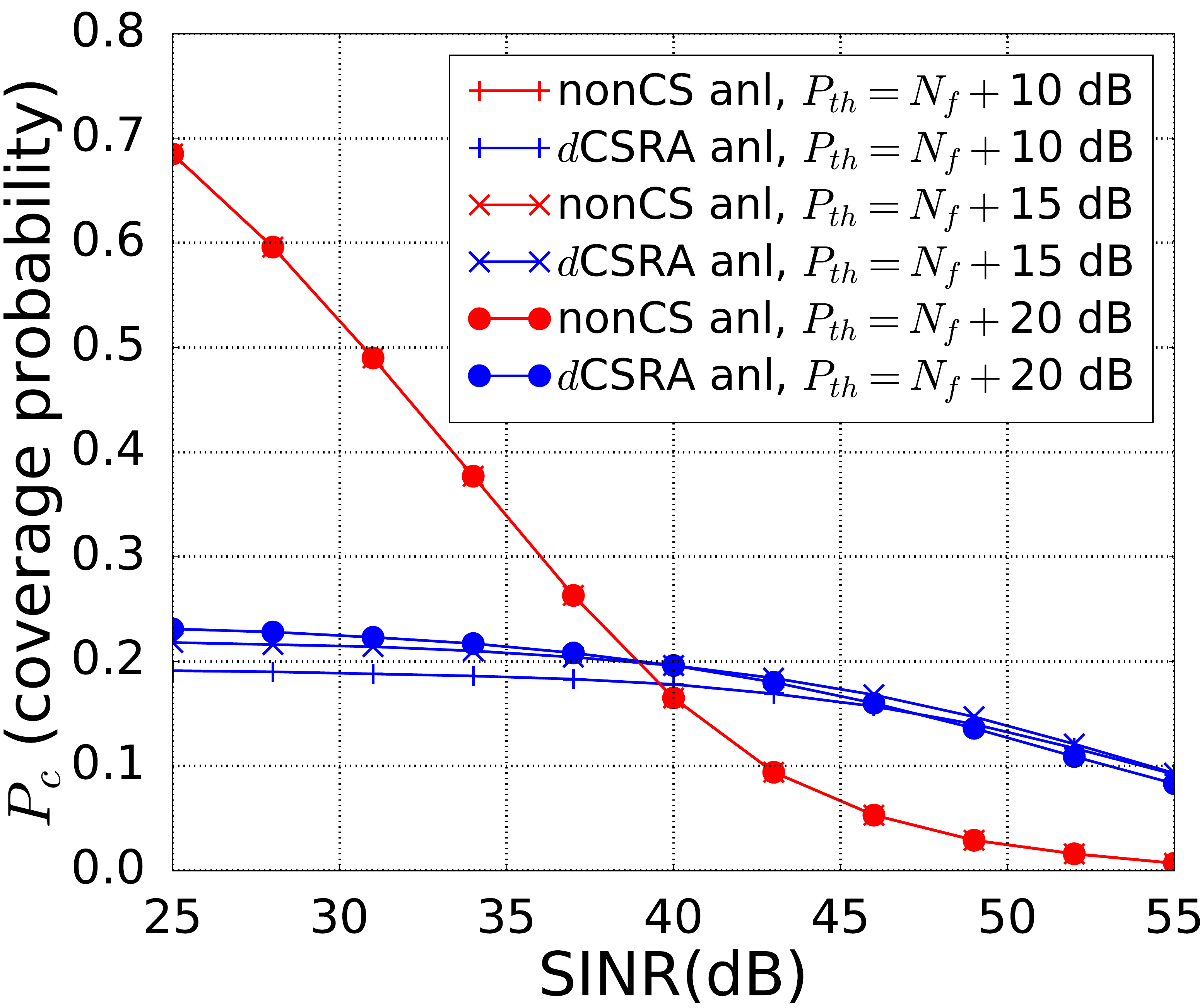}}
    \end{subfigure}
    \begin{subfigure}[Effect of $P_{th}^A$ on \textit{d}CSRA]
    {\label{fig:ptha_dcsra}
    \includegraphics[scale=0.161]{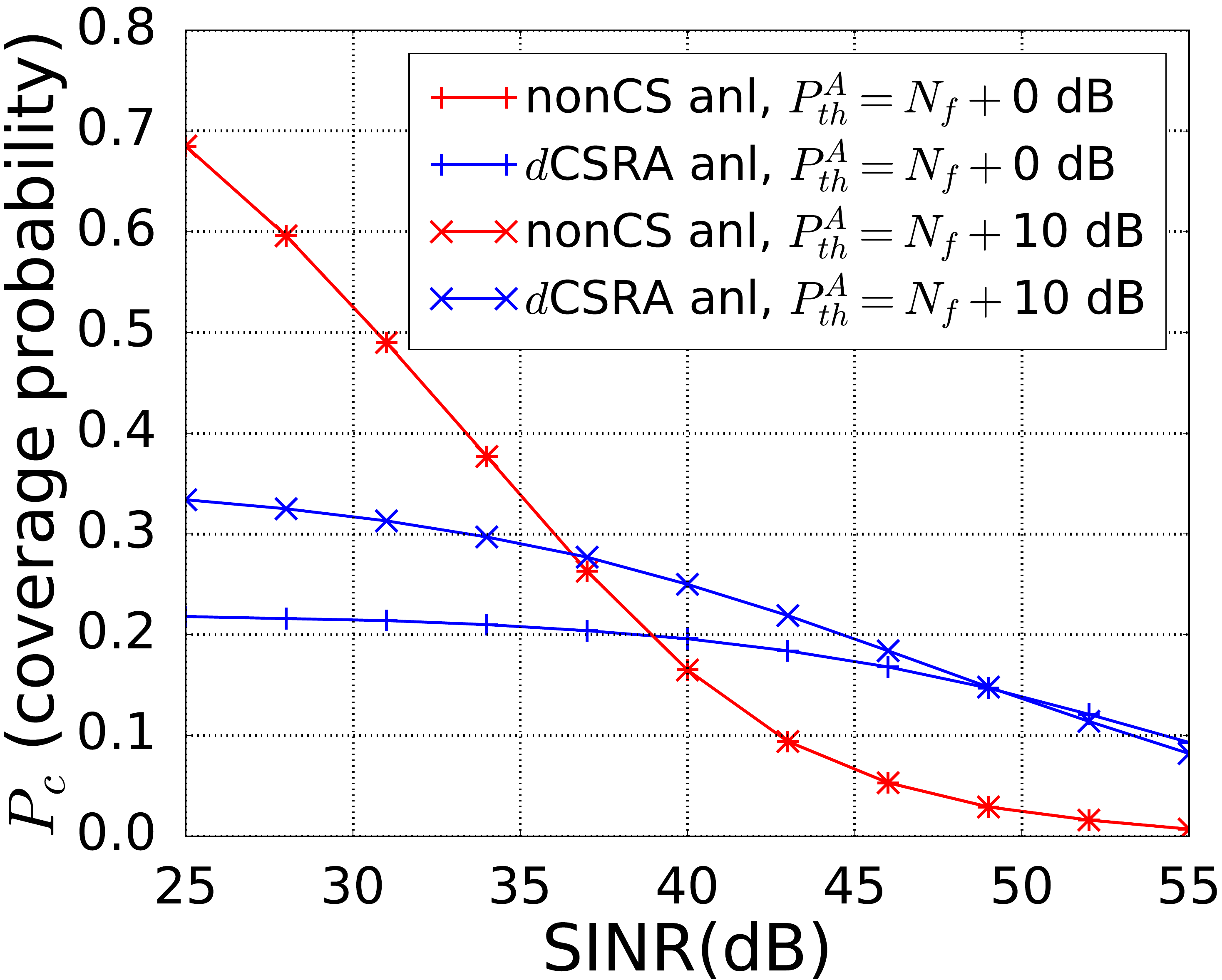}}
    \end{subfigure}
    \caption{(a), (b): Coverage probability with nonCS, \textit{d}CSR, and \textit{d}CSRA for different values of sensing threshold, $P_{th}$. (c): Coverage probability with \textit{d}CSRA for different values of $P_{th}^A$. In this figure, the overlap coefficient is $\rho = 0.5$, and the results are obtained by analysis.}
    \label{fig:pc_pth}
\end{figure*}

\textbf{CST vs. CSR}: We observe from Fig.~\ref{fig:pc} that 
$P_c(Z)$ with CSR is always better than that with CST (compare \textit{o}CSR versus \textit{o}CST and \textit{d}CSR versus \textit{d}CST in Fig.~\ref{fig:pc}). 
This happens primarily because of the drawbacks associated with CST, that were discussed in Section~\ref{protocols}.
Due to the inferior coverage probability with CST, we do not consider it for the subsequent evaluations, and focus on the CSR schemes.
Fig.~\ref{fig:pc} also shows that among the CSR protocols, \textit{d}CSR is the best choice in the middle SINR region (35 - 45 dB), and \textit{d}CSRA is the best choice in the higher SINR region (above 45 dB), in terms of $P_c(Z)$. \textit{d}CSRA is better than the other CSR protocols at higher SINR region because \textit{d}CSRA can eliminate the strong deaf interferers. However, if suffers from low transmission probability, as shown in Fig.~\ref{fig:tx_prob}, because of refrained transmissions due to active UE announcements. Recall from Lemma~\ref{lemma:tx_prob} that \textit{d}CSRA has additional contenders due to the announcement scheme, accounted by $\bar{N}_c^A$, which is not present for \textit{d}CSR and \textit{o}CSR.


\textbf{Validation of analytical coverage probability}: In Fig.~\ref{fig:pc_sim_vs_anl}, we use simulations to validate the expressions for coverage probability with nonCS and CSR schemes, derived in Section~\ref{cov_prob_for_protos}. In this figure, the analytical expression based curves are shown using solid lines and are labelled as `anl'.
The simulation-based curves are shown using the markers and are labelled as `sim'.
We observe that with the nonCS scheme, the $P_c(Z)$ obtained from analysis has very good match with the $P_c(Z)$ obtained using simulations. For the CSR schemes, the match between the analysis-based $P_c(Z)$ and the simulation-based $P_c(Z)$ is also good, but there are slight differences.
For \textit{d}CSR and \textit{d}CSRA the slight mismatch is in the higher SINR region. This happens because of using average sensing distances for the LoS and NLoS hidden interferers, $h_L$ and $h_N$, respectively, as explained in Section~\ref{cov_prob_csr}. 
In contrast, for \textit{o}CSR, the slight mismatch between the simulation-based $P_c(Z)$ and the analysis-based $P_c(Z)$ is in the lower SINR region. 
In this case, the slight mismatch results from the difference in the analysis-based $p_{\mathbb{T}}$ and the simulation-based $p_{\mathbb{T}}$, as shown in Fig.~\ref{fig:tx_prob_sim_vs_anl}. 
Finally, in Fig.~\ref{fig:tx_prob_sim_vs_anl}, the similarity between the simulation-based $p_{\mathbb{T}}$ and the analysis-based $p_{\mathbb{T}}$ for the CSR protocols shows that our approximations in the analysis of transmission probability (Section~\ref{tx_prob}) have minimal impact.
Fig.~\ref{fig:pc_sim_vs_anl} shows that the best protocol, in terms of coverage probability, varies with SINR. Thus, our coverage probability analysis framework and expressions serve as a useful tool for the network operators in deciding which protocol to use under different SINR conditions without running extensive simulations.

\textbf{Effect of BSs' site overlap}: Fig.~\ref{fig:pc_vs_overlap} shows the effect of BSs' site overlap on the coverage probability for three different values of $\rho$. Recall from Section~\ref{cov_prob_for_protos} that $\rho$ 
captures the spatial correlation between the BS sites of the two operators. $\rho = 0$ implies no BS site sharing and $\rho = 1$ implies all the BS sites are shared. We observe that, as $\rho$ increases, the $P_c(Z)$ with the nonCS method dips for the higher values of SINR. This happens because with a higher value of $\rho$, the possibility of strong interference from a co-located BS also increases, and the nonCS method has no interference protection from the co-located BSs. In contrast, the $P_c(Z)$ with the \textit{d}CSR (Fig.~\ref{fig:pc_dcsr}) and \textit{d}CSRA (Fig.~\ref{fig:pc_dcsra}) schemes are not affected by the increase of $\rho$ (in the higher SINR region) as CSR schemes can tackle interference from the co-located BSs. 
For lower values of SINR, the $P_c(Z)$ of \textit{d}CSR improves as $\rho$ increases. This is because with \textit{d}CSR, $p_{\mathbb{T}}$ improves with an increase in $\rho$, as shown in Fig.~\ref{fig:tx_prob_dcsr_dcsra}.
However, that is not the case with \textit{d}CSRA, as shown in Fig.~\ref{fig:tx_prob_dcsr_dcsra}.
Hence, unlike \textit{d}CSR, the $P_c(Z)$ of \textit{d}CSRA 
is unaffected by change in $\rho$.


\textbf{Effect of sensing threshold}: In Fig.~\ref{fig:pc_pth}, we show that the sensing threshold, $P_{th}$, plays an important role for the CS protocols. 
Both from Fig.~\ref{fig:pth_dcsr} and Fig.~\ref{fig:pth_dcsra} we observe that as $P_{th}$ is increased 
the $P_c(Z)$ with the CSR protocols improve in the lower SINR region.
This happens because a higher value of $P_{th}$ implies a smaller sensing region. 
A smaller sensing region reduces the number of contenders, and, in turn, improves the transmission probability of the CSR protocols. 
As explained in the context of Fig.~\ref{fig:compare_pc}, in the lower SINR region, the transmission probability is the dominant factor in the coverage probability with the CS protocols. 
However, as we increase $P_{th}$ 
the advantage of the CSR schemes gradually diminishes in the higher SINR region. 
This happens because increasing $P_{th}$ implies allowing higher interference, which plays a more dominant role over the transmission probability in the higher SINR region, as explained in the context of Fig.~\ref{fig:compare_pc}. 
In Fig.~\ref{fig:ptha_dcsra}, we show that, similar to $P_{th}$, careful selection of $P_{th}^A$ is also important, when dCSRA is used.
Finally, with the nonCS scheme, $P_c(Z)$ is unaffected by the change of $P_{th}$ or $P_{th}^A$ because it does not use any CS.


\section{Conclusions and future work} \label{conclusions}
We investigated CS for distributed interference management in a mmWave network where multiple non-coordinating operators share spectrum and BS sites. 
We argued that CST has several drawbacks, specifically in the context of our shared mmWave network, and proposed the use of CSR.
Since CSR cannot tackle the deaf interferers, we proposed \textit{d}CSRA, which can prevent interference from hidden interferers and most of the deaf interferers. 
We developed a framework for downlink coverage probability analysis of a UE in our shared mmWave network in the presence of CS. Furthermore, we derived the coverage probability expressions for the CSR schemes and the nonCS scheme using our framework. We validated our coverage probability analysis using simulations.
Using both simulations and numerical evaluations, we demonstrated the superiority of our 
CS schemes, over no CS, for the higher values of SINR. 
We also showed that the sensing threshold plays an important role in determining the coverage probability in the performance of CS. Finally, 
we showed that for the lower values of SINR not using any CS is the best strategy in terms of coverage probability.

The 
CS schemes suffer from low transmission probability; but, once the channel is assessed free, a downlink transmission is (almost) interference free. In contrast, without CS, 
the downlink transmissions would experience interference. In the future, we will investigate how the quality of downlink transmission (e.g., collisions, jitters) differs for the nonCS and the CS schemes.

\appendices
\section{Proof of Lemma~\ref{lemma:conditional_cp}: Conditional coverage probability} \label{appendix:conditional_cp}
The conditional coverage probability can be written as $ P \big (\text{SINR} > Z \cap \mathbb{A}_{\tau_{\mathcal{S}}} \big{|} \mathbb{T}, R = r \big) = P \Big (\frac{C_{\tau(b,n)}F_{b,n}G_{b,n}r^{-\alpha_{\tau(b,n)}}}{\sigma^2 + I(\tau_{\mathcal{S}})} > Z \Big{|} \mathbb{T}, R = r \Big)$, where $F_{b,n}$ is the fading loss between the typical UE and its associated BS and $I(\tau_{\mathcal{S}})$ is the interference to the typical UE. 
Following our convention regarding $\tau$, mentioned in Section~\ref{coverage_prob},
if $\tau_{\mathcal{S}} = L_{\mathcal{S}}$, then $C_{\tau(b,n)} = C_{L(b,n)}$ and $\alpha_{\tau(b,n)} = \alpha_{L(b,n)}$; otherwise, if $\tau_{\mathcal{S}} = N_{\mathcal{S}}$, then $C_{\tau(b,n)} = C_{N(b,n)}$ and $\alpha_{\tau(b,n)} = \alpha_{N(b,n)}$. The above expression for conditional coverage probability can be written as:
\begin{equation*}
\begin{aligned}
 & P \bigg(F_{b,n} > \frac{r^{\alpha_{\tau(b,n)}}Z\big(\sigma^2 + I(\tau_{\mathcal{S}})\big)}{C_{\tau(b,n)} G_{b,n}} \Big{|} \mathbb{T}, R = r \bigg) \\[-3pt]
 & \stackrel{\text{(i)}}= \int_u P \bigg(F_{b,n} > \frac{r^{\alpha_{\tau(b,n)}}Z(\sigma^2 + u)}{C_{\tau(b,n)} G_{b,n}} \Big{|} \mathbb{T}, R = r, I(\tau_{\mathcal{S}}) = u \bigg) f_{I(\tau_{\mathcal{S}})} (u) du \\[-4pt]
 & \stackrel{\text{(ii)}}= \mathbb{E}_{I(\tau_{\mathcal{S}})|R,\mathbb{T}} \bigg[P \bigg(F_{b,n} > \frac{r^{\alpha_{\tau(b,n)}}Z\big(\sigma^2 + I(\tau_{\mathcal{S}})\big)}{C_{\tau(b,n)} G_{b,n}} \bigg) \bigg]  \stackrel{\text{(iii)}}= \mathbb{E}_{I(\tau_{\mathcal{S}})|R,\mathbb{T}} \bigg[e^{ \Big(-{\frac{r^{\alpha_{\tau(b,n)}}Z\big(\sigma^2 + I(\tau_{\mathcal{S}})\big)}{C_{\tau(b,n)} G_{b,n}}} \Big)} \bigg] \\[-4pt]
 & = e^{ \Big(-{\frac{r^{\alpha_{\tau(b,n)}}Z\sigma^2}{C_{\tau(b,n)} G_{b,n}}} \Big)} \mathbb{E}_{I(\tau_{\mathcal{S}})|R,\mathbb{T}} \bigg[e^{ \Big(-{\frac{r^{\alpha_{\tau(b,n)}}Z I(\tau_{\mathcal{S}})}{C_{\tau(b,n)} G_{b,n}}} \Big)} \bigg]  \stackrel{\text{(iv)}}= e^{-\sigma^2 s_{\tau}} \mathcal{L}_{I(\tau_{\mathcal{S}})|R,\mathbb{T}} (s_{\tau})
\end{aligned}
\end{equation*}
where $f_{I(\tau_{\mathcal{S}})} (u)$ is the PDF of $I(\tau_{\mathcal{S}})$. In~\eqref{eq:convert_to_laplace}, $(i)$ follows from the total probability theorem, 
$(ii)$ follows from using the definition of expectation. $(iii)$ follows from our Rayleigh fading assumption, i.e., 
$f_{F_{b,n}} (x) = e^{-x}$ for $x \geq 0$ and $0$ otherwise. 
Finally, $(iv)$ follows from the definition of Laplace transform of a random variable \cite{elsawy2013stochastic}.

\section{Proof of Lemma~\ref{lemma:laplace_los_asscn}: Laplace transform for $I(L_{\mathcal{S}})$} \label{appendix:laplace_los}
$I(L_{\mathcal{S}})$ can be written as $I(L_{\mathcal{S}}) = I_{L}(L_{\mathcal{S}}) + I_{N}(L_{\mathcal{S}})$, where $I_L(L_{\mathcal{S}})$ and $I_N(L_{\mathcal{S}})$ are the interference caused by the LoS and NLoS interferers, respectively. Since the interference from the LoS interferers is independent of the interference from the NLoS interferers, 
\begin{equation*}
\mathcal{L}_{I(L_{\mathcal{S}})} = \mathcal{L}_{I_L(L_{\mathcal{S}})} . \mathcal{L}_{I_N(L_{\mathcal{S}})}
\end{equation*}
where $\mathcal{L}_{I(L_{\mathcal{S}})}$, $\mathcal{L}_{I_L(L_{\mathcal{S}})}$, and $\mathcal{L}_{I_N(L_{\mathcal{S}})}$ are the Laplace transform of $I(L_{\mathcal{S}})$, $I_L(L_{\mathcal{S}})$, and $I_N(L_{\mathcal{S}})$, respectively. Further, $I_L(L_{\mathcal{S}})$ can be written as $I_L(L_{\mathcal{S}}) = I_{L,h}(L_{\mathcal{S}}) + I_{L,d}(L_{\mathcal{S}})$, where $I_{L,h}(L_{\mathcal{S}})$ and $I_{L,d}(L_{\mathcal{S}})$ are parts of $I_L(L_{\mathcal{S}})$, that are caused by LoS interferers that act as hidden interferers and deaf interferers, respectively. Since the interference from the hidden interferers is independent of the interference from the deaf interferers,
\begin{equation*}
    \mathcal{L}_{I_L(L_{\mathcal{S}})} = \mathcal{L}_{I_{L,h}(L_{\mathcal{S}})} . \mathcal{L}_{I_{L,d}(L_{\mathcal{S}})}
\end{equation*}
where $\mathcal{L}_{I_{L,h}(L_{\mathcal{S}})}$ and $\mathcal{L}_{I_{L,d}(L_{\mathcal{S}})}$ are the Laplace transforms of $I_{L,h}(L_{\mathcal{S}})$ and $I_{L,d}(L_{\mathcal{S}})$, respectively. For $I_N(L_{\mathcal{S}})$, we can write, $I_N(L_{\mathcal{S}}) = I_{N,h}(L_{\mathcal{S}}) + I_{N,d}(L_{\mathcal{S}})$, where $I_{N,h}(L_{\mathcal{S}})$ and $I_{N,d}(L_{\mathcal{S}})$ are parts of $I_N(L_{\mathcal{S}})$, that are caused by NLoS interferers that act as hidden interferers and deaf interferers, respectively. Using a reasoning similar to the one used for $I_L(L_{\mathcal{S}})$, we can write,
\begin{equation*}
    \mathcal{L}_{I_N(L_{\mathcal{S}})} = \mathcal{L}_{I_{N,h}(L_{\mathcal{S}})} . \mathcal{L}_{I_{N,d}(L_{\mathcal{S}})}
\end{equation*} 
where $\mathcal{L}_{I_{N,h}(L_{\mathcal{S}})}$ and $\mathcal{L}_{I_{N,d}(L_{\mathcal{S}})}$ are the Laplace transforms of $I_{N,h}(L_{\mathcal{S}})$ and $I_{N,d}(L_{\mathcal{S}})$, respectively. Now, using the expression for $\mathcal{L}_{I_L(L_{\mathcal{S}})}$ and $\mathcal{L}_{I_N(L_{\mathcal{S}})}$ in $\mathcal{L}_{I(L_{\mathcal{S}})} = \mathcal{L}_{I_L(L_{\mathcal{S}})} . \mathcal{L}_{I_N(L_{\mathcal{S}})}$, we get 
\begin{equation*} 
    \mathcal{L}_{I(L_{\mathcal{S}})} (s_L) = 
    \mathcal{L}_{I_{L,h}(L_{\mathcal{S}})} (s_L) . \mathcal{L}_{I_{L,d}(L_{\mathcal{S}})} (s_L) . \mathcal{L}_{I_{N,h}(L_{\mathcal{S}})} (s_L) . \mathcal{L}_{I_{N,d}(L_{\mathcal{S}})} (s_L)
\end{equation*}
Using the expressions for each of the four terms on the right hand side of the above equation (derived below), we get the result of Lemma~\ref{lemma:laplace_los_asscn}.

We first consider the term $I_{L,h}(L_{\mathcal{S}})$, which can be written as:
\begin{equation*}
    \begin{aligned}
    I_{L,h}(L_{\mathcal{S}}) & = \sum_{\substack{\mathcal{S}^{''} \in \mathcal{P(O)} \\ n \notin \mathcal{S}^{''}}} \sum_{X_i \in L_{\mathcal{S}^{''}}} \sum_{l \in \mathcal{S}^{''}}  C_{L(i,l)} K_{i,l} F_{i,l} G_{i,l} ||X_i||^{-\alpha_{L(i,l)}} . \mathbf{1}_{X_i \in \mathcal{R}_{L,h}^c} \\[-5pt]
    & \hspace{0.3in} + \sum_{\substack{\mathcal{S}^{'} \in \mathcal{P(O)} \\ n \in \mathcal{S}^{'}}} \sum_{X_j \in L_{\mathcal{S}^{'}}} \sum_{m \in \mathcal{S}^{'}} C_{L(j,m)} K_{j,m} F_{j,m} G_{j,m} ||X_j||^{-\alpha_{L(j,m)}} . \mathbf{1}_{X_j \in B_0^c(r) \cap \mathcal{R}_{L,h}^c}  \\[-5pt]
    & \hspace{0.3in} + \hspace{0.2in} \sum_{\substack{q \in \mathcal{S}; \mathcal{S} \in \mathcal{P(O)} \\ q \neq n}} C_{L(0,q)} K_{0,q} F_{0,q} G_{0,q} r^{-\alpha_{L(0,q)}} .  
    \mathbf{1}_{X_{b,n} \in \mathcal{R}_{L,h}^c}
    \end{aligned}
\end{equation*}
Here we make a slight abuse of notation, and use $X_{i}$ and $X_{j}$ for BS locations belonging to $L_{\mathcal{S^{''}}}$ and $L_{\mathcal{S^{'}}}$, respectively. This will help us in conveniently applying the properties of PPPs to $X_{i}$ and $X_{j}$, as $L_{\mathcal{S^{''}}}$ and $L_{\mathcal{S^{'}}}$ are PPPs. The first term in the above equation is the interference from the inter-operator BSs, that do not share BS sites with operator $n$. These interferers can be anywhere except $\mathcal{R}_{L,h}$, which is the interference exclusion zone from LoS hidden interferers due to CS; $\mathcal{R}_{L,h}^c$ is the region outside $\mathcal{R}_{L,h}$. Due to CS, there is no LoS hidden interferer in $\mathcal{R}_{L,h}$; otherwise the BS at $X_{b,n}$ would not be engaged in downlink transmission to the typical UE (recall from Section~\ref{laplace} that we are analyzing the interference, conditioned on the event $\mathbb{T}$). With CST, $\mathcal{R}_{L,h}$ is a region around the sensing BS and with CSR, $\mathcal{R}_{L,h}$ is a region around sensing UE. 
The second term in $I_{L,h}(L_{\mathcal{S}})$ is the interference from the BS sites that have BS of operator $n$ as well as BSs of other operators. Thus, this term comprises of both intra and inter operator interference.
In this term, due to our association rule, there is no LoS interferer of operator $n$ inside $B_0(r)$. Accordingly, the BSs that are co-located with the BSs of operator $n$ cannot be inside $B_0(r)$. Further, the interfering BSs in this term must also be outside $\mathcal{R}_{L,h}$. The interference from the BSs co-located with $X_{b,n}$ is represented by the third term in $I_{L,h}(L_{\mathcal{S}})$. In this term the `$0$'s in the subscripts are just placeholders. Similar to the first two terms, the BSs co-located with $X_{b,n}$ interfere the typical UE only if they are outside $\mathcal{R}_{L,h}$. 

For a BS at $X_{j,m}$ to be a hidden interferer, we must have $X_{j,m} \in \mathcal{B}_{h,1}$ (see Section~\ref{hidden_def}). We use $K_{j,m}$, a random variable, to capture whether $X_{j,m} \in \mathcal{B}_{h,1}$. Specifically, $K_{j,m} = 1$ if $X_{j,m} \in \mathcal{B}_{h}$ and the BS is active. Without loss of generality, we assume $X_{j,m} \in \mathcal{B}_h$ with probability 0.5; otherwise, $X_{j,m} \in \mathcal{B}_d$ (see Fig.~\ref{fig:hidden_and_deaf}). The activity of the BS at $X_{j,m}$ is captured by its transmission probability, $p_{\mathbb{T}}$. Thus, we have the following distribution for $K_{j,m}$:
\begin{equation} \label{eq:active_prob}
    K_{j,m} = 
    \begin{cases}
    1 \text{ w.p. } 0.5 \times p_{\mathbb{T}} \\[-7pt]
    0 \text{ w.p. } 1 - (0.5 \times p_{\mathbb{T}})
    \end{cases}
\end{equation}

By taking the Laplace transform of $I_{L,h}(L_{\mathcal{S}})$, 
we get:
\begin{equation*}
    \begin{aligned}
    \mathcal{L}_{I_{L,h} (L_{\mathcal{S}})} (s_L) = & \prod_{\mathclap{\substack{\mathcal{S}^{''} \in \mathcal{P(O)} \\ n \notin \mathcal{S}}}} \mathbb{E}\Big[ \exp\Big(-s_L\sum_{\mathclap{\substack{X_i \in L_{\mathcal{S}^{''}} \\ l \in \mathcal{S}^{''}}}} C_{L(i,l)} Y_{i,l} ||X_i||^{-\alpha_{L(i,l)}} . \mathbf{1}_{X_i \in \mathcal{R}_{L,h}^c}\Big)\Big] \\[-4pt]
    & \hspace{0.2in} . \prod_{\mathclap{\substack{\mathcal{S}^{'} \in \mathcal{P(O)} \\ n \in \mathcal{S}^{'}}}} \mathbb{E}\Big[\exp\Big(-s_L \sum_{\mathclap{\substack{X_j \in L_{\mathcal{S}^{'}} \\ j \in \mathcal{S}^{'}}}} C_{L(j,m)} Y_{j,m} ||X_j||^{-\alpha_{L(j,m)}} . \mathbf{1}_{X_j \in B_0^c(r) \cap \mathcal{R}_{L,h}^c}\Big)\Big] \\[-4pt]
    & \hspace{0.2in} .  \prod_{\substack{q \in \mathcal{S}; \mathcal{S} \in \mathcal{P(O)} \\ q \neq n}} \mathbb{E}\Big[\exp\Big(-s_L C_{L(0,q)} K_{0,q} F_{0,q} G_{0,q} r^{-\alpha_{L(0,q)}} . \mathbf{1}_{r \in \mathcal{R}_{L,h}^c}\Big)\Big]
    \end{aligned}
\end{equation*}
In the above equation, we use the fact that the interference from the three terms in $I_{L,h}(L_{\mathcal{S}})$ are independent, and the elements of $\{\Phi_{\mathcal{S}}\}$ are also independent. Further, the interference from the BSs co-located with $X_{b,n}$ are independent too. In the above equation, $Y_{j,m} = K_{j,m} F_{j,m} G_{j,m}$. Using the probability generating function (PGFL) property of PPPs \cite{elsawy2013stochastic} for the first two terms of $\mathcal{L}_{I_{L,h} (L_{\mathcal{S}})}$, we can write:
\begin{equation*}
    \begin{aligned}
    \mathcal{L}_{I_{L,h} (L_{\mathcal{S}})} (s_L) = & \prod_{\mathclap{\substack{\mathcal{S}^{''} \in \mathcal{P(O)} \\ n \notin \mathcal{S}^{''}}}}  \exp\bigg(-\lambda_{\mathcal{S}^{''}} \int\limits_{\theta = 0} ^ {2\pi} \int\limits_{t = 0}^{\infty} \Big(1 - \mathbb{E}_Y \prod_{l \in \mathcal{S}^{''}} e^{-s_L C_L Y_l t^{-\alpha_L} . \mathbf{1}_{t \in \mathcal{R}_{L,h}^c}}\Big) t p_L(t) dt d\theta \bigg)\\[-12pt]
    & . \prod_{\mathclap{\substack{\mathcal{S}^{'} \in \mathcal{P(O)} \\ n \in \mathcal{S}^{'}}}}  \exp\bigg(-\lambda_{\mathcal{S}^{'}} \int\limits_{\theta = 0} ^ {2\pi} \int\limits_{t = r}^{\infty} \Big(1 - \mathbb{E}_Y \prod_{m \in \mathcal{S}^{'}} e^{-s_L C_L Y_m t^{-\alpha_L} . \mathbf{1}_{t \in \mathcal{R}_{L,h}^c}}\Big) t p_L(t) dt d\theta \bigg)\\[-12pt]
    & . \hspace{0.1in} \prod_{\substack{q \in \mathcal{S}; \mathcal{S} \in \mathcal{P(O)} \\ q \neq n}} \mathbb{E}_{K_q} \Big[ \mathbb{E}_{G_{q} | K_q}\Big[\mathcal{L}_{F_q|K_q, G_q}\Big(-s_L C_L K_q G_{q} r^{-\alpha_L} . \mathbf{1}_{r \in \mathcal{R}_{L,h}^c}\Big)\Big]\Big] 
    \end{aligned}
\end{equation*}
Note that the PGFL property is applied for the expectation with respect to the $X_i$ and $X_j$s. Thus, using the fact that $X_i$ and $X_j$s are independent of $Y_l$ and $Y_m$s, respectively, the expectations in the first two terms of the above equation are with respect to $Y_l = K_lF_lG_l$ and $Y_m = K_mF_mG_m$. We removed $i$ and $j$ from the subscripts of $Y_l$ and $Y_m$ as these random variables are independent of $X_i$ and $X_j$s. Similarly, we removed the subscripts $(i,l), (j,m)$, and $(0,q)$ from $C_L$ and $\alpha_L$ as these are constants. 
For the third term, we removed the `0' from the subscripts as the `0' was just a placeholder, and use the fact that $K_q$, $F_q$, and $G_q$ and independent. Let us denote the expression inside the product of the third term as $u_{L,h}(s_L,r)$. Next, using the independence of the $Y_l$s and $Y_m$s 
we can write: 
\begin{equation*}
    \begin{aligned}
    & \mathcal{L}_{I_{L,h} (L_{\mathcal{S}})} (s_L) =  \prod_{\mathclap{\substack{\mathcal{S}^{''} \in \mathcal{P(O)} \\ n \notin \mathcal{S}}}}  \exp\bigg(- 2\pi \lambda_{\mathcal{S}^{''}}  \int\limits_{t = 0}^{\infty} \Big(1 - \prod_{\mathclap{l \in \mathcal{S}^{''}}} \mathbb{E}_{Y_l} \Big[ e^{-s_L  C_L Y_l t^{-\alpha_L} . \mathbf{1}_{t \in \mathcal{R}_{L,h}^c}} \Big] \Big) t p_L(t) dt \bigg)\\[-10pt]
    & . \prod_{\mathclap{\substack{\mathcal{S}^{'} \in \mathcal{P(O)} \\ n \in \mathcal{S}^{'}}}}  \exp\bigg(- 2\pi \lambda_{\mathcal{S}^{'}} \int\limits_{t = r}^{\infty} \Big(1 - \prod_{\mathclap{m \in \mathcal{S}^{'}}} \mathbb{E}_{Y_m} \Big[ e^{-s_L  C_L Y_m t^{-\alpha_L} . \mathbf{1}_{t \in \mathcal{R}_{L,h}^c}} \Big] \Big) t p_L(t) dt \bigg) .  u_{L,h}(s_L,r)^{|\mathcal{S}| - 1}
    \end{aligned}
\end{equation*}
In the above equation, each of the expectations in the first two terms can be written as $u_{L,h}(s_L,t)$. Thus, the final expression for $\mathcal{L}_{I_{L,h} (L_{\mathcal{S}})}$ is: 
\begin{equation} \label{eq:laplace_los_ass_los_hidden}
    \begin{aligned}
    & \mathcal{L}_{I_{L,h} (L_{\mathcal{S}})} (s_L)= \prod_{\mathclap{\substack{\mathcal{S}^{''} \in \mathcal{P(O)} \\ n \notin \mathcal{S}}}}  \exp\bigg(-2\pi \lambda_{\mathcal{S}^{''}} \int\limits_{t = 0}^{\infty} \Big(1 - u_{L,h}(s_L,t) ^ {|\mathcal{S}^{''}|}\Big) t p_L(t) dt \bigg) \\[-10pt]
    & . \prod_{\mathclap{\substack{S^{'} \in \mathcal{P(O)} \\ n \in \mathcal{S}^{'}}}}  \exp\bigg(- 2\pi \lambda_{\mathcal{S}^{'}} \int\limits_{t = r}^{\infty} \Big(1 - u_{L,h}(s_L,t) ^ {|S^{'}|}\Big) t p_L(t) dt \bigg) .  u_{L,h}(s_L,r)^{|\mathcal{S}| - 1}
    \end{aligned}
\end{equation}

Next, we consider the term $I_{L,d}(L_{\mathcal{S}})$ and find its Laplace transform. The analysis for $I_{L,d}(L_{\mathcal{S}})$ is almost same as that for $I_{L,h}(L_{\mathcal{S}})$, with the only exception that the interference exclusion zone of the LoS deaf interferers is $\mathcal{R}_{L,d}$, not $\mathcal{R}_{L,h}$. Accordingly the Laplace transform of $I_{L,d}(L_{\mathcal{S}})$ is:
\begin{equation*}
    \begin{aligned}
    & \mathcal{L}_{I_{L,d} (L_{\mathcal{S}})} (s_L)= \prod_{\mathclap{\substack{\mathcal{S}^{''} \in \mathcal{P(O)} \\ n \notin \mathcal{S}}}}  \exp\bigg(- 2\pi \lambda_{\mathcal{S}^{''}} \int\limits_{t = 0}^{\infty} \Big(1 - u_{L,d}(s_L,t) ^ {|\mathcal{S}^{''}|}\Big) t p_L(t) dt \bigg) \\[-10pt]
    & . \prod_{\mathclap{\substack{\mathcal{S}^{'} \in \mathcal{P(O)} \\ n \in \mathcal{S}^{'}}}}  \exp\bigg(- 2\pi \lambda_{\mathcal{S}^{'}} \int\limits_{t = r}^{\infty} \Big(1 - u_{L,d}(s_L,t) ^ {|\mathcal{S}^{'}|}\Big) t p_L(t) dt \bigg) .  u_{L,d}(s_L,r)^{|\mathcal{S}| - 1}
    \end{aligned}
\end{equation*}
where $u_{L,d}(s_L,r) = \mathbb{E}_{K_q} \big[ \mathbb{E}_{G_{q} | K_q}\big[\mathcal{L}_{F_q|K_q, G_q}\big(-s_L C_L K_q G_{q} r^{-\alpha_L} . \mathbf{1}_{r \in \mathcal{R}_{L,d}^c}\big)\big]\big]$. 

Next, we find the Laplace transform of $I_{N,h}(L_{\mathcal{S}})$. $I_{N,h}(L_{\mathcal{S}})$ can be written as:
\begin{equation*}
    \begin{aligned}
    I_{N,h} (L_{\mathcal{S}}) = & \sum_{\substack{\mathcal{S}^{''} \in \mathcal{P(O)} \\ n \notin \mathcal{S}}} \sum_{X_i \in N_{\mathcal{S}^{''}}} \sum_{l \in \mathcal{S}^{''}} C_N K_{i,l} F_{i,l} G_{i,l} ||X_i||^{-\alpha_N} . \mathbf{1}_{X_i \in \mathcal{R}_{N,h}^c} \\[-5pt]
    & \hspace{0.1in} + \sum_{\substack{\mathcal{S}^{'} \in \mathcal{P(O)} \\ n \in \mathcal{S}^{'}}} \sum_{X_j \in N_{\mathcal{S}^{'}}} \sum_{m \in \mathcal{S}^{'}} C_N K_{j,m} F_{j,m} G_{j,m} ||X_j||^{-\alpha_N} . \mathbf{1}_{X_j \in B_0^c(D_N(r)) \cap \mathcal{R}_{N,h}^c}
    \end{aligned}
\end{equation*}
Compared to the expression for $I_{L,h} (L_{\mathcal{S}})$, we do not have the third term in $I_{N,h} (L_{\mathcal{S}})$ because the typical UE has a LoS link with its associated BS. Hence, other BSs, that are co-located with the typical UE's associated BS cannot have NLoS links with the typical UE. $\mathcal{R}_{N,h}$ is the interference exclusion zone from NLoS hidden interferers due to CS. Also, in this case the interference exclusion zone due to our association rule is $B_0(D_N(r))$ as the interferers are NLoS and the typical UE is associated with a BS via a LoS link (see~\eqref{eq:exclusion_zone}). Using steps similar to the ones used in the case of $I_{L,h} (L_{\mathcal{S}})$, we can write the Laplace transform of $I_{N,h} (L_{\mathcal{S}})$ as: 
\begin{equation*}
    \begin{aligned}
    \mathcal{L}_{I_{N,h} (L_{\mathcal{S}})} (s_L)= & \prod_{\mathclap{\substack{\mathcal{S}^{''} \in \mathcal{P(O)} \\ n \notin \mathcal{S}^{''}}}}  \exp\bigg(- 2\pi \lambda_{\mathcal{S}^{''}} \int\limits_{t = 0}^{\infty} \Big(1 - u_{N,h}(s_L,t) ^ {|\mathcal{S}^{''}|}\Big) t p_N(t) dt \bigg) \\[-10pt]
    & \hspace{0.1in} . \prod_{\mathclap{\substack{\mathcal{S}^{'} \in \mathcal{P(O)} \\ n \in \mathcal{S}^{'}}}}  \exp\bigg(- 2\pi \lambda_{\mathcal{S}^{'}} \int\limits_{\mathclap{t = D_N(r)}}^{\infty} \Big(1 - u_{N,h}(s_L,t) ^ {|\mathcal{S}^{'}|}\Big) t p_N(t) dt \bigg)
    \end{aligned}
\end{equation*}
Similarly, the Laplace transform $I_{N,d} (L_{\mathcal{S}})$ is:
\begin{equation*}
    \begin{aligned}
    \mathcal{L}_{I_{N,d} (L_{\mathcal{S}})} (s_L)= & \prod_{\mathclap{\substack{\mathcal{S}^{''} \in \mathcal{P(O)} \\ n \notin \mathcal{S}^{''}}}}  \exp\bigg(- 2\pi \lambda_{\mathcal{S}^{''}} \int\limits_{t = 0}^{\infty} \Big(1 - u_{N,d}(s_L,t) ^ {|\mathcal{S}^{''}|}\Big) t p_N(t) dt \bigg) \\[-10pt]
    & \hspace{0.1in} . \prod_{\mathclap{\substack{\mathcal{S}^{'} \in P(O) \\ n \in \mathcal{S}^{'}}}}  \exp\bigg(- 2\pi \lambda_{\mathcal{S}^{'}} \int\limits_{\mathclap{t = D_N(r)}}^{\infty} \Big(1 - u_{N,d}(s_L,t) ^ {|\mathcal{S}^{'}|}\Big) t p_N(t) dt \bigg)
    \end{aligned}
\end{equation*}
where $u_{N,h}(s_L,t) = \mathbb{E}_{K_q} \big[ \mathbb{E}_{G_{q} | K_q}\big[\mathcal{L}_{F_q|K_q, G_q}\big(-s_L C_N K_q G_{q} t^{-\alpha_N} . \mathbf{1}_{t \in \mathcal{R}_{N,h}^c}\big)\big]\big] $ and $u_{N,d}(s_L,t) = \mathbb{E}_{K_q} \big[ \mathbb{E}_{G_{q} | K_q}\big[\mathcal{L}_{F_q|K_q, G_q}\big(-s_L C_N K_q G_{q} t^{-\alpha_N} . \mathbf{1}_{t \in \mathcal{R}_{N,d}^c}\big)\big]\big]$.
The expressions for $u_{L,h}(s_L,t)$, $u_{L,d}(s_L,t)$, $u_{N,h}(s_L,t)$, and $u_{N,d}(s_L,t)$ are derived in Appendix~\ref{appendix:u_def}.

\section{Proof of Lemma~\ref{lemma:laplace_nlos_asscn}: Laplace transform for $I(N_{\mathcal{S}})$} \label{appendix:laplace_nlos}
In this case, the total interference to the typical UE is $I(N_{\mathcal{S}}) = I_{L}(N_{\mathcal{S}}) + I_{N}(N_{\mathcal{S}})$, where $I_L(N_{\mathcal{S}})$ and $I_N(N_{\mathcal{S}})$ are the interference caused by the LoS and NLoS interferers, respectively.
Using a set of arguments similar to the ones used for $ \mathcal{L}_{I(L_{\mathcal{S}})} (s_L)$, we can write, 
\begin{equation*} 
    \mathcal{L}_{I(N_{\mathcal{S}})} (s_N) =
    \mathcal{L}_{I_{L,h}(N_{\mathcal{S}})} (s_N) . \mathcal{L}_{I_{L,d}(N_{\mathcal{S}})} (s_N). \mathcal{L}_{I_{N,h}(N_{\mathcal{S}})} (s_N). \mathcal{L}_{I_{N,d}(N_{\mathcal{S}})} (s_N)
\end{equation*}
where $I_{L,h}(N_{\mathcal{S}})$ and $I_{L,d}(N_{\mathcal{S}})$ are parts of $I_L(N_{\mathcal{S}})$, that are caused by LoS interferers that act as hidden interferers and deaf interferers, respectively. $\mathcal{L}_{I_{L,h}(N_{\mathcal{S}})}$ and $\mathcal{L}_{I_{L,d}(N_{\mathcal{S}})}$ are the Laplace transforms of $I_{L,h}(N_{\mathcal{S}})$ and $I_{N,d}(N_{\mathcal{S}})$, respectively. $I_{N,h}(N_{\mathcal{S}})$ and $I_{N,d}(N_{\mathcal{S}})$ are parts of $I_N(N_{\mathcal{S}})$, that are caused by NLoS interferers that act as hidden interferers and deaf interferers, respectively. $\mathcal{L}_{I_{N,h}(N_{\mathcal{S}})}$ and $\mathcal{L}_{I_{N,d}(N_{\mathcal{S}})}$ are the Laplace transforms of $I_{N,h}(N_{\mathcal{S}})$ and $I_{N,d}(N_{\mathcal{S}})$, respectively. The constituent Laplace transforms in $\mathcal{L}_{I(N_S)}$ can be obtained using the approach described in Appendix~\ref{appendix:laplace_los}. For the sake of brevity, we omit the derivation and provide the final expressions. Using the following expressions for $\mathcal{L}_{I_{L,h}(N_S)}$, $\mathcal{L}_{I_{L,d}(N_S)}$, $\mathcal{L}_{I_{N,h}(N_S)}$, and $\mathcal{L}_{I_{N,d}(N_S)}$ in the above equation, we can obtain the result in Lemma~\ref{lemma:laplace_nlos_asscn}.
\begin{equation*}
    \begin{aligned}
    \mathcal{L}_{I_{N,h} (N_{\mathcal{S}})} (s_N) = & \prod_{\mathclap{\substack{\mathcal{S}^{''} \in \mathcal{P(O)} \\ n \notin \mathcal{S}^{''}}}}  \exp\bigg(- 2\pi \lambda_{\mathcal{S}^{''}} \int\limits_{t = 0}^{\infty} \Big(1 - u_{N,h}(s_N,t) ^ {|\mathcal{S}^{''}|}\Big) t p_N(t) dt \bigg) \\[-10pt]
    & . \hspace{0.2in} \prod_{\mathclap{\substack{\mathcal{S}^{'} \in \mathcal{P(O)} \\ n \in \mathcal{S}^{'}}}}  \exp\bigg(- 2\pi \lambda_{\mathcal{S}^{'}} \int\limits_{t = r}^{\infty} \Big(1 - u_{N,h}(s_N,t) ^ {|\mathcal{S}^{'}|}\Big) t p_N(t) dt \bigg) .  u_{N,h}(s_N,r)^{|\mathcal{S}| - 1} \\[-10pt]
    \mathcal{L}_{I_{N,d} (N_{\mathcal{S}})} (s_N) = & \prod_{\mathclap{\substack{\mathcal{S}^{''} \in \mathcal{P(O)} \\ n \notin \mathcal{S}^{''}}}}  \exp\bigg(- 2\pi \lambda_{\mathcal{S}^{''}} \int\limits_{t = 0}^{\infty} \Big(1 - u_{N,d}(s_N,t) ^ {|\mathcal{S}^{''}|}\Big) t p_N(t) dt \bigg) \\[-10pt]
    & . \hspace{0.2in} \prod_{\mathclap{\substack{\mathcal{S}^{'} \in \mathcal{P(O)} \\ n \in \mathcal{S}^{'}}}}  \exp\bigg(- 2\pi \lambda_{\mathcal{S}^{'}} \int\limits_{t = r}^{\infty} \Big(1 - u_{N,d}(s_N,t) ^ {|\mathcal{S}^{'}|}\Big) t p_N(t) dt \bigg) .  u_{N,d}(s_N,r)^{|\mathcal{S}| - 1}
    \end{aligned}
\end{equation*}
\begin{equation*}
    \begin{aligned}
    \mathcal{L}_{I_{L,h} (N_{\mathcal{S}})} (s_N) = & \prod_{\mathclap{\substack{\mathcal{S}^{''} \in \mathcal{P(O)} \\ n \notin \mathcal{S}^{''}}}}  \exp\bigg(- 2\pi \lambda_{\mathcal{S}^{''}} \int\limits_{t = 0}^{\infty} \Big(1 - u_{L,h}(s_N,t) ^ {|\mathcal{S}^{''}|}\Big) t p_L(t) dt \bigg) \\[-10pt]
    & . \hspace{0.2in} \prod_{\mathclap{\substack{\mathcal{S}^{'} \in \mathcal{P(O)} \\ n \in \mathcal{S}^{'}}}}  \exp\bigg(- 2\pi \lambda_{\mathcal{S}^{'}} \int\limits_{\mathclap{t = D_L(r)}}^{\infty} \Big(1 - u_{L,h}(s_N,t) ^ {|\mathcal{S}^{'}|}\Big) t p_L(t) dt \bigg) \\[-10pt]
    \mathcal{L}_{I_{L,d} (N_{\mathcal{S}})} (s_N) = & \prod_{\mathclap{\substack{\mathcal{S}^{''} \in \mathcal{P(O)} \\ n \notin \mathcal{S}^{''}}}}  \exp\bigg(- 2\pi \lambda_{\mathcal{S}^{''}} \int\limits_{t = 0}^{\infty} \Big(1 - u_{L,d}(s_N,t) ^ {|\mathcal{S}^{''}|}\Big) t p_L(t) dt \bigg) \\[-5pt]
    & . \hspace{0.2in} \prod_{\mathclap{\substack{\mathcal{S}^{'} \in \mathcal{P(O)} \\ n \in \mathcal{S}^{'}}}}  \exp\bigg(- 2\pi \lambda_{\mathcal{S}^{'}} \int\limits_{\mathclap{t = D_L(r)}}^{\infty} \Big(1 - u_{L,d}(s_N,t) ^ {|\mathcal{S}^{'}|}\Big) t p_L(t) dt \bigg)
    \end{aligned}
\end{equation*}

\section{Expressions for $u_{L,h}(s_L,t)$, $u_{L,d}(s_L,t)$, $u_{N,h}(s_L,t)$, and $u_{N,d}(s_L,t)$} \label{appendix:u_def}
These terms can be written as: 
$u_{L,h}(s_L,t) = 1$, if $t \in \mathcal{R}_{L,h}$, otherwise $u_{L,h}(s_L,t) = u_L(s_L,t)$; $u_{L,d}(s_L,t) = 1$ if $t \in \mathcal{R}_{L,d}$, otherwise $u_{L,d}(s_L,t) = u_L(s_L,t)$;
$u_{N,h}(s_L,t) = 1$ if $t \in \mathcal{R}_{N,h}$, otherwise $u_{N,h}(s_L,t) = u_N(s_L,t)$; $u_{N,d}(s_L,t) = 1$ if $t \in \mathcal{R}_{N,d}$, otherwise $u_{N,d}(s_L,t) = u_N(s_L,t)$, where $u_{L}(s_L,t) = \mathbb{E}_{K_n} \big[ \mathbb{E}_{G_{n} | K_n}\big[\mathcal{L}_{F_n|K_n, G_n}\big(-s_L C_L K_n G_{n} t^{-\alpha_L} \big)\big]\big]$
and $u_{N}(s_L,t)$ is same as $u_{L}(s_L,t)$ with $C_L$ and $\alpha_L$ replaced by $C_N$ and $\alpha_N$. Thus, it suffices to find the closed form expressions for $u_{L}(s_L,t)$, as shown below.
\begin{equation*}
\begin{aligned}
    & u_{L}(s_L,t) \stackrel{\text{(i)}} = \mathbb{E}_{K_n} \Big[ \mathbb{E}_{G_{n} | K_n}\Big[\mathbb{E}_{F_n|K_n, G_n}\Big(e^{-s_L C_L K_n F_n G_{n} t^{-\alpha_L}} \Big)\Big]\Big]  \\
    & \stackrel{\text{(ii)}}  = \mathbb{E}_{K_n} \Big[ \mathbb{E}_{G_{n} | K_n}\Big[\frac{1}{1 + s_L C_L K_n G_{n} t^{-\alpha_L}}\Big]\Big] \\
    & \stackrel{\text{(iii)}}  = \mathbb{E}_{K_n} \Big[ \frac{(\theta_{BS}/2 \pi)(\theta_{UE}/ 2 \pi)}{1 + s_L C_L K_n M_{BS}M_{UE} t^{-\alpha_L}} + \frac{(\theta_{BS} /2 \pi)(1 - \theta_{UE} /2 \pi)}{1 + s_L C_L K_n M_{BS}m_{UE} t^{-\alpha_L}} \\
    & \hspace{0.7in} + \frac{(1 - \theta_{BS} /2 \pi)(\theta_{UE} /2 \pi)}{1 + s_L C_L K_n m_{BS}M_{UE} t^{-\alpha_L}}
     + \frac{(1 - \theta_{BS}/2 \pi)(1 - \theta_{UE}/2 \pi)}{1 + s_L C_L K_n m_{BS}m_{UE} t^{-\alpha_L}} \Big] \\
    & \stackrel{\text{(iv)}}  = \Big(1 - \frac{p_T}{2}\Big) . \bigg[\frac{\theta_{BS}}{2 \pi} . \frac{\theta_{UE}}{2 \pi} + \frac{\theta_{BS}}{2 \pi} . \Big(1 - \frac{\theta_{UE}}{2 \pi}\Big) + \Big(1 - \frac{\theta_{BS}}{2 \pi}\Big) . \frac{\theta_{UE}}{2 \pi} + \Big(1 - \frac{\theta_{BS}}{2 \pi}\Big) . \Big(1 - \frac{\theta_{UE}}{2 \pi} \Big) \bigg] \\
    & \hspace{0.2in} + \frac{p_T}{2} . \bigg[ \frac{(\theta_{BS}/2 \pi)(\theta_{UE}/2 \pi)}{1 + s_L C_L M_{BS}M_{UE} t^{-\alpha_L}} + \frac{(\theta_{BS}/2 \pi)(1 - \theta_{UE}/2 \pi)}{1 + s_L C_L M_{BS}m_{UE} t^{-\alpha_L}} \\
    & \hspace{1.0in} + \frac{(1 - \theta_{BS}/2 \pi)(\theta_{UE}/2 \pi)}{1 + s_L C_L m_{BS}M_{UE} t^{-\alpha_L}} + \frac{(1 - \theta_{BS}/2 \pi)(1 - \theta_{UE}/2 \pi) }{1 + s_L C_L  m_{BS}m_{UE} t^{-\alpha_L}} \bigg]
\end{aligned}
\end{equation*}
where $(i)$ follows from the definition of Laplace transform, $(ii)$ from the PDF of $F_n$, $(iii)$ from the PDF of $G_n$, and $(iv)$ from the PDF of $K_n$.

\section{Proof for Lemma~\ref{lemma:tx_prob}: Average number of contenders} \label{appendix:tx_prob}
Let us assume that the CS node (BS or UE) is located at $Z_{cs}$. We define $ \stackrel \frown Z_{cs}(R_{cs,L}, \theta_{cs})$ as an arc of radius $R_{cs,L}$ and width $\theta_{cs}$, centered at $Z_{cs}$ with directionality the same as the CS node's sensing directionality. We use $\stackrel \frown {Z_{cs}^c} (r_{cs,L}, \theta_{cs})$ for the arc that is complement of $ \stackrel \frown Z_{cs}(R_{cs,L}, \theta_{cs})$, but with a different radius, $r_{cs,L}$. The LoS contenders of the CS node are present inside the arcs  $ \stackrel \frown Z_{cs}(R_{cs,L}, \theta_{cs})$ and $\stackrel \frown {Z_{cs}^c} (r_{cs,L}, \theta_{cs})$. Thus, $\bar{N}_{c,L}$ can be written as:
\begin{equation*}
    \bar{N}_{c,L} = \mathbb{E} \Big[ \sum_{\mathclap{\substack{\mathcal{S}^{''} \in \mathcal{P(O)} \\ n \notin \mathcal{S}^{''}}}} \Big( N_{c,L}^{\mathcal{S}^{''}, \stackrel \frown Z_{cs}(R_{cs,L}, \theta_{cs})}  + N_{c,L}^{\mathcal{S}^{''}, \stackrel \frown {Z_{cs}^c} (r_{cs,L}, \theta_{cs})} \Big) + \sum_{\mathclap{\substack{\mathcal{S}^{'} \in \mathcal{P(O)} \\ n \in \mathcal{S}^{'}}}} \Big( N_{c,L}^{\mathcal{S}^{'}, \stackrel \frown Z_{cs}(R_{cs,L}, \theta_{cs})}  + N_{c,L}^{\mathcal{S}^{'}, \stackrel \frown {Z_{cs}^c} (r_{cs,L}, \theta_{cs})} \Big) \Big]
\end{equation*}
where the expectation is over the operator PPPs and the arc lengths. The arc lengths are random variables because they depend on the antenna gain between a CS node and its contender, which is not deterministic (see~\eqref{eq:ocst_gain},~\eqref{eq:dcst_gain},~\eqref{eq:ocsr_gain}).
$N_{c,L}^{\mathcal{S}^{''}, \stackrel \frown Z_{cs}(R_{cs,L}, \theta_{cs})}$ and $N_{c,L}^{\mathcal{S}^{''}, \stackrel \frown {Z_{cs}^c} (r_{cs,L}, \theta_{cs})}$ are the number of LoS BSs of operator $\mathcal{S}^{''}; n \notin \mathcal{S}^{''}$ inside the arcs $ \stackrel \frown Z_{cs}(R_{cs,L}, \theta_{cs})$ and $\stackrel \frown {Z_{cs}^c} (r_{cs,L}, \theta_{cs})$, respectively. Similarly, $N_{c,L}^{\mathcal{S}^{'}, \stackrel \frown Z_{cs}(R_{cs,L}, \theta_{cs})}$ and $N_{c,L}^{\mathcal{S}^{'}, \stackrel \frown {Z_{cs}^c} (r_{cs,L}, \theta_{cs})}$  are the number of LoS BSs of operator $\mathcal{S}^{'}; n \in \mathcal{S}^{'}$ inside the arcs $ \stackrel \frown Z_{cs}(R_{cs,L}, \theta_{cs})$ and $\stackrel \frown {Z_{cs}^c} (r_{cs,L}, \theta_{cs})$, respectively. Since the number of contenders belonging to different $\mathcal{S}^{''} \text{or } \mathcal{S}^{'} \in \mathcal{P(O)}$ are independent, and $N_{c,L}^{\mathcal{S}^{''}, \stackrel \frown Z_{cs}(R_{cs,L}, \theta_{cs})}$, $N_{c,L}^{\mathcal{S}^{''}, \stackrel \frown {Z_{cs}^c} (r_{cs,L}, \theta_{cs})}$, $N_{c,L}^{\mathcal{S}^{'}, \stackrel \frown Z_{cs}(R_{cs,L}, \theta_{cs})}$, and $N_{c,L}^{\mathcal{S}^{'}, \stackrel \frown {Z_{cs}^c} (r_{cs,L}, \theta_{cs})}$ are also independent of each other, we can write:
\begin{equation} \label{eq:sum_of_contenders}
\begin{aligned}
    \bar{N}_{c,L} = & \sum_{\mathclap{\substack{\mathcal{S}^{''} \in \mathcal{P(O)} \\ n \notin \mathcal{S}^{''}}}} \Big( \mathbb{E} \Big[ N_{c,L}^{\mathcal{S}^{''}, \stackrel \frown Z_{cs}(R_{cs,L}, \theta_{cs})} \Big] + \mathbb{E} \Big[ N_{c,L}^{\mathcal{S}^{''}, \stackrel \frown {Z_{cs}^c} (r_{cs,L}, \theta_{cs})} \Big] \Big) \\[-6pt]
    & \hspace{0.2in} + \sum_{\mathclap{\substack{\mathcal{S}^{'} \in \mathcal{P(O)} \\ n \in \mathcal{S}^{'}}}} \Big( \mathbb{E} \Big[ N_{c,L}^{\mathcal{S}^{'}, \stackrel \frown Z_{cs}(R_{cs,L}, \theta_{cs})} \Big] + \mathbb{E} \Big[ N_{c,L}^{\mathcal{S}^{'}, \stackrel \frown {Z_{cs}^c} (r_{cs,L}, \theta_{cs})} \Big] \Big) 
\end{aligned}
\end{equation}
Using Campbell's theorem (see Theorem 2 in \cite{bai2014coverage}), and the fact that $r_{cs,L}$ and $R_{cs,L}$ are independent of the PPPs, we can write:
\begin{equation*}
\begin{aligned}
    \mathbb{E} \Big[ N_{c,L}^{\mathcal{S}^{''}, \stackrel \frown Z_{cs}(R_{cs,L}, \theta_{cs})} \Big] = \lambda_{\mathcal{S}^{''}} \mathbb{E}_{R_{cs,L}} \Big[ \int_{\theta = 0} ^ {\theta_{cs}} \int_{t = 0} ^ {R_{cs,L}}  p_L(t) t dt d\theta \Big] \\[-2pt]
    \mathbb{E} \Big[ N_{c,L}^{\mathcal{S}^{''}, \stackrel \frown {Z_{cs}^c} (r_{cs,L}, \theta_{cs})} \Big] = \lambda_{\mathcal{S}^{''}} \mathbb{E}_{r_{cs,L}} \Big[ \int_{\theta = \theta_{cs}} ^ {2\pi} \int_{t = 0} ^ {r_{cs,L}}  p_L(t) t dt d\theta \Big]
    \end{aligned}
\end{equation*}
Similarly, we can write, 
\begin{equation*}
\begin{aligned}
    \mathbb{E} \Big[ N_{c,L}^{\mathcal{S}^{'}, \stackrel \frown Z_{cs}(R_{cs,L}, \theta_{cs})} \Big]  = \lambda_{\mathcal{S}'} \mathbb{E}_{R_{cs,L}} \Big[ \int_{\theta = 0} ^ {\theta_{cs}} \int_{t = D(\bar{R})} ^ {R_{cs,L}}  p_L(t) t dt d\theta \Big] \\[-2pt]
    \mathbb{E} \Big[ N_{c,L}^{\mathcal{S}^{'}, \stackrel \frown {Z_{cs}^c} (r_{cs,L}, \theta_{cs})} \Big] = \lambda_{\mathcal{S}'} \mathbb{E}_{R_{cs,L}} \Big[  \int_{\theta = \theta_{cs}} ^ {2\pi} \int_{t = D(\bar{R})} ^ {r_{cs,L}}  p_L(t) t dt d\theta \Big]
 \end{aligned}
\end{equation*}
where we use the fact that $D(\bar{R})$ is the radius of the interference exclusion zone (circular) around the CS node where LoS and NLoS interferers (thus, contenders) belonging to $\mathcal{S}^{'} \in \mathcal{P(O)}; n \in \mathcal{S}^{'}$ cannot be present. Due to our association criteria, there is an interference exclusion zone around the typical UE, where no interferers belonging to network $n$ (typical UE's subscribed network) would be present \cite{jurdi2018modeling}. Accordingly, in this interference exclusion zone there would be no interfering BSs belonging to operators other than $n$, that share BS sites with that of network $n$ \cite{jurdi2018modeling}. Thus, when CSR is used (which happens after the association), no contender belonging to $\mathcal{S}^{'} \in \mathcal{P(O)}; n \in \mathcal{S}^{'}$ would be present in this interference exclusion zone. This interference exclusion zone depends on the association distance and the association link type. Specifically, if a UE at $U_{x,y}$ associates with a BS using a LoS link of distance $r$ meters, then the LoS interference exclusion zone is $B_{x,y}(r)$ (ball of radius $r$ centered at $U_{x,y}$), and the NLoS interference exclusion zone is $B_{x,y}(D_N(r))$. If the UE associates with a BS using a NLoS link of distance $r$ meters, then the NLoS interference exclusion zone is $B_{x,y}(r)$, and the LoS interference exclusion zone is $B_{x,y}(D_L(r))$. $D_L(r)$ and $D_N(r)$ have been defined in~\eqref{eq:exclusion_zone}. Since we are using the average association distance, $\bar{R}$, for the transmission probability analysis, we assume that the interference exclusion radius for both LoS and NLoS interferers are $\bar{R}$. The fact that, in general, $D_N(r) < r$ and $D_L(r) > r$, makes this a reasonable assumption for the average analysis scenario. For the CST schemes, $D(\bar{R}) = 0$ as the interference exclusion zones due to our association criteria are around the UEs, not the BSs.
Finally, using the above expressions for $\mathbb{E} \big[ N_{c,L}^{\mathcal{S}^{''}, \stackrel \frown Z_{cs}(R_{cs,L}, \theta_{cs})} \big]$, $\mathbb{E} \big[ N_{c,L}^{\mathcal{S}^{''}, \stackrel \frown {Z_{cs}^c} (r_{cs,L}, \theta_{cs})} \big]$, $\mathbb{E} \big[ N_{c,L}^{\mathcal{S}^{'}, \stackrel \frown Z_{cs}(R_{cs,L}, \theta_{cs})} \big]$, and $\mathbb{E} \big[ N_{c,L}^{\mathcal{S}^{'}, \stackrel \frown {Z_{cs}^c} (r_{cs,L}, \theta_{cs})} \big]$ in~\eqref{eq:sum_of_contenders} we get~\eqref{eq:los_contenders} in Lemma~\ref{lemma:tx_prob}. Proceeding in a manner similar to the case of $\bar{N}_{c,L}$, we can obtain the expression for $\bar{N}_{c,N}$, given in ~\eqref{eq:nlos_contenders}.

The expression for $\bar{N}_c^A$ in~\eqref{eq:announcement_contenders} can also be obtained in a similar way, while taking into account a couple of differences. First, the announcements are listened by the BS, thus, there is no interference exclusion zone. Note that, while listening for the announcements, the contenders to a BS are the UEs sending out announcements. However, we use the fact that the number of contending UEs is the same as the number of BSs in the sensing region of the announcements (each BS serves one UE at a time). Thus, we use $\lambda_{\mathcal{S}^{'''}}; \mathcal{S}^{'''} \in \mathcal{P(O)}$ in~\eqref{eq:announcement_contenders}. Second, $R_{A,L}$, $R_{A,N}$, $r_{A,L}$ and $r_{A,N}$ are deterministic because the announcements are sent out omnidirectionally. Thus there are no expectations with respect to $R_{A,L}$, $R_{A,N}$, $r_{A,L}$ and $r_{A,N}$ in the expression for $N_c^A$.

\section{Distributions of sensing distances for different CS protocols} \label{appendix:tx_prob_dist}
The sensing distance of a CS node is the distance of the farthest contender that the CS node can sense. Based on this definition, the distributions of $R_{cs,L}$, $R_{cs,N}$, $r_{cs,L}$ and $r_{cs,N}$ for different protocols, and the expressions for $R_{A,L}$, $R_{A,N}$, $r_{A,L}$ and $r_{A,N}$ (introduced in Section~\ref{tx_prob}) are given below. 
In the following, the distributions vary with protocols as $\theta_{cs}$ is different for different protocols.

\noindent \textbf{\textit{o}CST}: In this case, $\theta_{cs} = 2\pi$ and the sensing node is a BS. Thus, 
\begin{equation*}
    R_{cs,L} = r_{cs,L} =
    \begin{cases}
            \big(\frac{P_X C_{L} M_{BS} \times 10^{-0.7} \times M_{BS}}{P_{th}}\big) ^ {\frac{1}{\alpha_{L}}} \text{ w.p. } \frac{\theta_{BS}}{2\pi}  \\[-7pt]  
            \big(\frac{P_X C_{L} M_{BS} \times 10^{-0.7} \times m_{BS}}{P_{th}}\big) ^ {\frac{1}{\alpha_{L}}} \text{ w.p. } \big(1 - \frac{\theta_{BS}}{2\pi}\big)
    \end{cases}
\end{equation*}

\noindent \textbf{\textit{d}CST}: In this case, $\theta_{cs} = \theta_{BS}$, and the sensing node is a BS. Thus, 
\begin{equation*}
\hspace{0.01in} R_{cs,L} = 
    \begin{cases}
         \big(\frac{P_X C_{L} M_{BS} M_{BS}}{P_{th}}\big) ^ {\frac{1}{\alpha_{L}}} \text{ w.p. } \frac{\theta_{BS}}{2\pi}  \\[-7pt]
        \big(\frac{P_X C_{L} M_{BS} m_{BS}}{P_{th}}\big) ^ {\frac{1}{\alpha_{L}}} \text{ w.p. } \big(1 - \frac{\theta_{BS}}{2\pi}\big)
    \end{cases}
\text{; } r_{cs,L} = 
        \begin{cases}
         \big(\frac{P_X C_{L} m_{BS} M_{BS}}{P_{th}}\big) ^ {\frac{1}{\alpha_{L}}} \text{ w.p. } \frac{\theta_{BS}}{2\pi}  \\[-7pt]
        \big(\frac{P_X C_{L} m_{BS} m_{BS}}{P_{th}}\big) ^ {\frac{1}{\alpha_{L}}} \text{ w.p. } \big(1 - \frac{\theta_{BS}}{2\pi}\big)
    \end{cases}
\end{equation*}

\noindent \textbf{\textit{o}CSR}: In this case, $\theta_{cs} = 2\pi$, and the sensing node is a UE. Thus,
\begin{equation*}
   R_{cs,L} = r_{cs,L} = 
   \begin{cases}
        \big(\frac{P_X C_{L} M_{UE} \times 10^{-0.7} \times M_{BS}}{P_{th}}\big) ^ {\frac{1}{\alpha_{L}}} \text{ w.p. } \frac{\theta_{BS}}{2\pi}\\[-7pt]
        \big(\frac{P_X C_{L} M_{UE} \times 10^{-0.7} \times m_{BS}}{P_{th}}\big) ^ {\frac{1}{\alpha_{L}}} \text{ w.p.} \big(1 - \frac{\theta_{BS}}{2\pi}\big)
   \end{cases}
\end{equation*}

\noindent \textbf{\textit{d}CSR}: In this case, $\theta_{cs} = \theta_{UE}$, and the sensing node is a UE.
\begin{equation*}
R_{cs,L} =  
    \begin{cases}
         \big(\frac{P_X C_{L} M_{UE} M_{BS}}{P_{th}}\big) ^ {\frac{1}{\alpha_{L}}} \text{ w.p. } \frac{\theta_{BS}}{2\pi} \\[-7pt]
         \big(\frac{P_X C_{L} M_{UE} m_{BS}}{P_{th}}\big) ^ {\frac{1}{\alpha_{L}}} \text{ w.p. }  \big(1 - \frac{\theta_{BS}}{2\pi}\big)
    \end{cases}
\text{; } r_{cs,L} =  
    \begin{cases}
         \big(\frac{P_X C_{L} m_{UE} M_{BS}}{P_{th}}\big) ^ {\frac{1}{\alpha_{L}}} \text{ w.p. } \frac{\theta_{BS}}{2\pi} \\[-7pt]
         \big(\frac{P_X C_{L} m_{UE} m_{BS}}{P_{th}}\big) ^ {\frac{1}{\alpha_{L}}} \text{ w.p. }  \big(1 - \frac{\theta_{BS}}{2\pi}\big)
    \end{cases}
\end{equation*}
Finally, for \textit{d}CSRA, the announcements are sent out omnidirectionally and listened by the BSs in a directional way. Thus,
\begin{equation*}
    R_{A,L} = \Big(\frac{P_{U} C_L M_{BS} M_{UE} \times 10^{-0.7}}{P_{th}^{A}}\Big)^{\frac{1}{\alpha_{L}}} ; \hspace{0.1in} r_{A,L} = \Big(\frac{P_{U} C_L m_{BS} M_{UE} \times 10^{-0.7} }{P_{th}^{A}}\Big)^{\frac{1}{\alpha_{L}}}
\end{equation*}
The expressions for $R_{cs,N}$, $r_{cs,N}$, $R_{A,N}$, and $r_{A,N}$ are the same as $R_{cs,L}$, $r_{cs,L}$, $R_{A,L}$ and $r_{A,L}$, respectively, with $C_L$ and $\alpha_L$ replaced by $C_N$  and $\alpha_N$, respectively.

\bibliographystyle{IEEEtran}
\bibliography{References}

\end{document}